\documentclass[useAMS,usenatbib]{mn2e}

\usepackage{epsfig}
\usepackage{myaasmacros}
\usepackage{amsmath}
\usepackage{amsfonts}
\usepackage{amssymb}
\usepackage{subfig}
\usepackage{comment}
\usepackage{color}

\title[The origin of population gradients] 
{The stellar accretion origin of stellar population gradients in
  massive galaxies at large radii}  
\author[Hirschmann et al.]{Michaela Hirschmann$^{1,2}$\thanks{E-mail:
hirschma@iap.fr}, Thorsten Naab$^{3}$,  Jeremiah P. Ostriker$^{4,5}$,
\newauthor Duncan A. Forbes$^{6}$, Pierre-Alain Duc$^{7}$, Romeel
Dav\'{e}$^{8,9,10}$, Ludwig Oser$^{3,5}$, \newauthor Emin Karabal$^{11,7}$ 
 \\
$^{1}$UPMC-CNRS, UMR7095, Institut d'~Astrophysique de Paris, F-75014
Paris, France\\ 
$^{2}$INAF-Osservatorio astronomico di Trieste, Via Tiepolo 11,
I-34131 Trieste, Italy\\
$^{3}$Max-Plank-Institut f\"ur Astrophysik, Karl-Schwarzschild
Strasse 1, D-85740 Garching, Germany\\
$^{4}$Department of Astrophysical Sciences, Princeton University,
Princeton, NJ 08544, USA\\ 
$^{5}$Department of Astronomy, Columbia University,
New York, NY 10027, USA\\ 
$^{6}$Centre for Astrophysics \& Supercomputing, Swinburne University,
Hawthorn, VIC 3122, Australia\\ 
$^{7}$Laboratoire AIM Paris-Saclay, CEA/IRFU/SAp, CNRS/INSU, 
UniversitÃ© Paris Diderot, 91191 Gif-sur-Yvette Cedex, France \\
$^8$University of the Western Cape, Bellville, Cape Town 7535, South Africa\\
$^9$ South African Astronomical Observatories, Observatory, Cape Town 7925, South Africa\\
$^{10}$African Institute for Mathematical Sciences, Muizenberg, Cape Town 7945, South Africa\\
$^{11}$ESO, Karl-Schwarzschild-Str. 2, D-85740 Garching, Germany\\
}

\begin{document}

\date{Accepted ???. Received ??? in original form ???}

\pagerange{\pageref{firstpage}--\pageref{lastpage}} \pubyear{2002}

\maketitle

\label{firstpage}

\begin{abstract}
We investigate the evolution of stellar population gradients from $z
=2$ to $z=0$ in massive galaxies at large radii ($r > 2
R_{\mathrm{eff}}$) using ten cosmological zoom simulations of halos
with  $6 \times 10^{12} M_{\odot} <  M_{\mathrm{halo}} < 2 \times
10^{13}M_{\odot}$. The simulations follow metal cooling and enrichment
from SNII, SNIa and AGB winds. We explore the differential impact of
an empirical model for galactic winds that reproduces the
mass-metallicity relation and its evolution with redshift. At larger 
radii the galaxies, for both models, become more dominated by stars
accreted from satellite galaxies in major and minor mergers. In the
wind model, fewer stars are accreted, but they are significantly more
metal poor resulting in steep global metallicity ($\langle \nabla
Z_{\mathrm{stars}} \rangle= -0.35$~dex/dex) and color (e.g. $\langle \nabla
g-r \rangle = -0.13$~dex/dex) gradients in agreement with
observations. In contrast, colour and metallicity gradients of the models
without winds are inconsistent with observations. Age gradients are in general
mildly positive at $z=0$ ($\langle \nabla Age_{\mathrm{stars}} \rangle=
0.04$~dex/dex) with significant differences between the models at
higher redshift. We demonstrate that for the wind model, stellar
accretion is steepening existing in-situ metallicity gradients by
about 0.2~dex by the present day and helps to match observed gradients
of massive early-type galaxies at large radii. Colour and metallicity
gradients are significantly steeper for systems which have accreted
stars in minor mergers, while galaxies with major mergers have
relatively flat gradients, confirming previous results. The effect of
stellar migration of in-situ formed stars to large radii is
discussed. This study highlights the importance of stellar accretion
for stellar population properties of massive galaxies at large radii,
which can provide important constraints for formation models.  
\end{abstract}


%*****************************************************************************************************
%*****************************************************************************************************
\section{Introduction}\label{intro}
%*****************************************************************************************************
%*****************************************************************************************************

Metallicity gradients in galaxies contain important informations.
It is a natural prediction of modern hierarchical cosmological models
that the assembly of massive galaxies involves major and minor mergers
although most stars in most galaxies have been made in-situ from
accreted or recycled gas. Nonetheless, these mergers are expected to
play a significant role for the structural and morphological evolution
of the massive early-type galaxy population
(e.g. \citealp{1996MNRAS.283L.117K, 1996MNRAS.281..487K,
  2006MNRAS.366..499D, 2006ApJ...648L..21K, 2007MNRAS.375....2D,
  2008MNRAS.384....2G, 2009ApJS..182..216K, 2010ApJ...715..202H,
  Oser10, 2013arXiv1311.0284N}). During mergers, gas and stars that
have formed in other, typically smaller galaxies are added to the main
galaxies, their stellar populations are mixed and new stars can form,
predominantly in the central regions. In the absence of a significant
cold gas component, the mixing of the stellar populations is entirely
determined by stellar dynamics. 

Two important structural galaxy properties, which are thought to be
strongly influenced by mergers, are the (in general negative)
abundance and colour gradients observed early-on in
massive, present-day elliptical (e.g. \citealp{Vaucouleurs61,
  McClure69, Franx90, Peletier90, Davies93, Carollo93, Mehlert03,
  Wu05, LaBarbera05, Annibali07, Rawle10, Eigenthaler13}), but also in
disk galaxies (e.g. \citealp{Wyse89, Vila92, Zaritsky94, vanZee98,
  MacArthur04}), typically within $1 R_{\mathrm{eff}}$. 

Thanks to improved and more elaborated observational techniques,
present-day metallicity and colour gradients can nowadays be measured
out to much larger radii (beyond $1  R_{\mathrm{eff}}$), occasionally out to
even $8 R_{\mathrm{eff}}$, (e.g. \citealp{Sanchez07,LaBarbera09,
  Foster09, Weijmans09, Spolaor10, Coccato10, LaBarbera12,
  Greene12, Greene13, Mihos13, Pastorello14, Raskutti14, DSouza14},
Duc et al. 2014)
and future observational surveys will allow observations at higher
redshifts.  For the gaseous component, observations, interestingly,
suggest an inversion at high redshifts ($z \sim 2$), i.e. positive
\textit{gaseous} metallicity gradients, which could be caused by
gas dilution due to cold gas infall or on-going mergers
(\citealp{Cresci10, Queyrel12, Jones13}).   

Theoretically, the evolution of such abundance gradients has been in
the focus already of the very first numerical N-Body simulations of
mergers of spheroidal galaxies later on also focusing on shapes and
kinematics, scaling relations, sizes and dark  matter fractions
\citep{1978MNRAS.184..185W, 1979MNRAS.189..831W, 1997ApJ...481...83M,
  2005MNRAS.362..184B, 2006ApJ...636L..81N, 2006MNRAS.369.1081B,
  2008MNRAS.383...93B, 2009A&A...501L...9D, 2009ApJ...703.1531N,
  2012MNRAS.422.1714N}.   
Already the simulations by \citet{1978MNRAS.184..185W,
  1980MNRAS.191P...1W} clearly indicated significant mixing and severely
flattened population gradients in equal mass (or major) merger events
(see also \citealp{2004MNRAS.347..740K} including a more detailed
treatment for chemical enrichment and the more recent studies by
\citealp{Rupke10} and \citealp{Navarro13}). \citet{DiMatteo09}, for
example, find with such simulations a flattening of the metallicity
gradient depending on the strength of the initial gradients (before
the merger). It has also been predicted early-on that minor mergers
show a different mixing behaviour. The stars of the merging galaxies
are predominantly added at large radii and - assuming lower
metallicity for the satellites - lead to abundance (colour) gradients
at large radii \citep{1983MNRAS.204..219V}.     

In the same spirit, \citet{2012MNRAS.425.3119H, 2013MNRAS.429.2924H}
have recently re-investigated in detail the collisionless dynamics of
major and minor mergers of spheroidal galaxies embedded in massive
extended dark matter halos in direct comparison to analytic predictions
\citep{Cole00, 2009ApJ...699L.178N, 2009ApJ...697.1290B}. They confirm
the previous findings and also demonstrate how violent relaxation in
major mergers mixes dark matter into the  central regions
\citep{2005MNRAS.362..184B, 2006MNRAS.369.1081B}.
However, in minor mergers with mass-ratios of 1:5 to 1:10, violent
relaxation is less important and the influence of a dark matter
component is significant. Stars of the merging satellites are
stripped and added to the host system at even larger radii dominated
by dark matter (see also \citealp{2012MNRAS.424..747L,
  2013MNRAS.428..641O, Laporte13, Cooper13}) . As a result, the
effective radii of the galaxies as well as the dark matter fractions
grow more rapidly  and the Sersic index of the surface density
profiles increase significantly \citep{2013MNRAS.429.2924H}. 

The above experiments result in a clear prediction. Early-type
galaxies with major collisionless mergers dominating the formation
history have flat population gradients, galaxies with a significant
contribution from minor mergers (and no subsequent major merger) have
significant population gradients at large radii, where the stars of
the accreted satellites assemble  - a process promoted by dark
matter. These considerations are by no means purely 
theoretical as 'dry' mergers of gas poor early-type galaxies have been
directly observed up to high redshift \citep{2005AJ....130.2647V,
  2005ApJ...627L..25T, 2006ApJ...640..241B,
  2008ApJ...672..177L, 2009ApJ...697.1971J,   2012ApJ...746..162N,
  2012ApJ...744...85M}.

Hierarchical cosmological models in comparison with observations
provide more realistic models for the assembly of massive galaxies -
beyond a simple sequence of binary mergers. Direct observations
indicate that a significant fraction of massive early-type galaxies
($\sim 10^{11} 
M_{\odot}$) were more compact in the past  \citep{2005ApJ...626..680D,
  2005ApJ...631..145V, 2006ApJ...650...18T, 2007MNRAS.374..614L, 
2007ApJ...671..285T, 2008ApJ...687L..61B, 2008ApJ...677L...5V,
2008ApJ...688...48V, 2008A&A...482...21C, 2008ApJ...688..770F,
2009ApJ...695..101D, 2009ApJ...696L..43C,2009ApJ...697.1290B, 
2010ApJ...709.1018V, 2011ApJ...736L...9V} and have predominantly grown
in mass and size by the assembly (and not the formation) of stellar
mass at larger radii \citep{2009MNRAS.398..898H, 2010ApJ...709.1018V,
  2012ApJ...749..121S, 2012MNRAS.422.3107S, Patel13}. 

A plausible formation scenario bears a two-phase characteristic
\citep{2007MNRAS.375....2D, Naab07,2008MNRAS.384....2G, 2008ApJ...688..789G,
  2010ApJ...709..218F, Oser10, 2011ApJ...736...88F, Johansson12,
  2013MNRAS.436.2929H}. The early formation of (compact) massive
galaxies is dominated by dissipative processes (i.e. significant
radiative energy losses) and in-situ star formation. Instead, towards
lower redshifts in-situ star formation becomes less important and the
galaxies grow in mass and size by the accretion of stars formed in
other galaxies and added in major and minor mergers with dissipation
becoming less and less important. 
Such a scenario is  supported by direct numerical simulations
\citep{2009ApJ...697L..38J, 2010ApJ...712...88L, 2012MNRAS.424..747L,
  Oser12, 2012MNRAS.425..641L, Gabor12}, semi-analytical
models \citep{2006MNRAS.366..499D, 2007MNRAS.375....2D,
  2008MNRAS.384....2G, Hirschmann12}  and independent estimates from
abundance matching techniques \citep{2013MNRAS.428.3121M,
  2013ApJ...770...57B, Yang13}. All models predict that the accretion
of stars is more important for massive galaxies than for less massive
ones. The amount of accreted stars for massive galaxies hosted in
massive halos $M_{\mathrm{vir}} \sim  10^{13}M_{\odot}$ varies,
however, from about 40\% to 60\%. 

Instead, for lower mass (e.g. Milky-Way-like) galaxies neither major
nor minor mergers dominate the stars found in the bulk of the
systems at any time and dissipative processes are still important at
low redshifts. It is generally thought that roughly $\sim 95$~\% of
the stars in a Milky-Way-like galaxy are formed in-situ from infalling 
gas. This particularly implies that (in-situ formed) stellar
metallicity gradients of disk dominated galaxies originate from a
different process than discussed so far (not the accretion of stars, see
e.g. \citealp{Tortora10}): by the continuous infall of metal-poor gas
with higher and higher angular momentum towards low redshift which can
be then turned into stars at larger and larger radii (inside-out
formation).. Various models (chemical evolution and hydrodynamical
simulations) can predict gradients in disk galaxies
(e.g. \citealp{Matteucci89, Steinmetz94}) and all models indicate that 
with evolving time the gradients become flatter as the disks
grow (e.g. \citealp{Molla97, Prantzos98, Chiappini01, Molla05, Naab06,
  Spitoni11, Pilkington12, Fu13}).      

Apart from the stellar mass dependence (and from galaxy-to-galaxy
variations), the accreted and in-situ formed stellar fractions also
strongly depend on the exact galaxy formation model, and it has been
shown that physical processes like feedback from supernovae, young
massive stars and AGN can significantly influence the relative
importance of in-situ formation versus accretion of stars
\citep{2012MNRAS.425..641L, Hirschmann12, 2013MNRAS.436.2929H,
  Dubois13}. Therefore, such processes most likely also affect the
evolution of in-situ formed metallicity gradients and the relevance of
the contribution of accreted stars (and thus, the overall
gradient). \citet{Pilkington12} have, for example, investigated the
evolution of metallicity gradients in disk galaxies using
hydrodynamical simulations and find that different sub-grid and
feedback schemes can significantly alter the in-situ formed metallicity
gradients, while the merger history has only a minor effect.

As stated earlier -- the origin of metallicity and colour
gradients of \textit{disk} galaxies was investigated in detail by 
observations and chemical evolution models or cosmological
simulations. Even if very basic, the formation and evolution of
metallicity gradients of early-type galaxies is,however, less well
explored in literature, despite some recent and on-going progress
from the observational side (e.g. \citealp{Spolaor10, LaBarbera09,
  Tortora10, Greene12, LaBarbera12, Greene13, Raskutti14,
  Pastorello14}, Duc et al. 2014, in prep.). Previous studies mainly
focused on \textit{inner} stellar population gradients (up to
3~$R_{\mathrm{eff}}$) at relatively low resolution (softening length
$\gtrsim 1$~kpc in a cosmological set-up) either performing isolated
merger simulations (\citealp{DiMatteo09, Rupke10}) or cosmological
simulations (\citealp{Kawata03, 2004MNRAS.347..740K, Navarro13}). They 
all agree on a flattening of the gradients due to major mergers (depending
on the exact mass-ratio though). \citet{2004MNRAS.347..740K}
additionally state that inner gradients can be regenerated when strong
central star formation is induced. \citet{Navarro13} mention that
minor mergers (accretion of stellar systems) bring in older and less
metallic stars which can steepen metallicity gradients. But they
investigate only inner gradients (up to 2~$R_{\mathrm{eff}}$) at
comparably poor spatial resolution (2.7~kpc). 

Therefore, in this work, we focus on the stellar accretion
origin of metallicity and colour gradients in simulated massive
galaxies \textit{at large radii} ($2R_{\mathrm{eff}} < r <
6R_{\mathrm{eff}}$) in a full cosmological context. We particularly
intend to explore the combined effect of strong galactic winds and of
the individual merger and accretion histories on the in-situ formed
and accreted stellar fractions and on the steepening/flattening of the
metallicity, age and colour gradients at these large radii. We
explicitly distinguish between metallicity gradients formed by in-situ
star formation and by accretion of stars. Our results give a possible
explanation for the origin and evolution of abundance gradients of
massive galaxies in hierarchical cosmological galaxy formation
models. In addition, they provide theoretical predictions for future
observational surveys measuring abundance gradients at such large
radii and/or at high redshifts.  

For this analysis, we consider the 10 most massive high-resolution,
cosmological zoom simulations (covering a mass range of  $6\times
10^{12} <  M_{\mathrm{halo}} < 2 \times 10^{13} M_\odot$) presented in
\citet{2013MNRAS.436.2929H} including a treatment for metal enrichment 
(SNII, SNIa and AGB stars) and a phenomenological feedback scheme for
galactic winds.

These cosmological zoom simulations were shown to be successful in
suppressing early star formation at $z>1$, in predicting reasonable
star formation histories for galaxies in present day halos of $\sim
10^{12} M_\odot$, in producing galaxies with high cold gas fractions
(30 - 60 per cent) at high redshift, and in significantly reducing the
baryon conversion efficiencies for halos ($M_{\mathrm{halo}} <
10^{12}M_\odot$) at all redshifts in overall good agreement with
observational constraints. Due to the delayed  onset of star formation
in the wind models, the metal enrichment of  gas and stars is delayed
and is also found to agree well with observational constraints. In
addition, galactic winds increase the  ratio of in-situ formed to
accreted stars -- the relative importance of dissipative
vs. dissipationless assembly. These cosmological zoom simulations
provide, therefore, a good basis for our analysis of the spatial
distribution of metallicity and colours in massive galaxies.   

The paper is organised as follows. Section \ref{model} provides an
introduction into our simulation set-up and the construction,
quantification and visualisation of the merger histories. In section
\ref{massmet} we show the redshift evolution of the mass-metallicity
and the mass-age relations. We further show the evolution of the
(outer) metallicity and age gradients in Sections \ref{metgradients}
and \ref{agegradients}, distinguishing between the metallicity and the 
stellar age of the accreted and in-situ formed stellar component and
discussing the effect of galactic winds. Using stellar population
models, we investigate the evolution of color gradients in Section
\ref{colorgradients}. Section \ref{Mergerhistory} provides a discussion
about the effect of the merger history on the individual gradients and
section \ref{Discussion} also a comparison of our simulated
gradients with currently available observational data. A final summary
of this work is given in Section \ref{summary}.

%*****************************************************************************************************
%*****************************************************************************************************
\section{High-resolution simulations of individual galaxy halos}\label{model} 
%*****************************************************************************************************
%*****************************************************************************************************

%*****************************************************************************************************
\subsection{Simulation setup}\label{setup}
%*****************************************************************************************************

For this analysis, we have selected the ten most massive halos from
the high-resolution, cosmological zoom simulation set presented in
\citet{2013MNRAS.436.2929H} (M0163, M0209, M0215, M0224, M0227, M0259, 
M0290, M0300, M0305, M0329). They are based on the initial conditions
described in detail in \citet{Oser10, Oser12}. We briefly review the
simulation setup here, but refer the reader to the original papers for
more details.  

The dark matter halos for further refinement were selected from a dark
matter only N-body simulation with a co-moving periodic box length of
$L=100\ \mathrm{Mpc}$ and $512^3$ particles (\citealp{Moster10}). We
assume a $\Lambda$CDM cosmology based on the WMAP3 measurements (see  
e.g. \citealp{Spergel03}) with $\sigma_8 = 0.77$, $\Omega_{m}=0.26$,
$\Omega_{\Lambda}=0.74$, and $h=H_0/(100\
\mathrm{kms}^{-1})=0.72$. The simulation was started at $z=43$ and run
to $z=0$ with a fixed co-moving softening length of $2.52\ h^{-1} 
\mathrm{kpc}$ and a dark matter particle mass of $M_{\mathrm{DM}} = 2
\times 10^8 M_{\odot}\ h^{-1}$. 
From this simulation, we picked different halos identified with the
halo finder algorithm $FOF$ at $z=0$. To construct the high-resolution
initial conditions for the re-simulations, we traced back in time all
particles closer than $2 \times r_{200}$ to the centre of the halo in
any snapshot and  replaced them with dark matter as well as gas
particles at higher resolution ($\Omega_b=0.044,
\Omega_{DM}=0.216$). In the high resolution region the dark matter
particles have a mass resolution of $m_{\mathrm{dm}} = 2.5\cdot 10^7
M_{\odot}h^{-1}$, which is 8 times higher than in the original
simulation, and the gas particle masses are $m_{\mathrm{gas}} =
m_{\mathrm{star}} = 4.2\cdot 10^6 M_{\odot}h^{-1}$. The co-moving
gravitational softening length for the gas and star particles is $400\
h^{-1} \mathrm{pc}$ and $890\ h^{-1} \mathrm{pc}$ for the
high-resolution dark matter particles.  

\begin{figure*}
\centering
\epsfig{file=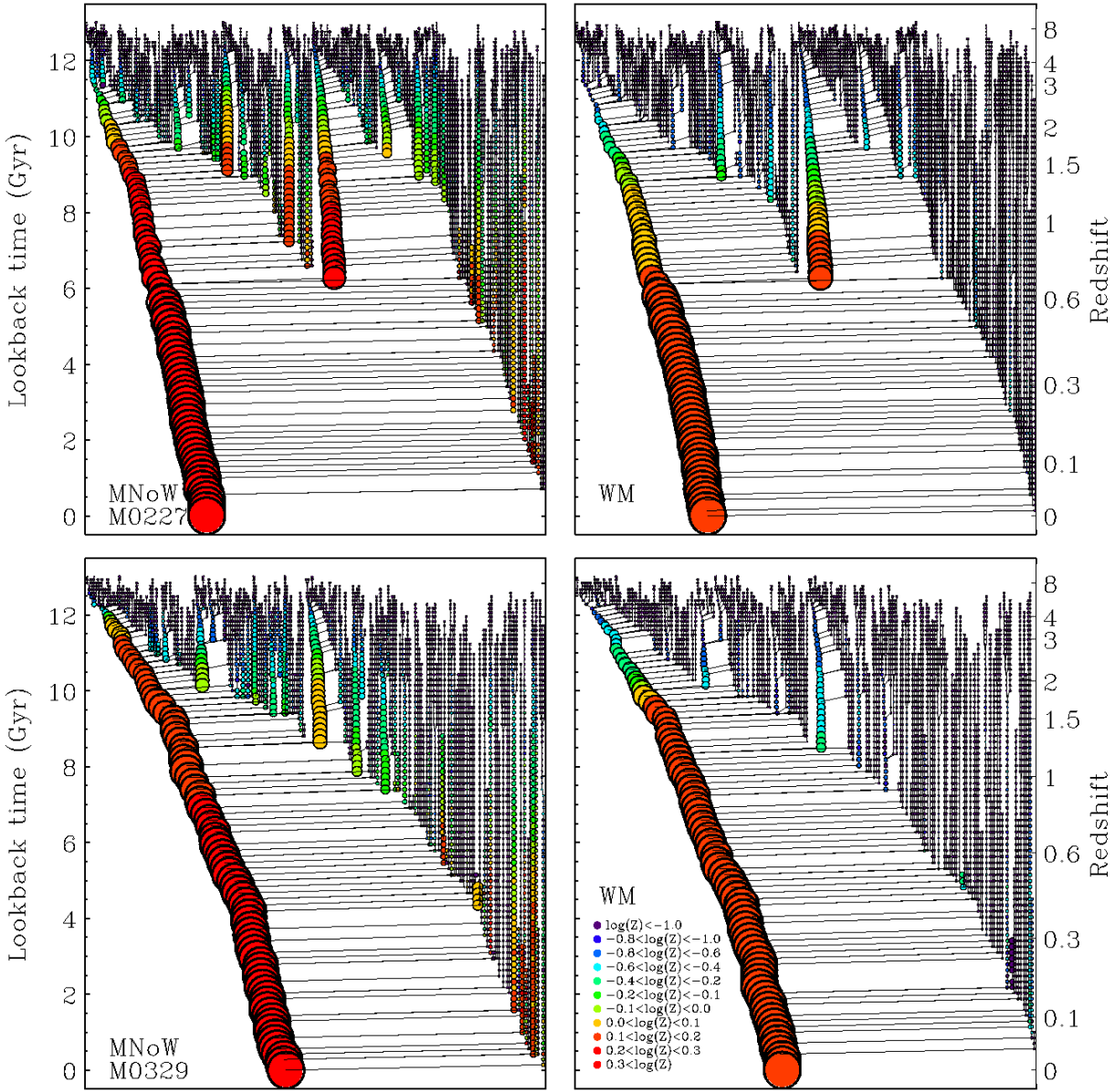, width=0.95\textwidth}
 \caption{Halo-based merger trees for the two example galaxies M0227
   (top panels) and M0329 (bottom panels) for the model without winds
   (MNoW, left column) and with winds (WM, right column). The size of
   the circles is scaled to the mass of the central galaxy in the
   respective halos, the colour is coded by the metallicity of the
   stellar component from blue (low metallicity) to red (high
   metallicity). In the no wind models (MNoW, left panels) the
   satellite halos host more massive galaxies with more metal rich
   stellar components, in particular at low redshifts. M0227 has a
   major merger with galaxy of comparable metallicity at $z \sim
 0.7$, whereas M0329 grows by minor mergers alone. The infalling
 systems are particularly metal poor in the wind simulations (WM,
 bottom right panel). Not all galaxies of the halos indicated here
 merge with the central galaxy by $z=0$ (see Fig. \ref{Minmajmerg_evol}).}  
 {\label{TreeVis}}
\end{figure*}

\begin{table*}
\centering
\begin{tabular}{ | p{0.55cm} || p{0.35cm} p{0.35cm} p{0.35cm} p{0.35cm} p{0.3cm} p{0.3cm}
    p{0.55cm} p{0.6cm} p{0.55cm} p{0.6cm} p{0.55cm} p{0.6cm} p{0.55cm}
    p{0.6cm} p{0.55cm} p{0.6cm} p{0.55cm} p{0.6cm} |} 
\hline \multicolumn{18}{c}{ {\bf{MNoW}} }\\ \hline
{\bf{ID}} & $M_{\mathrm{vir}}$ & $M_{\mathrm{*}}$ & $M_{\mathrm{ins}}$ &
$R_{\mathrm{eff}}$ & $N_{\mathrm{maj}}$ & $N_{\mathrm{min}}$ &
\multicolumn{2}{c}{$\nabla (Z_*,z0)$} & 
\multicolumn{2}{c}{$\nabla (Z_*,z1)$} & 
\multicolumn{2}{c}{$\nabla (Z_*,z2)$} &  
\multicolumn{2}{c}{$\nabla (Age_*,z0)$} & 
\multicolumn{2}{c}{$\nabla (Age_*,z1)$} & 
\multicolumn{2}{c}{$\nabla (Age_*,z2)$}\\ \vspace{-0.5cm}
 & \multicolumn{3}{c}{ {\scriptsize $10^{10} $} } & {\scriptsize kpc}
 & & &
 {\scriptsize{1e-1}} &  {\scriptsize{1e-2}}  & 
 {\scriptsize{1e-1}} &  {\scriptsize{1e-2}}  & 
  {\scriptsize{1e-1}} &  {\scriptsize{1e-2}}  & 
{\scriptsize{1e-2}} & {\scriptsize{1e-3}} &
{\scriptsize{1e-2}} & {\scriptsize{1e-3}} &
{\scriptsize{1e-2}} & {\scriptsize{1e-3}} \\
 & \multicolumn{3}{c}{ {\scriptsize $h^{-1} M_\odot$} } & {\scriptsize
   $h^{-1}$} & & & 
 {\scriptsize{dex/dex}} & {\scriptsize{dex/kpc}} &
 {\scriptsize{dex/dex}} & {\scriptsize{dex/kpc}} &
 {\scriptsize{dex/dex}} & {\scriptsize{dex/kpc}} &
{\scriptsize{dex/dex}} & {\scriptsize{dex/kpc}} &
{\scriptsize{dex/dex}} & {\scriptsize{dex/kpc}} &
{\scriptsize{dex/dex}} & {\scriptsize{dex/kpc}} \\
\hline \hline 
{ M0163} & 914.6 & 35.81 & 9.70 & 6.0 & 1 & 29 & $-1.1$ & $-0.1$ & $-3.1$&
$-2.4$& $-8.9$& $-6.9$& $+2.6$& $+1.3$& $-7.3$& $-3.4$& $-28$& $-26$\\ 

{ M0209} & 678.6 & 20.74 & 9.33 & 3.1 & 2 & 13 & $-1.8$ & $-0.5$ & $-1.7$&
$-1.4$& $-3.4$& $-3.9$& $+0.03$& $+0.07$& $-3.6$& $-2.5$& $-5.7$& $-6.1$\\ 

{ M0215} & 659.2 & 28.20 & 15.50 & 4.0 & 0 & 19 & $-0.5$ & $-0.1$ & $-2.1$&
$-1.6$& $-1.6$& $-1.6$& $+3.8$& $+0.8$& $-11$& $-7.2$& $+7.4$& $+8.9$\\ 

{ M0224} & 621.4 & 22.52 & 6.76 & 4.3 & 2 & 29 & $+0.3$ & $+0.0$ & $-4.2$&
$-3.2$& $-0.7$& $-0.5$& $+2.3$& $+0.4$& $+2.0$& $+1.1$& $-9.9$& $-13$\\ 

{ M0227} & 700.0 & 30.33 & 7.58 & 4.6 & 1 & 23 & $-1.2$ & $-0.2$ & $-4.5$&
$-2.8$& $-6.2$& $-7.0$& $+4.3$& $+0.7$& $+14$& $+8.7$& $+27$& $+31$\\ 

{ M0259} & 525.2 & 19.95 & 10.98 & 3.0 & 0 & 5 & $-2.5$ & $-0.7$ & $-1.9$&
$-1.3$& $-5.4$& $-5.1$& $+7.0$& $+2.0$& $+13$& $+11$& $+7.9$& $+8.3$\\ 

{ M0290} & 544.4 & 20.03 & 9.01 & 4.1 & 2 & 16 & $-2.1$ & $-0.5$ & $-1.1$&
$-0.8$& $-4.6$& $-4.1$& $+7.9$& $+1.7$& $+9.7$& $+7.1$& $-4.4$& $-6.4$\\ 

{ M0300} & 495.1 & 17.98 & 8.99 & 4.9 & 2 & 10 & $-0.9$ & $-0.1$& $+0.7$&
$-0.4$& $-2.7$& $-3.2$& $+3.2$& $+0.6$& $+9.2$& $+7.0$& $+5.2$& $+7.2$\\ 

{ M0305} & 463.9 & 14.69 & 10.22 & 4.8 & 1 & 8 & $-0.2$ & $+0.0$ & $-1.5$&
$-1.0$& $-3.5$& $-4.0$& $-1.0$& $-0.1$& $+3.7$& $+2.6$& $-1.9$& $-2.5$\\ 

{ M0329} & 450.4 & 20.74 & 14.56 & 4.6 & 1 & 10 & $-0.8$ & $-0.2$& $-1.2$&
$-0.9$& $-1.8$& $-2.1$& $-0.0$& $-0.0$& $+0.5$& $+0.3$& $+9.7$& $+12$\\  \hline \hline
\multicolumn{7}{c}{\bf Mean gradients} & $-1.1$ & $-0.2$ & $-2.1$ &$-1.8$ &
$-3.6$ &$-3.8$ & $+3.0$ &$+0.7$ & $+3.0$ &$+2.5$ & $+7.3$ &$+13$  \\ \hline
\end{tabular}
\caption{Halo ID, halo virial mass ($M_{\mathrm{vir}}$), total
  ($M_{\mathrm{stellar}}$) and in-situ formed stellar mass
  ($M_{\mathrm{insitu}}$), the effective radius at $z=0$, the
  number of major and minor merger since $z=2$
  ($N_{\mathrm{major/minor}}$) and the slopes for the fitted
  metallicity $\nabla_{l/k}(Z_{\mathrm{stars}}) = \nabla_{l/k}(Z_*)$
  and age gradients $\nabla_{l/k}(Age_{\mathrm{stars}}) =
  \nabla_{l/k}(Age_*)$ at $z=0,1,2$ of the \textit{central} galaxies
  in the MNoW run. All masses are in units of $10^{10} h^{-1}
  M_\odot$, the radii in kpc~$h^{-1}$ and the gradients (fitted
  between $ 2-6 \times R_{\mathrm{eff}} $) are given in dex$/$dex and
  dex$/$kpc.}    
\label{sim_tab_MNoW}
\end{table*}

\begin{table*}
\centering
\begin{tabular}{ | p{0.55cm} || p{0.35cm} p{0.35cm} p{0.35cm}
    p{0.35cm} p{0.3cm} p{0.3cm} p{0.55cm} p{0.6cm} p{0.55cm} p{0.6cm}
    p{0.55cm} p{0.6cm} p{0.55cm} p{0.6cm} p{0.55cm} p{0.6cm} p{0.55cm}
    p{0.6cm} |}  
\hline \multicolumn{18}{c}{ {\bf{WM}} }\\ \hline
{\bf{ID}} & $M_{\mathrm{vir}}$ & $M_{\mathrm{*}}$ & $M_{\mathrm{ins}}$
& $R_{\mathrm{eff}}$ & $N_{\mathrm{maj}}$ & $N_{\mathrm{min}}$ &  
\multicolumn{2}{c}{$\nabla (Z_*,z0)$} & 
\multicolumn{2}{c}{$\nabla (Z_*,z1)$} & 
\multicolumn{2}{c}{$\nabla (Z_*,z2)$} &  
\multicolumn{2}{c}{$\nabla (Age_*,z0)$} & 
\multicolumn{2}{c}{$\nabla (Age_*,z1)$} & 
\multicolumn{2}{c}{$\nabla (Age_*,z2)$}\\ \vspace{-0.5cm} 
 & \multicolumn{3}{c}{ {\scriptsize $10^{10} $} } & {\scriptsize  kpc}
 & & & 
{\scriptsize{1e-1}} & {\scriptsize{1e-2}} &
{\scriptsize{1e-1}} & {\scriptsize{1e-2}} &
{\scriptsize{1e-1}} & {\scriptsize{1e-2}} &
{\scriptsize{1e-2}} & {\scriptsize{1e-3}} &
{\scriptsize{1e-1}} & {\scriptsize{1e-2}} &
{\scriptsize{1e-1}} & {\scriptsize{1e-2}} \\
 & \multicolumn{3}{c}{ {\scriptsize $h^{-1} M_\odot$} } & {\scriptsize
   $h^{-1}$} & & & 
 {\scriptsize{dex/dex}} & {\scriptsize{dex/kpc}} &
 {\scriptsize{dex/dex}} & {\scriptsize{dex/kpc}} &
 {\scriptsize{dex/dex}} & {\scriptsize{dex/kpc}} &
{\scriptsize{dex/dex}} & {\scriptsize{dex/kpc}} &
{\scriptsize{dex/dex}} & {\scriptsize{dex/kpc}} &
{\scriptsize{dex/dex}} & {\scriptsize{dex/kpc}} \\
\hline \hline 
M0163 & 904.7 & 36.58 & 18.51 & 4.8 & 1 & 24 & $-2.7$ & $-0.5$ & $-3.4$
&$-4.1$ &$-1.8$ &$-1.1$ &$+2.2$&$+0.0$ & $-0.8$ &$-0.5$ & $+3.6$&
$+4.5$\\  
M0209 & 681.3 & 19.67 & 18.07 & 2.6 & 1 & 6  & $-7.6$ & $-2.6$ & $-5.2$
&$-3.3$ &$-4.1$ &$-4.3$ &$+8.7$&$+2.0$ & $+4.7$ &$+3.9$ & $+1.0$&
$+1.9$\\  
M0215 & 662.1 & 26.92 & 21.54 & 3.4 & 0 & 11 & $-4.1 $ & $-1.0 $ & $-1.9$
& $-1.5$ &$-3.6$ &$-2.9$ &$+3.0$&$+5.0$ & $-0.5$ &$-0.2$ &
$+2.9$& $+3.3$\\  
M0224 & 640.3 & 24.56 & 19.16 & 3.1 & 1 & 10 & $-4.6$ & $-1.2 $ & $-4.4$
&$-5.2$ &$-1.6$ &$-0.7$ &$+5.4$&$+8.5$ & $+2.8$ &$+2.0$ & $-0.6$&
$-0.4$\\  
M0227 & 695.4 & 30.90 & 12.67 & 4.1 & 2 & 8  & $-1.4 $ & $-0.3 $ & $-3.8$ 
&$-2.7$ &$-4.0$ &$-3.6$ &$+0.7$&$-2.2$ & $+1.7$ &$+1.2$ &
$+0.6$& $+1.3$\\ 
M0259 & 555.2 & 17.80 & 16.02 & 3.1 & 1 & 2  & $-2.2 $ & $-0.6 $& $-7.7$
&$-1.7$ &$-0.34$ &$-2.2$ &$+7.4$&$+1.4$ & $+5.9$ &$+4.1$ & $+0.5$&
$+0.1$\\  
M0290 & 546.7 & 20.65 & 16.52 & 4.6 & 1 & 7 & $-5.8 $ & $-1.1$ & $-0.3$
&$-0.2$ &$-2.8$ &$-1.4$ &$+14$&$+2.7$ & $-1.7$ &$-0.9$ & $+0.5$&
$+0.1$\\  
M0300 & 504.4 & 17.01 & 7.31 & 3.5 & 2 & 7 & $-1.0 $ & $-0.3 $ & $-8.5$
&$-2.8$ &$-6.2$ &$-5.7$ &$-1.6$&$+1.1$ & $+4.2$ &$+2.9$ & $+4.1$&
$+0.8$\\  
M0305 & 465.6 & 20.37 & 13.85 & 4.0 & 1 & 3 & $-1.3 $ & $-0.3 $& $-4.1$
&$-3.5$ &$-4.1$ &$-4.2$ &$+1.3$&$-0.2$ & $+1.8$ &$+1.4$ & $-0.4$&
$-0.3$\\  
M0329 & 462.0 & 19.98 & 16.98 & 3.6 & 0 & 7 & $-3.9 $ & $-1.0 $ & $-6.0$
&$-4.7$ &$-7.0$ &$-6.1$ &$+1.7$&$+0.1$ & $+1.0$ &$+0.7$ & $+3.5$&
$+0.8$\\  \hline 
\hline
\multicolumn{7}{c}{\bf Mean gradients} & $-3.5$ & $-0.9$ & $-4.5$ & $-3.0$ &
$-3.9$ & $-3.2$ & $+4.3$ & $+1.8$ & $+1.9$ & $+1.5$ & $+1.5$ & $+1.2$ \\ \hline
\end{tabular}
\caption{Halo ID, halo virial mass ($M_{\mathrm{vir}}$), total
  ($M_{\mathrm{stellar}}$) and in-situ formed stellar mass
  ($M_{\mathrm{insitu}}$), the effective radius at $z=0$, the
  number of major and minor merger since $z=2$
  ($N_{\mathrm{major/minor}}$) and the slopes for the fitted
  metallicity $\nabla_{l/k}(Z_{\mathrm{stars}}) = \nabla_{l/k}(Z_*)$
  and age   gradients $\nabla_{l/k}(Age_{\mathrm{stars}}) =
  \nabla_{l/k}(Age_*)$ at $z=0,1,2$  of the\textit{central} galaxies
  in the WM run. All masses are in units of $10^{10} h^{-1} M_\odot$,
  the radii in kpc~$h^{-1}$ and the gradients (fitted between $2-6
  \times R_{\mathrm{eff}} $) are given in dex$/$dex and dex$/$kpc.}   
\label{sim_tab_WM}
\end{table*}

To model the gas component we use the entropy conserving formulation 
of SPH (\textsc{Gadget}-2, \citealp{Springel05}) with the extension of
\citet{Oppenheimer06, Oppenheimer08} including a prescription for
metal enrichment and momentum-driven winds. This version includes
ionisation and heating by a spatially uniform, redshift dependent
background radiation according to \citet{Haardt01}. Gas particles
undergo radiative cooling down to $10^4$K  under the assumption of
ionisation equilibrium; we account for metal-line cooling using the
collisional ionisation equilibrium tables of
\citet{Sutherland93}. 

\begin{figure*}
\begin{center}
\epsfig{file=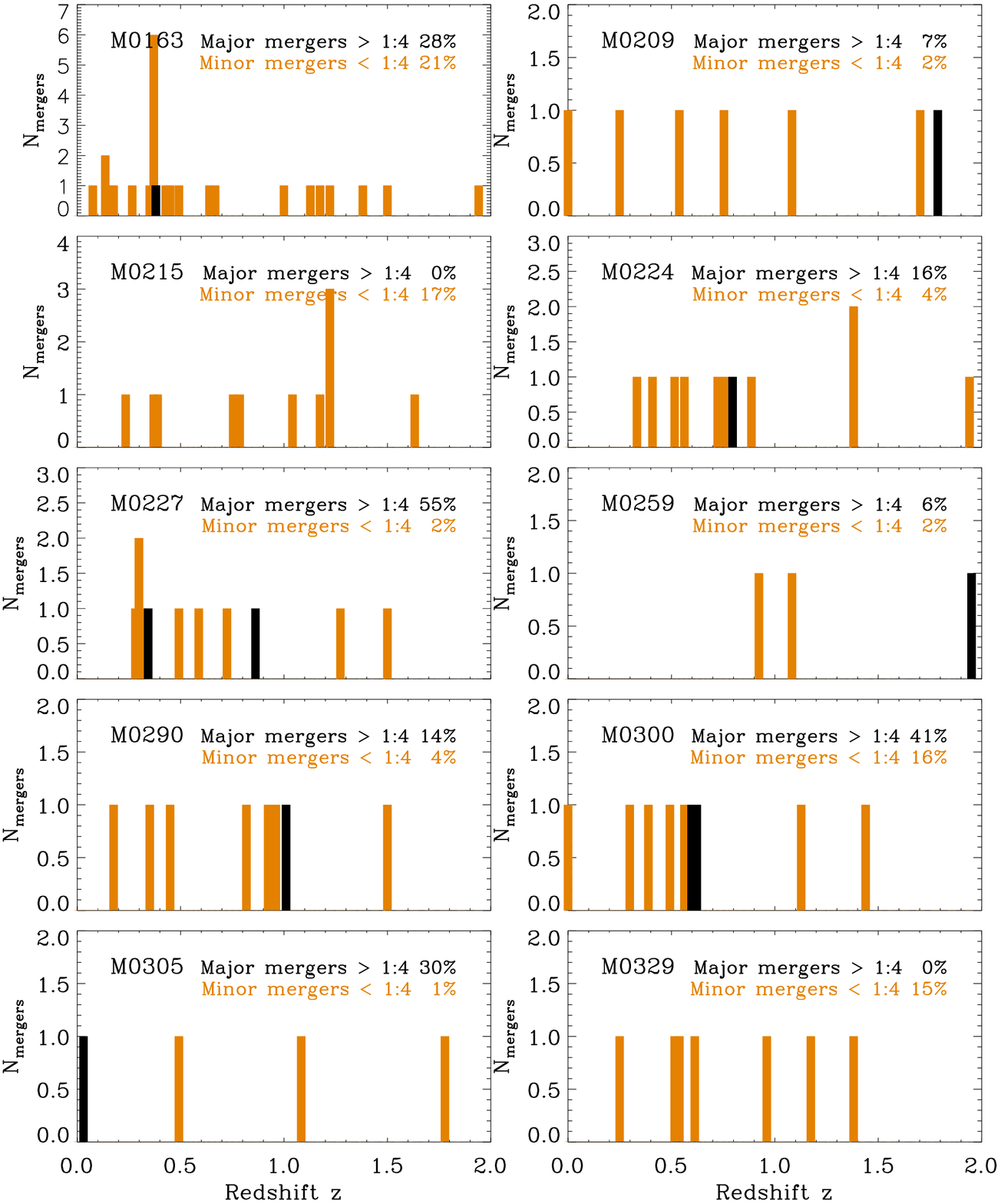, width=0.8\textwidth}
  \caption{\textit{Galaxy} merger histories of the WM simulations
    represented by the number $N_{\mathrm{merger}}$ of major (black,
    mass-ratio $>$ 1:4) and minor (orange, mass-ratio $<$ 1:4) stellar
    mergers, as a function  of redshift for M0163, M0209, M0215,
    M0224, M0227, M0259, M0290, M0300, M030, and M0329 (from top left
    to bottom right). The respective mass fraction of the final stellar galaxy
    mass added in major (black) and minor (orange) mergers is given in
    each panel.} 
 {\label{Minmajmerg_evol}}
\end{center}
\end{figure*}

Following \citet{Springel03}, stars are formed from dense gas clouds
using a sub-resolution multi-phase model which describes condensation
and evaporation in the interstellar medium (\citealp{McKee77}). We
have a density threshold for star formation of $n_{th} = 0.13\
\mathrm{cm}^{-3}$, which is calculated self-consistently in a way that
the equation of state is continuous at the onset of star
formation. Besides, we adopt a \citet{Chabrier03} IMF throughout this
study implying a fraction of stars, which results in type II
supernovae, of $f_{SN} = 0.198$. The model is tuned via a single
parameter, the star formation rate timescale, using simulations of
isolated disk galaxies to reproduce the observed Schmidt-Kennicut
relation. Note that in cosmological simulations the result may,
however, deviate from the observed relation (see \citealp{Hirschmann12}).    

Following \citet{Oppenheimer08}, we account for metal enrichment from
supernovae type II (SNII), type Ia (SNIa) and asymptotic giant branch
(AGB) stars and four elements (C, O, Si, Fe) are tracked individually.
The SNII enrichment follows \citet{Springel03} using an instantaneous
recycling approximation, but is modified by adopting metallicity
dependent yields from the nucleosynthesis calculations by
\citet{Limongi05}. The SNIa rate is modelled following the
two-component parametrisation from \citet{Scannapieco05}, where one
component is proportional to the stellar mass (slow, delayed
component) and the other to the SFRs (rapid component). Besides the
production of metals, each SNIa is assumed to deposit $10^{51}$ ergs of
energy, and this energy is added thermally directly to the
gas particle. AGB stars mainly provide feedback in the form of mass
(energy can be neglected as most mass leaves the AGB stars with
velocities far below $100$ km/s) and produce carbon and oxygen (while
silicon and iron remains almost unprocessed). Nevertheless, AGB stars
may have significant thermal impact (\citealp{Conroy14}) which we do
not include in our model.

The momentum-driven wind model is based on the wind model of
\citet{Springel03}: outflows are directly tied to the star formation
rates using the relation $\dot{M}_{\mathrm{wind}} = \eta
SFR$, where $\eta$ is defined as the mass loading 
factor. Star forming gas particles get stochastically kicked
vertically to the disc and are thus, launched in a wind with the
probability $\eta$. A selected wind particle is given the additional
velocity of $v_w$ in the direction of $\mathbf{v} \times \mathbf{a}$,
where $\mathbf{v}$ and $\mathbf{a}$ are the velocity and acceleration
of a particle, respectively (\citealp{Springel03}). Subsequently, the
gas particles are decoupled from hydrodynamics for a short time in
order to escape their dense, star-forming regions and eventually to
leave their galaxies (see e.g. \citealp{DallaVecchia08}). These
particles are only allowed to again interact hydrodynamically as soon
as they either reach a SPH density less than 10 per cent of the SF
density threshold or the time it takes to travel 30~kpc at the wind
velocity $v_w$. 

The values of $\eta$ and $v_w$ define the wind model: while
\citet{Springel03} used constant values for these parameters,
\citet{Oppenheimer06} adopt a momentum-driven wind model and 
introduce  a scaling with the velocity dispersion of the galaxies
motivated by observations of galactic super-winds of \citet{Martin05,
  Rupke05} and by analytical calculations of \citet{Murray05,
  Zhang12}. The mass-loading factor $\eta$ (i.e. the fraction of
star-forming particles, which get kicked) is calculated according to  
\begin{equation}\label{massload}
\frac{\dot{M}_{\mathrm{wind}}}{SFR} = \eta = \frac{\sigma_0}{\sigma},
\end{equation}
where $\sigma \propto M_{\mathrm{gal}}^{1/3}$ is the velocity
dispersion which is calculated from the galaxy mass using an
on-the-fly group finder. $\sigma_0$ is a constant which is set to
reproduce the overall evolution of the SFR density in
\citet{Oppenheimer06} (here: $\sigma_0 = 300\ \mathrm{km/s}$).

To disentangle the effect of galactic winds on the evolution of
metallicity and colour gradients we study their influence separately by
analysing two sets of simulations with two different models:  
\begin{itemize}
\item{{\bf{MNoW}}: Metal enrichment and metal cooling, but no
    galactic winds}
\item{{\bf{WM}}: Metal enrichment, metal cooling and momentum-driven  
    galactic winds}
\end{itemize}
Note that all simulations include thermal supernova feedback as
described in detail in \citet{Springel03}.
For more details on the simulation set-up, we refer
the reader to \citet{2013MNRAS.436.2929H}. As in this study, the
simulations including winds were shown to predict more realistic
galaxies in several aspects (even if some drawbacks remain as e.g. too
much late star formation as a consequence of missing AGN feedback), we
\textit{a priori} expect the metallicity and colour gradients of the WM
galaxies to be closer to reality.

Note that tables \ref{sim_tab_MNoW} and \ref{sim_tab_WM} summarise
different galaxy properties such as halo mass, total and in-situ formed
stellar mass, effective radii, the number of stellar minor and major
mergers and also the fitted metallicity and age gradients at
$z=0,1,2$ (which will be discussed in the course of this study) of the
10 selected halos in the MNoW and the WM re-simulations, respectively.

%*****************************************************************************************************
\subsection{Merger histories}\label{merger}
%*****************************************************************************************************

We extract the merger trees for the dark matter component directly
from the cosmological re-simulations as described in
\citet{Hirschmann10}. For every snapshot at a given redshift, we first  
identify individual dark matter haloes using a FOF
(Friends-of-Friends) algorithm with a linking length of $b=0.2$ ($ 
\approx 28 \mathrm{kpc}$, \citealp{Davis85}). In a second step we 
extract the subhalos of every FOF group using the \textsc{Subfind}
algorithm \citep{Springel01GAD}. The algorithm to connect the dark
matter halos between the snapshots at different redshifts is described
in detail in \citet{Maulbetsch07} and in \citet{Hirschmann10}. Note
that the tree-algorithm is only applied to the dark matter particles
-- star particles are not separately traced back in time and they are
assumed to follow the evolution of the dark matter. The stellar
particles associated within $1/10$ of the virial radius are defined as
the stellar mass of the central galaxy and these were considered to
calculate the stellar metallicity of the central galaxy.  

In Fig. \ref{TreeVis}, we visualise four merger trees of two different
re-simulated halos with a virial masses of $9.7 \times 10^{12} M_{\odot}$
(M0227, top row) and of $6.3 \times 10^{12} M_\odot$ (M0329, bottom
row) for the MNoW (left column) and the WM model (right column). The
sizes of the black circles approximate the stellar mass within 1/10 of
the dark matter virial radius and they scale with the square root of
mass normalised to the final stellar mass. The stellar metallicity of
these galaxies is colour-coded (from dark-blue, metal-poor to red,
metal-rich, as indicated by the legend). 

Fig. \ref{TreeVis} nicely illustrates that for the two MNoW runs, both
the central and satellite galaxies are enriched faster than in the
corresponding WM runs. In addition, in the MNoW model, more merger 
events occur and the accreted satellite galaxies have typically a
larger stellar mass and are more metal-rich than in the WM
model. Independently of the model, however, the M0329 halo always has a
rather quiet merger history, under-going only minor mergers, while the
M0227 halo experiences two major mergers since $z=2$, one at $z \sim 
1.5$ and one at $z \sim 0.7$. Note that not all galaxies of the halos
indicated here merge with the central galaxy by $z=0$.

As galaxy properties are often more tightly connected with the merger 
events of the \textit{stellar} component, which typically occur with
some time delay \textit{after} the dark matter halo merger, we have
also quantified stellar merger histories (for the central galaxy only)
using the approach of \citet{Oser12}. We identify every satellite
merging with the most massive progenitor of the central galaxy using
FOF run over the \textit{stellar} particles with a minimum number of
20 particles ($\sim 1.2 \times 10^8 M_\odot$) and with a linking
length 1.6~kpc/h (physical). The satellite galaxy is assumed to be
merged with the central galaxy as soon as the FOF algorithm cannot
separate both FOF-galaxies anymore (the satellite is part of the
central FOF group). At $z=2$, all galaxies are more massive than $\sim   
10^{10} M_\odot$, i.e. we resolve mergers at least down to a
mass-ratio of $1:100$. 

\begin{figure*}
\begin{minipage}[b]{1.0\textwidth}
\centering
  \epsfig{file=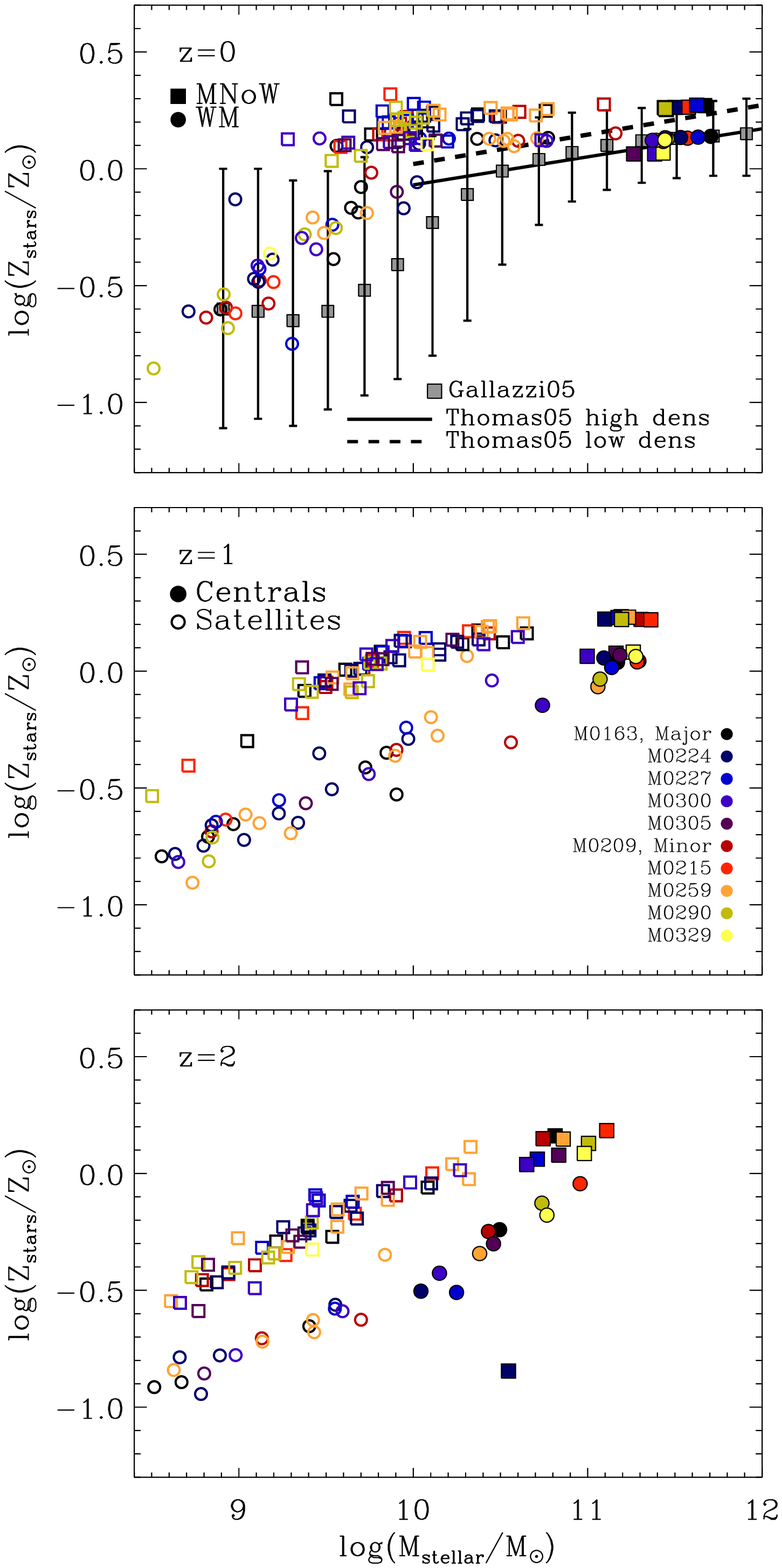, width=0.4\textwidth}
   \epsfig{file=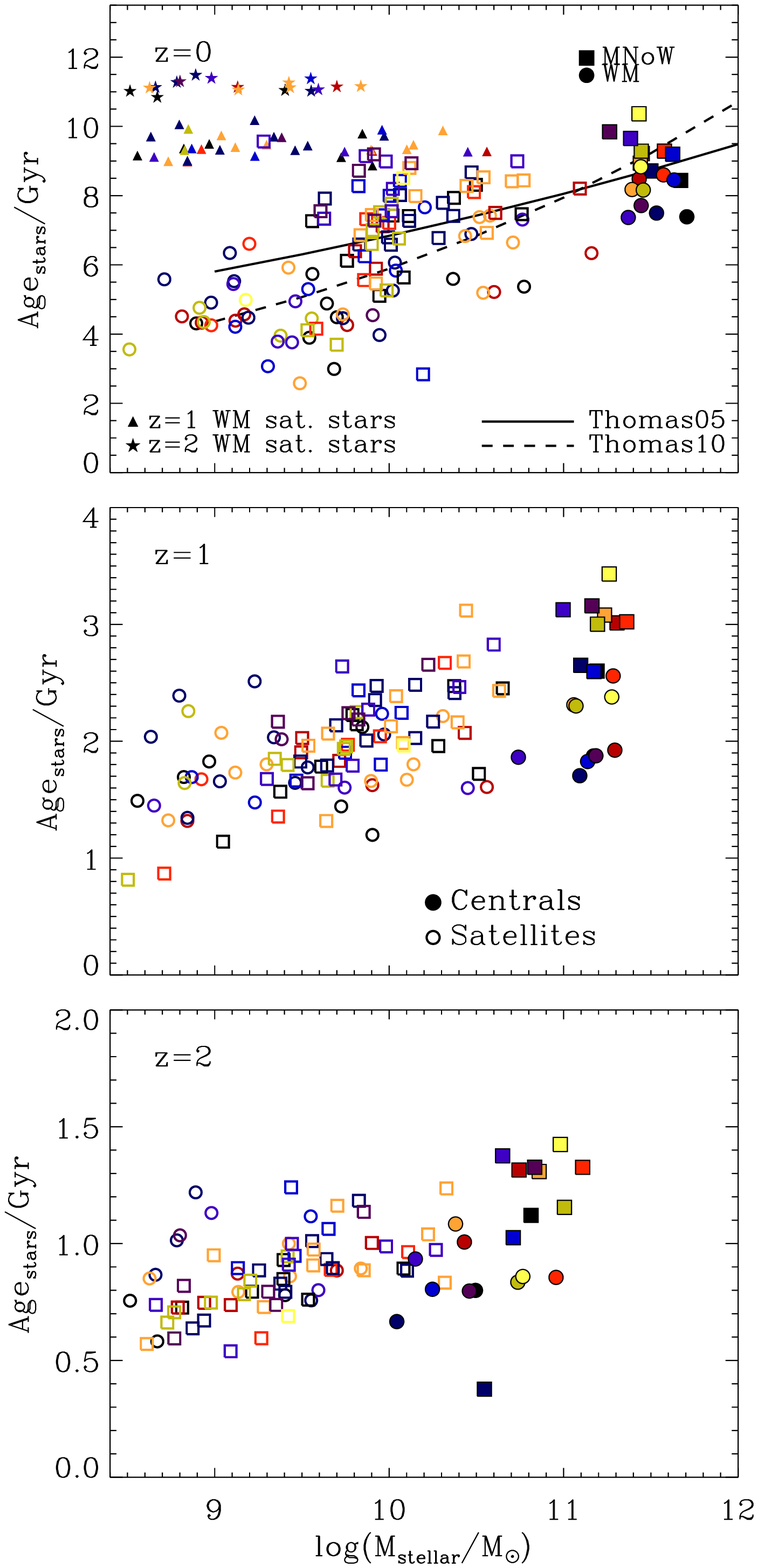, width=0.4\textwidth}
  \caption{\textit{Left column: }Mean stellar metallicity as a
    function of galaxy stellar mass at $z = 0, 1$and $2$ (from top to bottom)
    for ten massive galaxies (different colours) simulated with the
    MNoW model (filled squares) and the WM model (filled circles). The
    satellite galaxies residing in the respective galaxy halos (same
    colours) are indicated by open symbols.  At all redshifts, the
    satellite galaxies in the wind model (WM) have significantly lower
    metallicities than those simulated without winds (MNoW). For this
    model the $z=0$ metallicities agree with observational estimates
    (\citealp{Thomas05}, black lines; \citealp{Gallazzi05}, grey
    squares). 
    \textit{Right column: }Mean stellar age as a function of galaxy mass at $z
    = 0, 1$ and $2$ (from top to bottom) for the ten galaxies and
    their satellites as in the left column. Compared to observations
    (\citealp{Thomas05, Thomas10}), the central galaxies in the WM
    model are slightly too young by roughly 2~Gyrs. For the WM model,
    the small filled triangles and stars indicate the hypothetic
    present-day ages of the satellites at $z=1$ and $z=2$, most of
    them have merged with the central galaxies until $z=0$.}
\label{Mgal_sat_only10}
\end{minipage}
\end{figure*}

Fig. \ref{Minmajmerg_evol} shows the stellar galaxy merger histories
since $z=2$ of all ten WM simulations represented by the number of
major (black, mass-ratio $>$ 1:4) and minor mergers (orange,
mass-ratio $<$ 1:4, but larger than 1:100) as a function of
redshift (see also \citealp{Naab14}). The respective mass fraction of
the final stellar galaxy mass contributed by major (black) and minor
(orange) mergers is given in each panel. Some galaxies have a
significant contribution from major mergers (e.g. M0163, M0224, M0227,
M0300 and M0305), while others have a more quiet merger history and
encounter mainly or even only minor mergers (e.g. M0209, M0215, M0259,
M0290, M0329).  

The amount of stellar major and minor mergers for the ten halos in
both runs is summarised in tables \ref{sim_tab_MNoW} and
\ref{sim_tab_WM}. It is evident that the number of major and
particularly the one of minor mergers is significantly larger in the
MNoW run than in the WM run, as without winds, the star formation
rates peak at higher redshifts ($z \sim 3-4$, see Fig. 2 in
\citealp{2013MNRAS.436.2929H}) so that more (massive) galaxies can
form at earlier times and can subsequently merge together. How
  these stellar merger histories are connected with the steepness of the
  metallicity, age and colour gradients will be discussed in section
  \ref{Mergerhistory}.

\section{Redshift evolution of mass-metallicity and mass-age
  relations}  
\label{massmet}

Before investigating the radial metallicity and age distributions, we
start with discussing the evolution of integrated galaxy properties,
the stellar mass-metallicity and the stellar mass-age relation.

The right column of Fig. \ref{Mgal_sat_only10} shows the evolution of
the stellar mass-metallicity relation since redshift $z =2$ (shown is
$z=2, z=1$, and $z=0$ from bottom to top) for the main galaxies of the
model without winds (MNoW, filled squares) and the model including
winds (WM, filled circles).  The satellites in the respective halos are
indicated by open circles. These are (most of) the systems which are
going to merge with the central galaxies. Note that we have calculated
the stellar mass and metallicity within $1/10$ of the halo virial
radius, which typically captures its central galaxy. We have, however,
not performed any aperture correction.

At $z=2$ (bottom left panel) both models result in a tight
mass-metallicity relation below $\log(M_{\mathrm{stellar}}/M_\odot) =
10.5$. Already at this redshift, the WM galaxies and satellites are
offset by about a factor of two ($0.3$~dex) to lower
metallicities. Above this mass, the a 
mass-metallicity relation for MNoW seems to flatten out already at
slightly super-solar metallicities ($Z_{\mathrm{stars}}/Z_\odot \sim
0.2$).  If, at and below this redshift, satellites merge with their
host galaxies, they bring in stars with lower metallicity than the
main galaxies in both models. 

Already by $z=1$ (middle left panel of Fig. \ref{Mgal_sat_only10})
the situation has changed. The shape of the $z=1$ WM mass-metallicity
relation has changed little, the main galaxies and a few satellites
have increased their mass and metallicity. There is, however, a
dramatic change in the MNoW mass-metallicity relation. Already at
$z=1$, most of the satellites have masses above
$\log(M_{\mathrm{stellar}}/M_\odot) = 9.5$ and super-solar
metallicities of $Z_{\mathrm{stars}}/Z_\odot \sim 0.1 - 0.2$, very
similar to the main galaxies that have grown in mass to above
$\log(M_{\mathrm{stellar}}/M_\odot) = 11$. When these satellites merge
with the central galaxies there will be no difference, at least with
respect to  the metal content.

Towards $z=0$  (upper left panel in Fig. \ref{Mgal_sat_only10}),
the trend continues and the MNoW mass-metallicity relation has
completely flattened out and appears to be in conflict  with
observations \citep{Thomas05, Gallazzi05}. This is, however, a bit
inconclusive as we run out of low mass satellites in MNoW. The WM
main galaxies have slightly lower  metallicities than the MNoW
galaxies ($Z_{\mathrm{stars}}/Z_\odot \sim 0.1$ as opposed to
$Z_{\mathrm{stars}}/Z_\odot \sim 0.2$), but  the metallicities of the
satellites still decrease significantly for masses below
$\log(M_{\mathrm{stellar}}/M_\odot) = 10$, more consistent with the
\citet{Gallazzi05} observations. 

Also in the wind model, the metallicities of
  low-mass satellites are systematically higher by
  0.2-0.3~dex compared to the mean values of \citet{Gallazzi05}.  The
  galactic wind model might   still be not efficient enough in
  suppressing early star formation   (and thus, in delaying metal
  enrichment) as it was for example   indicated by the higher SFRs in
  low-mass systems   compared to abundance matching predictions (see
  top right panel in Fig. 2 in \citealp{2013MNRAS.436.2929H} and
  \citealp{Dave13} for further discussion).  However, the
  observational data of \citet{Gallazzi05} is dominated by central 
  galaxies, while in the simulations, we are considering satellite
  galaxies. Other observational studies revealed that low-mass
  satellite galaxies might have slightly larger stellar metallicities
  than central galaxies (by up to 0.2 dex larger as shown in Fig. 4 of
  \citealp{Pasquali10}) which could account for the systematic offset
  between the wind model and the observations of \citet{Gallazzi05}.
  In this respect, our simulated satellite galaxies in the wind model
  may already provide a fairly realistic stellar metallicity content.

Turning to (mass-weighted) stellar ages, the right column of
Fig. \ref{Mgal_sat_only10} shows the mass-age relation at
$z=0,1,2$ (panels from bottom to top)  for the main (filled symbols)
and satellite (open symbols) galaxies of the MNoW (squares) and the WM
model (circles).  
At $z=2$, central WM galaxies are mostly younger than those in
the MNoW model, while for satellites, the trend is just reversed. This
has the consequence that in WM model, mass-age relation is nearly flat
(or even slightly negative) with a large scatter in the stellar age at
the low mass end. Instead, without winds, already at $z=2$ a positive
correlation between mass and age has evolved. The reason for this
significantly different behaviour is most likely twofold: first, in
the WM model, massive galaxies at $z=2$ are highly star-forming
(delayed star formation peaking at $z=1-2$ due to the winds, see top
left panel of Fig. 2 in \citet{2013MNRAS.436.2929H}). Second, the
strong galactic winds still prevent and suppress star formation in less
massive galaxies at that redshift so that they can become equally or
even older than their central galaxies. 

Towards $z=1$, the trends seen for $z=2$ are still visible, even if
the mass-relation has now steepened in the MNoW model, and flattened
(from a slightly negative correlation) in the WM model. 

By the epoch $z=0$, however, the situation has changed: for the entire
mass range, the MNoW galaxies are on average older than the WM
galaxies. In  the WM model, stellar mass and age are now positively
correlated: less massive satellite galaxies are significantly younger
than their central galaxies, as the former are still highly
star-forming, particularly due to re-infall of previously ejected gas,
while the latter have already consumed a lot of their gas via previous
star formation. This trend is consistent with observations of close
pairs in the Survey for High-z Absorption Red and Dead Sources
(SHARDS, \citealp{Ferreras14}), also finding that satellites are
younger than the centrals at a given redshift up to $z=1$. Compared to
observations (\citealp{Thomas05, Thomas10}), WM  galaxies are a bit
young (by roughly 1-2~Gyrs), which is most likely a consequence of the
increased late in situ star formation in massive galaxies. Instead, a
number of MNoW galaxies tend to be too old (by 1-2~Gyrs) as they have
formed most of their stars too early (see SFR evolution in Fig.2 in
\citealp{2013MNRAS.436.2929H}). AGN feedback would likely reduce the
late star formation in those galaxies (\citealp{Choi14})

At $z=0$, the top right panel of Fig. \ref{Mgal_sat_only10} also
visualises the ``hypothetic'' present-day ages of the $z=1$ and $z=2$
satellite galaxies (small filled triangles and stars) in the WM model
which have until $z=0$ mostly merged with their central galaxies. This
demonstrates that such previously accreted stellar systems are by
about 1-3~Gyrs older compared to the stellar populations of
present-day central galaxies and will, thus, affect their radial
stellar age distributions at large radii (see section
\ref{agegradients}).  

\section{Redshift evolution of metallicity gradients}
\label{metgradients}

\begin{figure*}
\begin{center}
  \epsfig{file=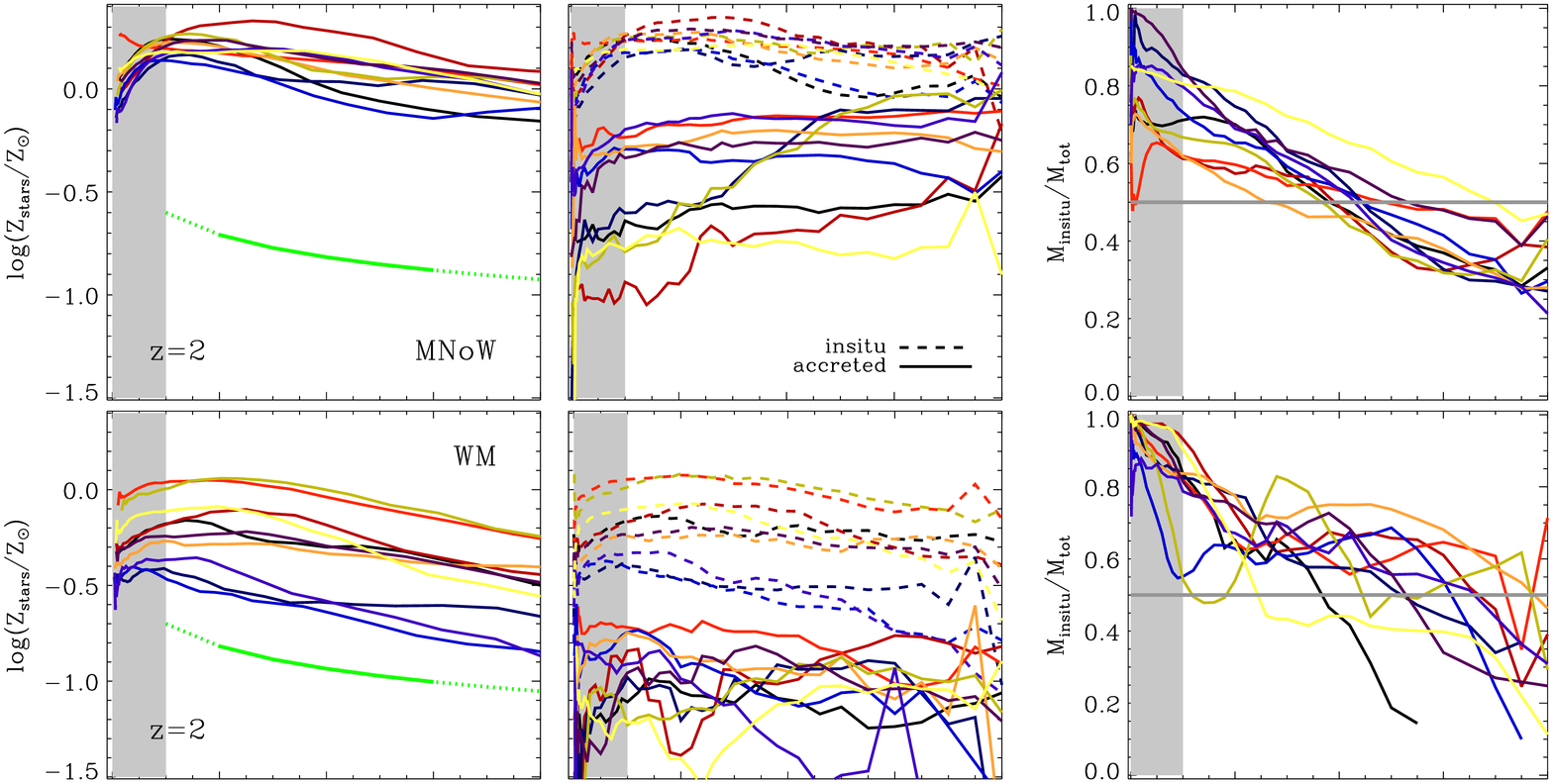, width=0.8\textwidth}\vspace{-0.5cm}
  \epsfig{file=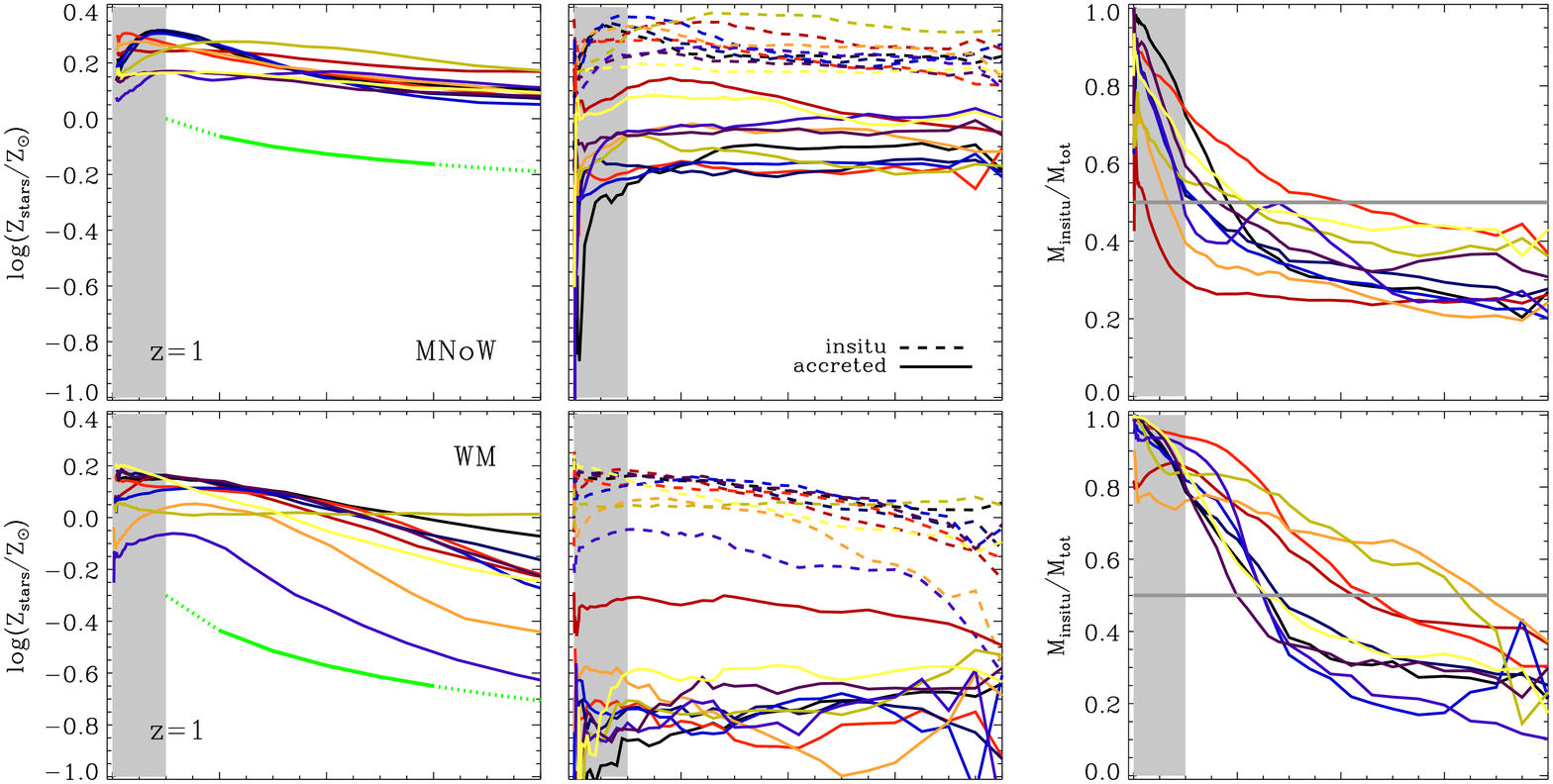, width=0.8\textwidth}\vspace{-0.5cm}
  \epsfig{file=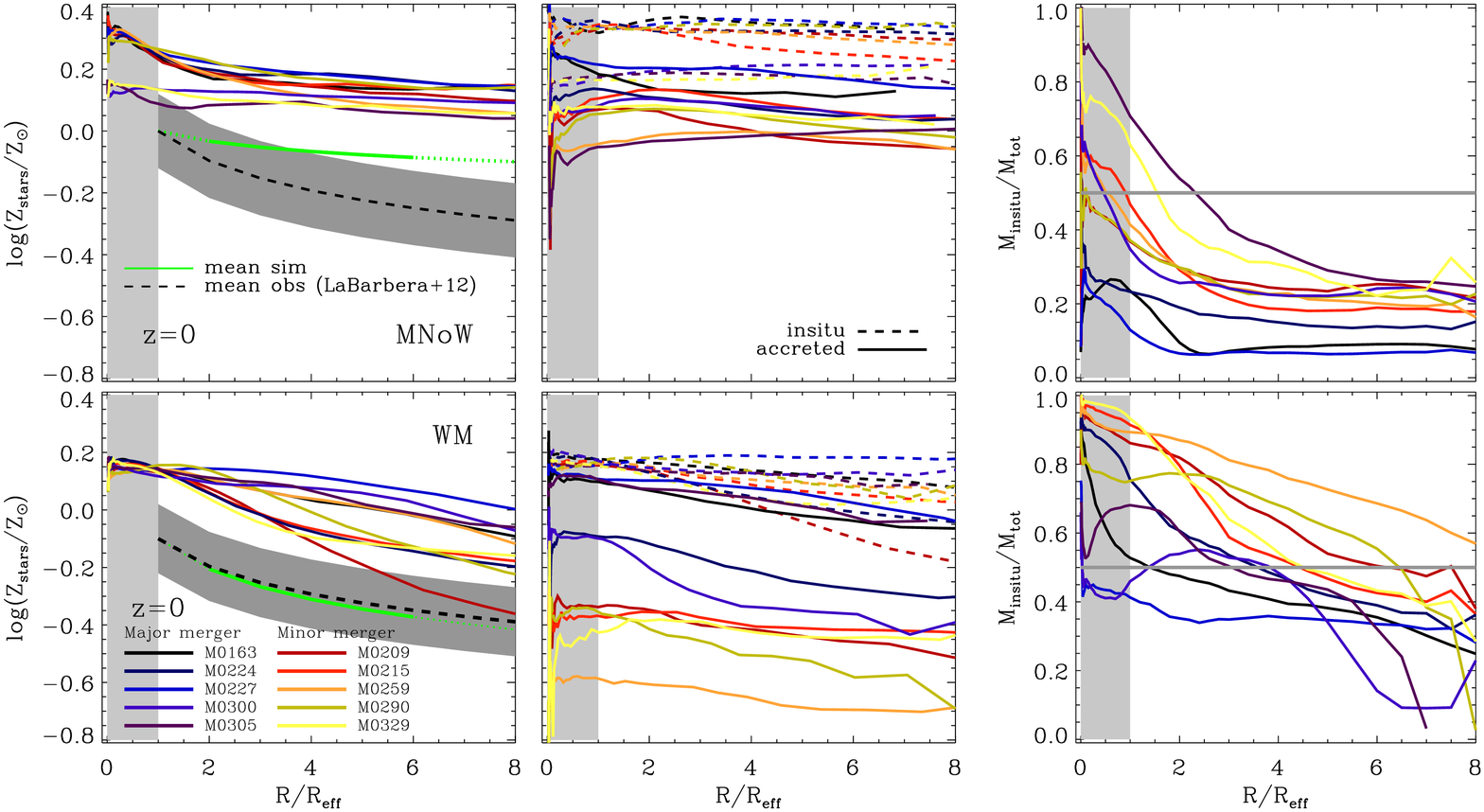, width=0.8\textwidth}\vspace{-0.2cm}
  \caption{{\it Left column:} Total stellar metallicity gradients (mass
    weighted) at $z=2,1,0$ for the ten main galaxies (different
    colours) simulated with the MNoW and WM model. The green
    solid lines indicate the average gradient at
    $2<R/R_{\mathrm{eff}}<6$. {\it Middle panels:} Metallicity
    gradients at $z=2,1,0$ separated into stars formed in-situ (dashed
    lines) and accreted stars (solid lines) for the ten main galaxies
    (different colours) simulated with the MNoW and WM model. {\it
      Right panels:} Fraction of in-situ to total stellar mass as a
    function of radius at $z=2,1,0$ for the ten main galaxies
    (different colours) simulated with the MNoW and WM model. The grey
    shaded in all panels indicate the inner regions
    ($R/R_{\mathrm{eff}}<1$) which will be most likely affected by
    missing AGN feedback.}  
\label{Metgrads_only10_insacc_mm_zall}
\end{center}
\end{figure*}

With the distinctively different evolution of the mass-metallicity
relation discussed above, we also expect a different evolution of
the radial stellar metallicity distribution of the central galaxies,
at least with respect to the stars added by the merged satellite
galaxies. To demonstrate this we have - for all central galaxies -
identified stars that have formed within the galaxies (in-situ) and
stars that have formed outside in other (satellite) galaxies that have
been added to the system by merging (accreted). 

In Fig. \ref{Metgrads_only10_insacc_mm_zall}, we show the total 
(mass-weighted) stellar metallicity gradients out to 8
$R_{\mathrm{eff}}$ for the main galaxies (left panels) at $z=2,1,0$
(panels from top to bottom) for the MNoW and the WM galaxies (as
indicated in the legend). We have calculated the effective radii
  at the corresponding redshift step as half-mass radii using the
  stellar mass within 1/10~$R_{\mathrm{eff}}$ \footnote{Note that
    calculating the half-mass radii using the total stellar mass
    within the virial radius or the half-light radius hardly changes
    our result.} which are summarized (for $z=0$ only) in tables
  \ref{sim_tab_MNoW} and \ref{sim_tab_WM}.  For the evolution of the
  effective radii, we refer the reader to Fig. 10 in
  \citet{2013MNRAS.436.2929H}. In this study it was also shown (their
  Fig. 9) that the present-day sizes of massive central galaxies can
  be smaller by a factor of 1.5 - 2 compared to observations. For a
  fair comparison to observations this has to be considered. This is
  discussed in more detail in section \ref{Observations}.   

The grey shaded area marks in each panel of
Fig. \ref{Metgrads_only10_insacc_mm_zall} the innermost region of a
galaxy ($r<R_{\mathrm{eff}}$), which should be handled with care as
our simulation do not include any models for AGN feedback, which,
however, is expected to strongly influence the innermost regions of
a galaxy. Beyond $2R_{\mathrm{eff}}$, AGN feedback should hardly
  have any direct impact on the stellar populations and thus, the 
metallicity. It can, however, indirectly affect stars that have formed
in the central regions and then migrate outwards (see also
discussion in section \ref{AGNfb}). Therefore, in the following we
will mainly focus on ``outer'' metallicity gradients.   

All the slopes $\nabla(Z_{\mathrm{stars}})$ of the fitted
metallicity gradients of the total stellar mass (within
$2R_{\mathrm{eff}} < r <  6R_{\mathrm{eff}}$) at the three redshifts
are summarised in tables 1 (MNoW galaxies) and 2 (WM galaxies) in
units of dex/kpc according to   
\begin{equation}
\log(Z_{\mathrm{stars}}) = \mathrm{offset} +
\nabla_k(Z_{\mathrm{stars}}) \times r [\mathrm{kpc}] 
\end{equation}
and in units of dex/dex\footnote{It is
  important to note that fitting the slope with a logarithmic x-axis is
  completely independent of the exact units, as $kpc$ or
  $r/R_{\mathrm{eff}}$} according to
\begin{equation}
\log(Z_{\mathrm{stars}}) = \mathrm{offset} +
\nabla_l(Z_{\mathrm{stars}}) \times \log(r/R_{\mathrm{eff}}). 
\end{equation}
The mean gradients are illustrated in the left panels of
Fig. \ref{Metgrads_only10_insacc_mm_zall} by the green solid lines
(the green dotted lines just show the extension towards smaller and 
larger radii beyond the fitted regime).

At $z=2$ (first two rows), the gradients for the different models are
very similar, and interestingly the inner gradients (inside 1-2
$R_{\mathrm{eff}}$) are positive in both cases. We suspect this is due
to rapid infall of not or only weakly enriched 'primordial' gas. The
MNoW galaxies, due to the missing galactic winds, which are expelling
metal rich gas, have already reached super-solar metallicities at
small radii and the metallicities drop a bit below the solar
metallicity at the larges radii. The WM galaxies have lower
metallicities with a much larger spread but a similar shape.   

Even by redshift $z=1$, the inner metallicity gradients (inside 1
$<R_{\mathrm{eff}}$)  are (at least partly) still positive (see third
and fourth left panels in
Fig. \ref{Metgrads_only10_insacc_mm_zall}). Interestingly, positive
metallicity gradients in the gas phase at $z \sim 3$ and $z\sim 1.2$
were detected by recent observational studies of \citet{Cresci10} and
\citet{Queyrel12}, respectively, explaining them by gas dilution due
to accretion of primordial gas at these high redshifts as predicted by
cold flow models. As stellar metallicity follows the gas-phase
metallicity, their high-z observations would be  qualitatively
consistent with our results.

\begin{figure}
\begin{center}
  \epsfig{file=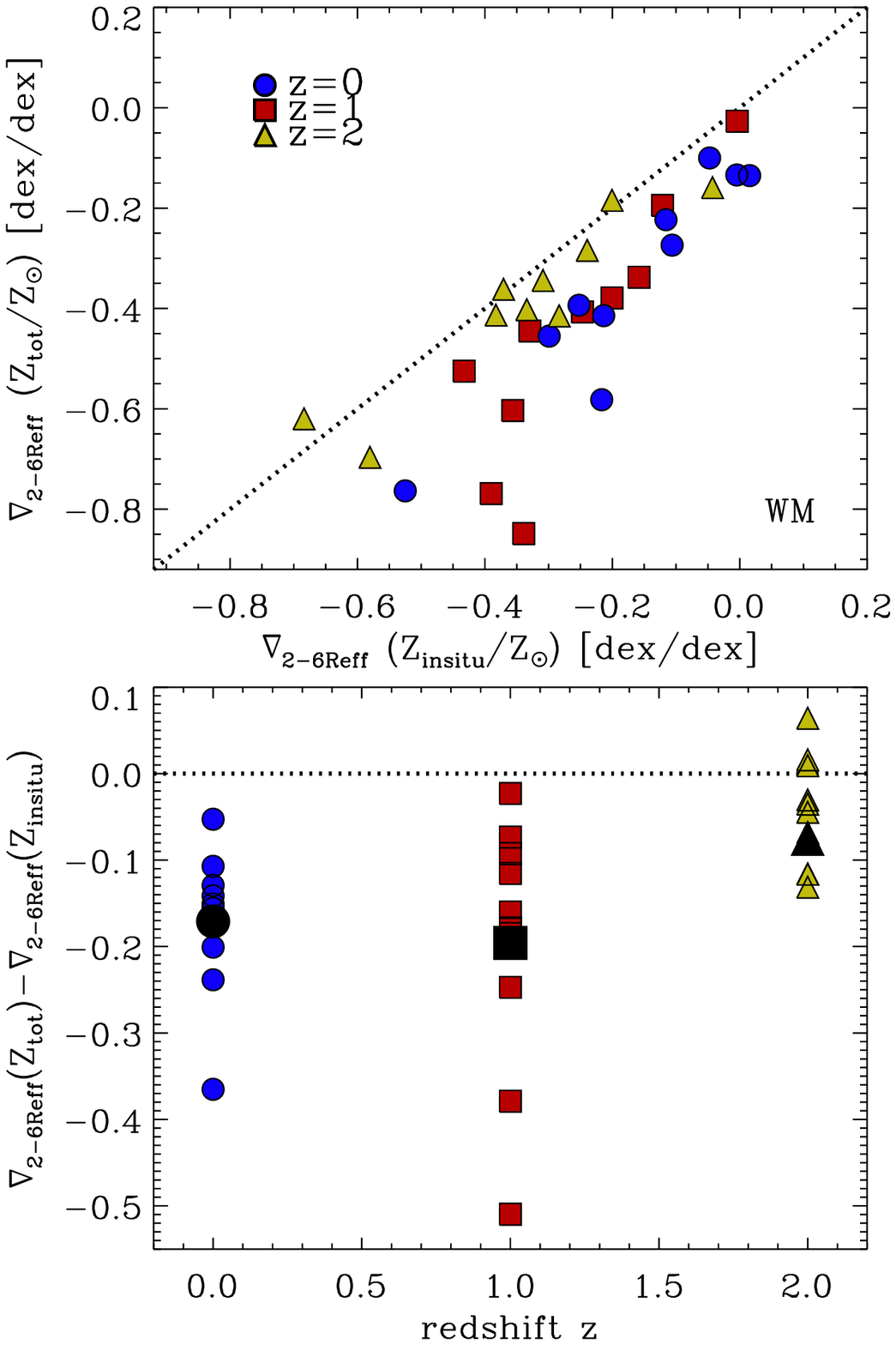, width=0.4\textwidth}
  \caption{\textit{Top panel:} Fitted metallicity gradient (at
    $2<R/R_{\mathrm{eff}}<6$) of the total stellar component versus
    the in-situ stellar component only at $z = 0, 1, $and $2$
    (differently coloured symbols) for the WM model. The black dotted
    line indicates equal gradients. \textit{Bottom panel:} Difference
    between total and in-situ metallicity gradient versus redshift. At
    $z=2$ the total gradients are similar to those of the in-situ
    stellar component, whereas towards lower redshifts, the total
    gradient is typically by $\sim 0.2$~dex lower than the in-situ
    gradient. Large filled symbols indicate the mean.} 
\label{Insitu_tot_metgrad}
\end{center}
\end{figure}

The first two panels in the middle column of
Fig. \ref{Metgrads_only10_insacc_mm_zall} illustrate the separation
into metallicity gradients of in-situ and accreted stars at $z=2$. We
have very few accreted stars at these high redshifts (see
\citealp{2013MNRAS.436.2929H}) so that the total gradients are almost
entirely dominated by in-situ stars. Nevertheless, we have a few
accreted stars, which all have lower metallicities with a similar
offset for both models as expected from the $z=2$ mass-metallicity
relation (see Fig. \ref{Mgal_sat_only10}, left bottom panel). Already
at this redshift, we can see the clear trend that accreted stars
become more important in  the outer regions of the galaxies as the
in-situ formed stellar fractions drop below 0.5 only at large radii
(first two panels in the right column in
Fig. \ref{Metgrads_only10_insacc_mm_zall}). Still, for both models the
galaxies are dominated by in-situ stars out to $4 R_{\mathrm{eff}}$ in  
the case of MNoW and $ \sim 6 R_{\mathrm{eff}}$ in the case of WM.     

By redshift $z=1$ the situation changes and the two models evolve
differently (see third and fourth row in
Fig. \ref{Metgrads_only10_insacc_mm_zall}). The MNoW metallicity
gradients have now become very shallow out to large radii at
metallicities of $Z/Z_\odot \sim 0.2$ at the centre and  $Z/Z_\odot
\sim 0.1$ at large radii (third left panel). The slopes (summarised in table
\ref{sim_tab_MNoW}) reach a minimum value of $-0.45$~dex/dex. The
accreted stars have on average a $0.4$~dex lower metallicity than the
in-situ stars (third middle panel) and now start dominating the
stellar budget at radii larger than $\sim 2 R_{\mathrm{eff}}$ (third
right panel). By comparison to the flat metallicity distribution of
the in-situ stars it is evident that the modest global gradients
originate from accreted stars.  

For the WM model, by $z=1$ the galaxies have developed clear outer
negative  stellar metallicity gradients (fourth left panel of Fig.
\ref{Metgrads_only10_insacc_mm_zall}. The slopes (summarised in  
table \ref{sim_tab_WM}) reach the maximum slope $=-0.85$~dex/dex,
which is twice as large as for the MNoW galaxies. The separation into
in-situ and accreted stars indicates that the negative slope partly
originates from the in-situ stars, but the steepening of the gradients
in the outer regions ($\gtrsim 3 R_{\mathrm{eff}}$) is particularly
supported by accreted stars which by now have - on average -  a 
significantly lower metallicity (about one order of magnitude) than
the in-situ stars and become important at radii larger than $2 - 6
R_{\mathrm{eff}}$ (fourth right panel of
Fig. \ref{Metgrads_only10_insacc_mm_zall}).  

We show the same analysis at $z=0$ in the two bottom rows of
Fig. \ref{Metgrads_only10_insacc_mm_zall}. For the MNoW galaxies, the
(most) central metallicity has now increased to $Z/Z_\odot \sim 0.4$ and
drops to $Z/Z_\odot \sim 0.1$ at large radii, the slopes reach a
minimum value of $-0.25$ dex/dex (see table \ref{sim_tab_MNoW}). This
gradient is mainly driven by the accreted stars (the in-situ
distributions are almost all nearly flat), which by now have reached
solar metallicity on average (fifth middle panel of
Fig. \ref{Metgrads_only10_insacc_mm_zall}) and dominate most systems
outside $1 R_{\mathrm{eff}}$ (fifth right panel of
Fig. \ref{Metgrads_only10_insacc_mm_zall}).    

The WM galaxies (bottom panels of
Fig. \ref{Metgrads_only10_insacc_mm_zall}) have lower central 
metallicities  ($Z/Z_\odot \sim 0.2$) with much steeper outer
gradients down to $-0.76$~dex/dex with a mean of $-0.35$~dex/dex (see
table \ref{sim_tab_WM}).   The reason for the steeper gradients in the
WM compared to the MNoW model is twofold: on the one hand, as
discussed,  the steeper gradients originate from the accretion of
metal-poorer stellar populations. On the other hand, also the in-situ
components show metallicity gradients contributing to the overall 
gradients. The latter is most likely due to infall of (particularly
re-infall of previously ejected) metal-poor gas onto the galaxy which 
can be then turned into metal-poor stars as a consequence of an
inside-out growth, the same process causing the metallicity gradients
in disk galaxies. Late re-accretion of previously ejected gas occurs
typically in the WM model due to the strong galactic winds, but not in
the MNoW model, where the in-situ gradients are, therefore, relatively
flat (see fifth middle panel of
Fig. \ref{Metgrads_only10_insacc_mm_zall}). 
{In addition, the WM galaxies have much more extended gas distributions 
(and maybe associated gas distributions) than the MNoW models in
better agreement with observations (e.g. Atlas3D, \citealp{Serra14}).}

To quantify the contribution of accretion of metal-poor stars for a 
given ``in-situ'' gradient, the top panel of
Fig. \ref{Insitu_tot_metgrad} shows the total gradients versus the
in-situ gradients in the WM model at $z=0,1,2$ (different
colours). While at $z=2$, the total gradients are very similar to the
in-situ ones, at $z=1$ and $z=0$, the total gradient is significantly
reduced compared to the in-situ gradient as a consequence of the
accretion of metal-poor stars. The lower panel of
Fig. \ref{Insitu_tot_metgrad} shows the difference between the two
gradients versus redshift. On average, the in-situ gradients at $z=0$
and $z=1$ (having average values of $-0.15$ and $-0.25$) are reduced
by $\sim 0.2$~dex (big open symbols) due to accretion of metal-poor
stellar population. A detailed comparison with observations will be
given in section \ref{Observations}. 

\begin{figure*}
\begin{center}
\epsfig{file=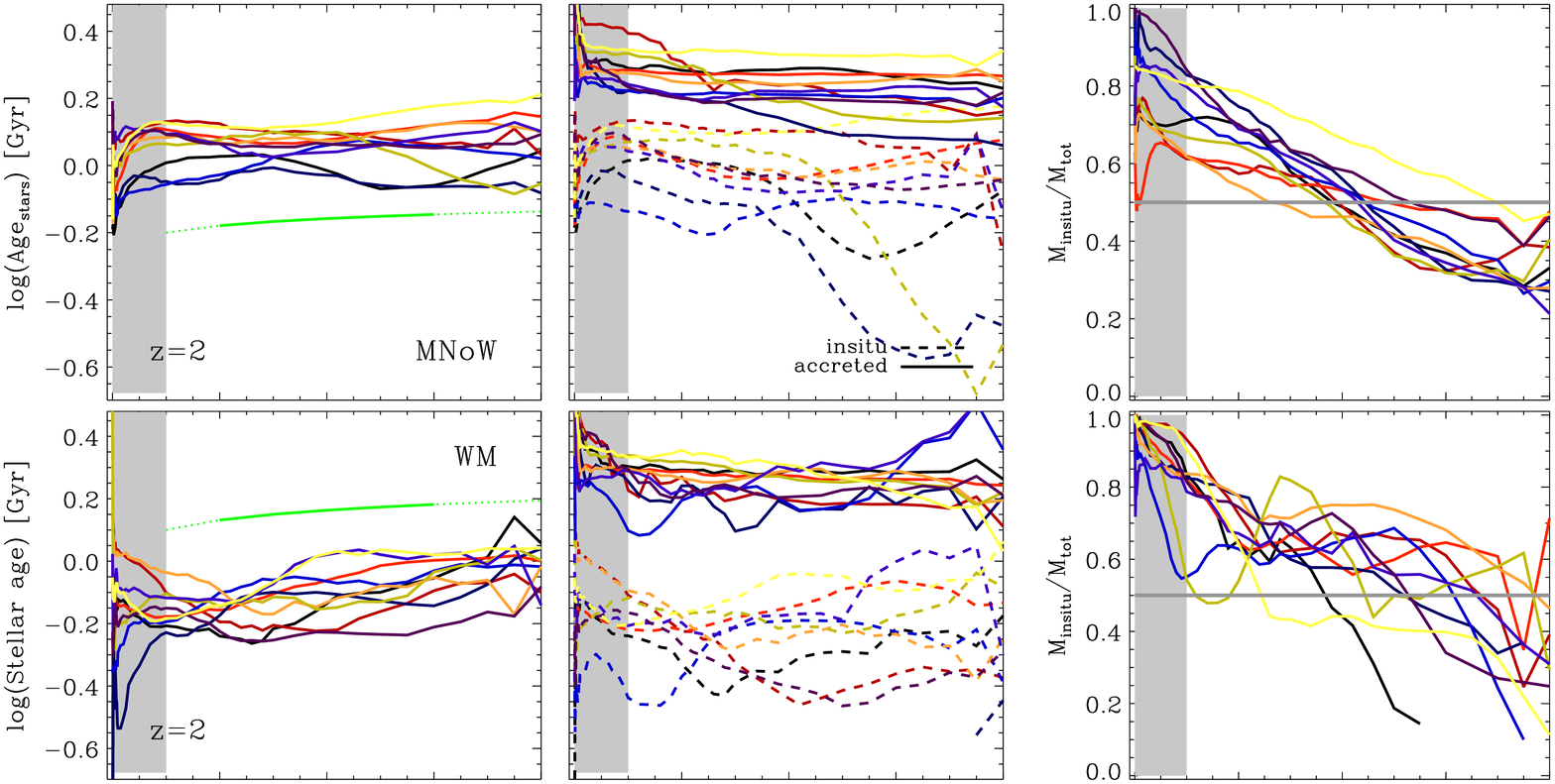, width=0.8\textwidth}\vspace{-0.5cm}
\epsfig{file=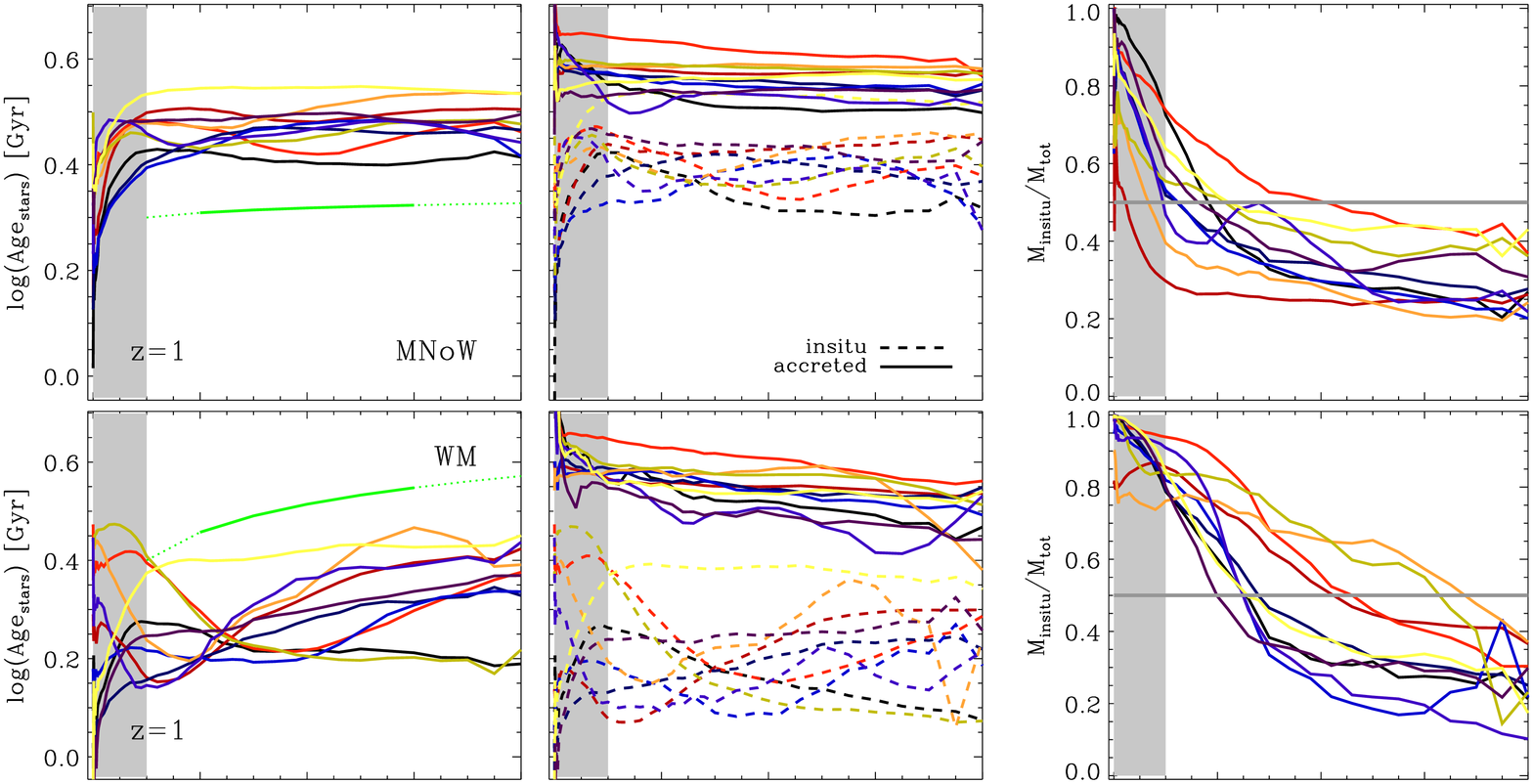, width=0.8\textwidth}\vspace{-0.5cm}
\epsfig{file=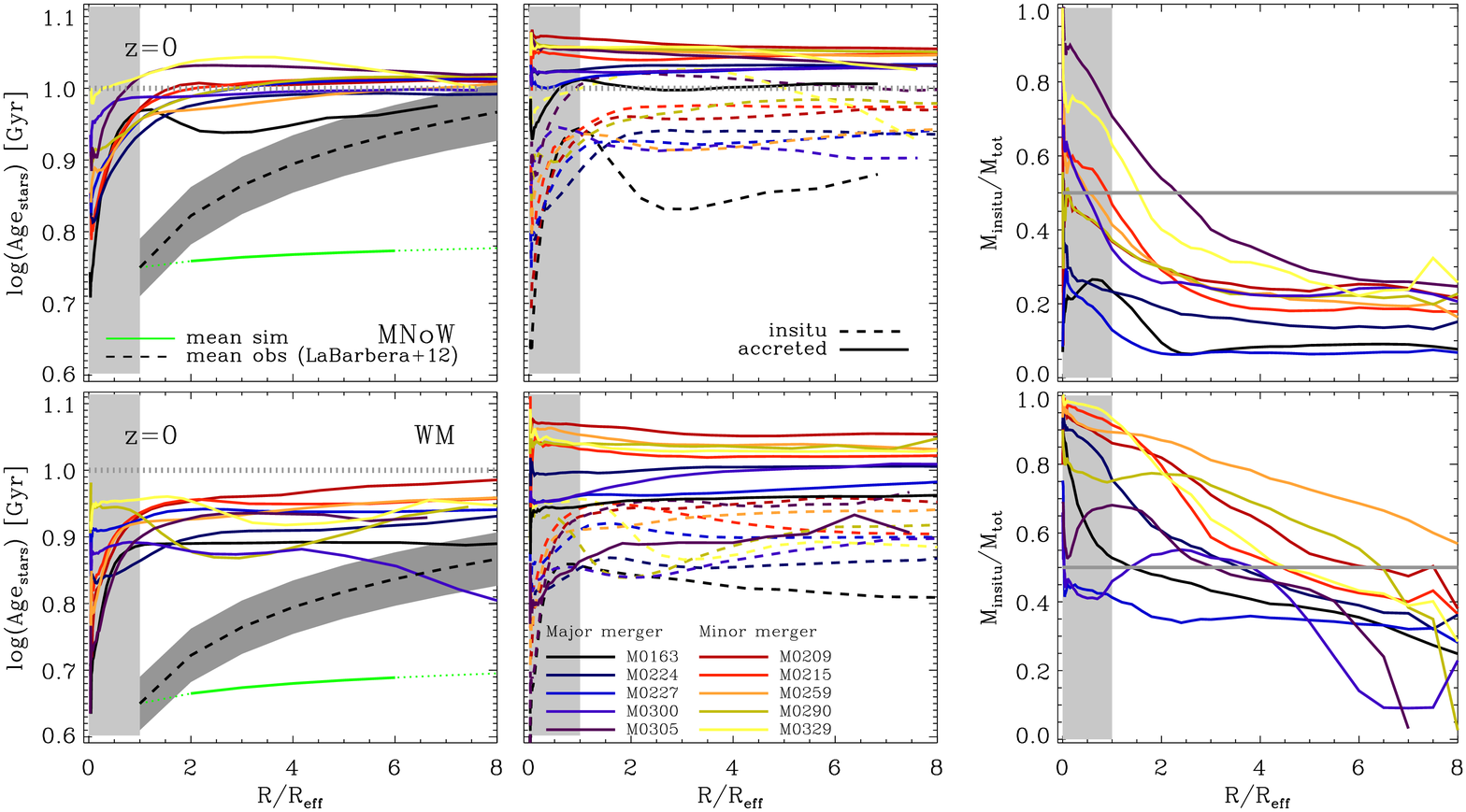, width=0.8\textwidth}
  \caption{Same as Fig. \ref{Metgrads_only10_insacc_mm_zall}, but now
    for stellar age gradients at $z=2,1,0$ (from top to bottom) for 
    the ten galaxies in the MNoW and WM models. } 
 {\label{Agegrads_only10_insacc_mm_zall}}
\end{center}
\end{figure*}

\section{Redshift evolution of stellar age gradients}\label{agegradients}

In Fig. \ref{Agegrads_only10_insacc_mm_zall}, we show the same
analysis as for the metallicity in Fig. \ref
{Metgrads_only10_insacc_mm_zall}, but now for the stellar ages -- the
second important property determining a stellar population. 

Almost independent of the model and redshift, the central
regions of the galaxies (illustrated by the grey shaded areas) have
steep positive age gradients (left column in Fig.
\ref{Agegrads_only10_insacc_mm_zall}), i.e. the core of the  galaxies
($r<1 R_{\mathrm{eff}}$) is (probably unrealistically) young,
particularly at $z=0$. This effect is most likely due to missing AGN 
feedback in our simulations which would be expected to 
suppress star formation in the centre of a galaxy stronger than
extended SF, and thus, leading to older central stellar populations. 

Turning to the outer age gradients ($r>2 R_{\mathrm{eff}}$, left
column), the WM galaxies are on average younger by $0.1-0.2$~dex than 
the MNoW galaxies due to delayed star formation in the wind model (and
at lower redshifts, also to due to re-accretion of previously expelled gas). 

At $z=2$, the WM galaxies show slightly positive age gradients, while
the ones of the MNoW galaxies are almost flat (green line indicates
the average gradient). In both models, the ages of the accreted
stellar population are typically older than that of the in 
situ  formed stellar component (first and second middle panels in
Fig. \ref{Agegrads_only10_insacc_mm_zall}). The accreted stars have a
similar age in both models, while the in-situ formed stars are much
younger in the wind model leading to a significantly larger difference
between in-situ and accreted stellar populations for the WM galaxies
(0.5~dex, i.e. 1.4~Gyr) than for the MNoW galaxies (0.2~dex,
i.e. 0.74~Gyr). This supports the (positively) steeper age gradients
in the WM model, even if at that redshift, the stellar mass assembly
in both models is clearly dominated by in-situ star formation (right
first and second panels of
Fig. \ref{Agegrads_only10_insacc_mm_zall}). The younger in-situ formed  
stellar populations are a consequence of the delayed and enhanced star
formation in the WM model due to late infall of previously ejected gas
(see also the discussion of the inverted mass-age relation at $z=2$ in
section \ref{massmet}).  
 
Towards $z=1$, the WM and MNoW galaxies are even more different: the
age gradients in the MNoW model stay nearly flat, some have a slightly
positive slope though (third left panel in 
Fig. \ref{Agegrads_only10_insacc_mm_zall}). The accreted stars are on
average only 1.5~Gyr older than the in-situ formed ones (third middle
panel in Fig. \ref{Agegrads_only10_insacc_mm_zall}). This 
supports the slightly positive but shallow age gradients, since at that
redshift, at $r>4R_{\mathrm{eff}}$ accretion of stars starts dominating
over in-situ star formation (third right panel of
Fig. \ref{Agegrads_only10_insacc_mm_zall}). 

Instead for the WM model, the age gradients become strongly positive
with a high average value of 1.91~dex/dex (see green line in the
fourth left panel in Fig. \ref{Agegrads_only10_insacc_mm_zall}).  This
a consequence of both the (at least partly) steep positive in-situ age
gradients and the late accretion of old stellar systems compared to
the young in-situ stellar component (see fourth middle panel in
Fig. \ref{Agegrads_only10_insacc_mm_zall}). The latter is now more
relevant than at $z=2$ as the outer parts of the galaxies are
dominated by accretion of stars (see fourth right panel of
Fig. \ref{Metgrads_only10_insacc_mm_zall}). 

In contrast, at $z=0$, at large radii ($\gtrsim 2 R_{\mathrm{eff}}$),
the average age of the stellar populations is again only very slightly
increasing for some galaxies, generally somewhat stronger for the WM
(0.04~dex/dex) galaxies than for the MNoW galaxies (0.03~dex/dex,
fifth and sixth left panels in Fig. \ref{Agegrads_only10_insacc_mm_zall}). 
Nevertheless, as for $z=1$, the shallow outer increase is partly
driven by slightly positive in-situ age gradients, but also by the
growing importance of the older accreted stars (right fifth and sixth
panel of Fig. \ref{Agegrads_only10_insacc_mm_zall}). 

\begin{figure}
\begin{center}
  \epsfig{file=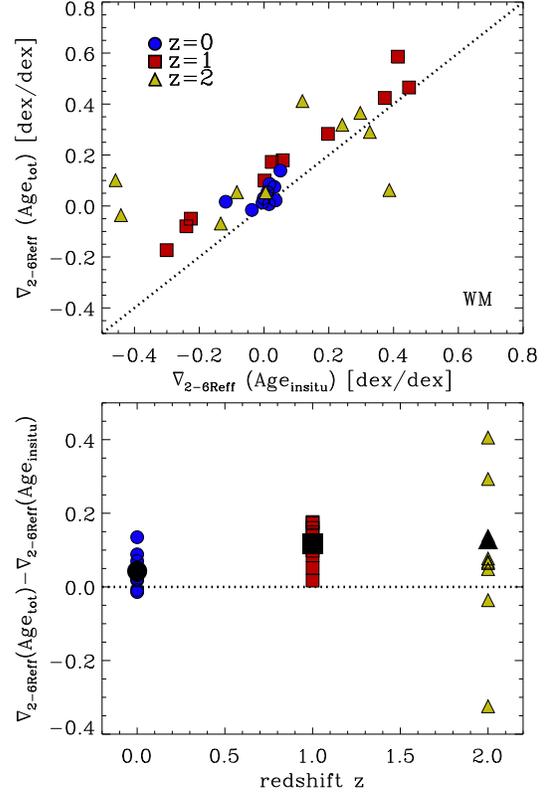, width=0.4\textwidth}
  \caption{\textit{Top panel:} Fitted age gradients (at
    $2<R/R_{\mathrm{eff}}<6$) of the total stellar component versus
    those of the in-situ stellar component at $z = 0, 1, $and $2$
    (differently coloured symbols) for ten massive galaxies for the WM
    model. The black dotted line indicates equal total and in-situ
    gradients. \textit{Bottom panel:} Difference between the total and
    in-situ age gradients versus redshift. On average, the total
    gradient is typically by $0.05-0.1$~dex higher than the in-situ
    gradient in the wind model (see black symbols in the bottom
    panel).} 
\label{Insitu_tot_agegrad}
\end{center}
\end{figure}

To quantify by how much the accretion of old stellar systems
contributes to the steepening of the positive age gradients in the WM
model, the top panel of Fig.  \ref{Insitu_tot_agegrad} illustrates the
fitted total gradients versus that of in-situ formed stars at $z=0,1,2$
(different symbols and colors) and the bottom panel explicitly shows
the difference between those gradients versus redshift. At $z=1,2$,
there is a huge scatter for the in-situ gradients (reaching very positive
slopes), while at $z=0$, they are mainly flat. From
Fig. \ref{Insitu_tot_agegrad} it is evident that the positive in-situ
age gradients are indeed steepened due to accretion of older systems
(by 0.08~dex at $z=2$, by 0.12~dex at $z=1$ and by 0.04~dex at $z=0$,
black symbols in the bottom panel), but the effect is rather weak,
particularly at $z=0$. For the MNoW model, such an 
effect is even less pronounced (not explicitly shown). 
Overall, we can conclude that positive age gradients originate
from the accretion of older stars at large radii. 

\section{Redshift evolution of color gradients}\label{colorgradients}

An important set of observables for galaxies are their colours, which
are observationally more easy to measure (from images) than age and
metallicity (requiring spectroscopic data). Colours are (degenerately)
dependent on the intrinsic metallicity and ages of their stellar
populations. To facilitate a comparison of the simulated radial
distributions of galaxy colours to present-day, observed (outer)
\textit{color} gradients and in particular, to provide predictions for 
future high-redshift surveys (with a good spatial resolution), we 
quantify in this section our simulated rest-frame (without extinction
effects) colour gradients at $z=0, 1$ and $2$. Neglecting the effect
of dust on color gradients is \textit{observationally} motivated as
early-type galaxies exhibit radial gradients of metal absorption
features which cannot be attributed to dust (e.g. \citealp{Gonzalez93,
Mehlert00}).  

We use the metal- and age-dependent models for the spectral evolution
of stellar populations of \citet{Bruzual03}, assuming a Chabrier IMF
to compute some photometric properties (g-i, g-r and u-g colours) of
our simulated galaxies. Note that massive ellipticals are very likely
to have steeper than Chabrier IMFs. We treat every star particle as a single
stellar population with a given mass, metallicity and age so that we
can compute for each star particle the flux in a given band. Summing
up over the fluxes in a given radial bin allows to compute the radial
distribution of the magnitudes and thus, of the g-i, g-r and u-g
colours.

\begin{figure*}
  \begin{minipage}[b]{0.9\linewidth}\vspace{-0.3cm}
\centering
\epsfig{file=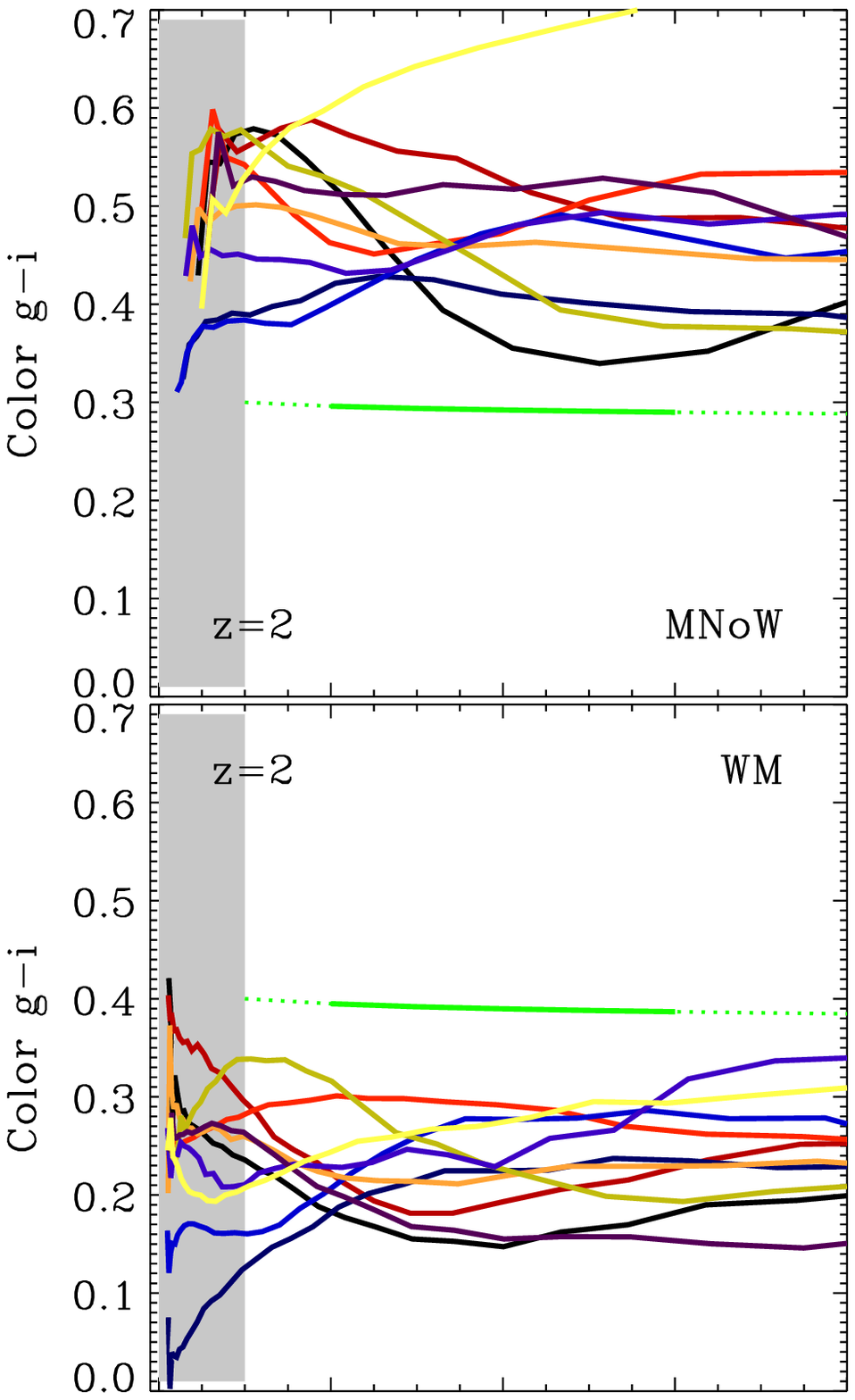, width=0.31\textwidth}\hspace{-0.5cm}
\epsfig{file=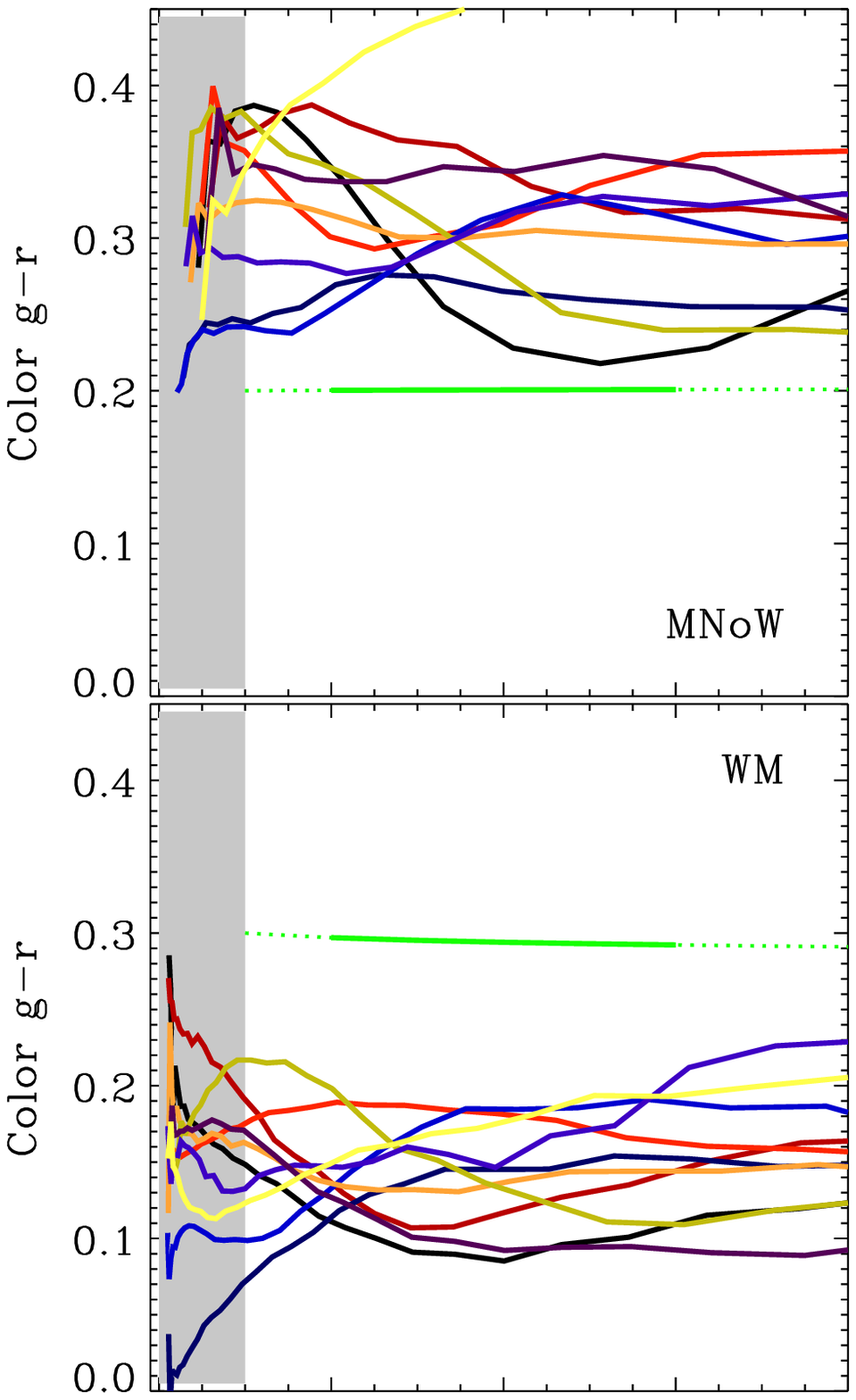, width=0.31\textwidth}\hspace{-0.6cm}
\epsfig{file=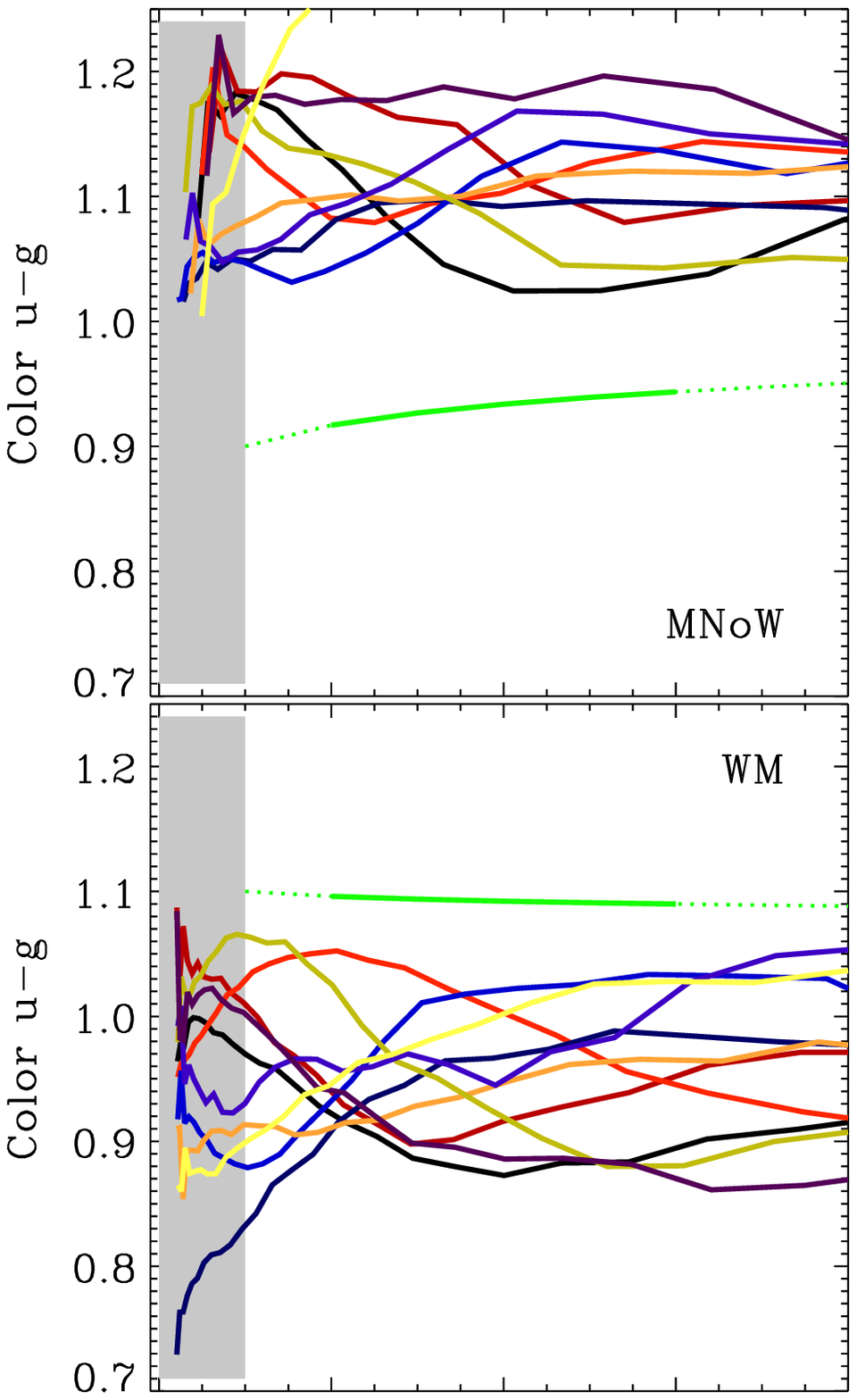, width=0.31\textwidth}
\end{minipage}
\begin{minipage}[b]{0.9\linewidth}\vspace{-0.8cm}
\centering
\epsfig{file=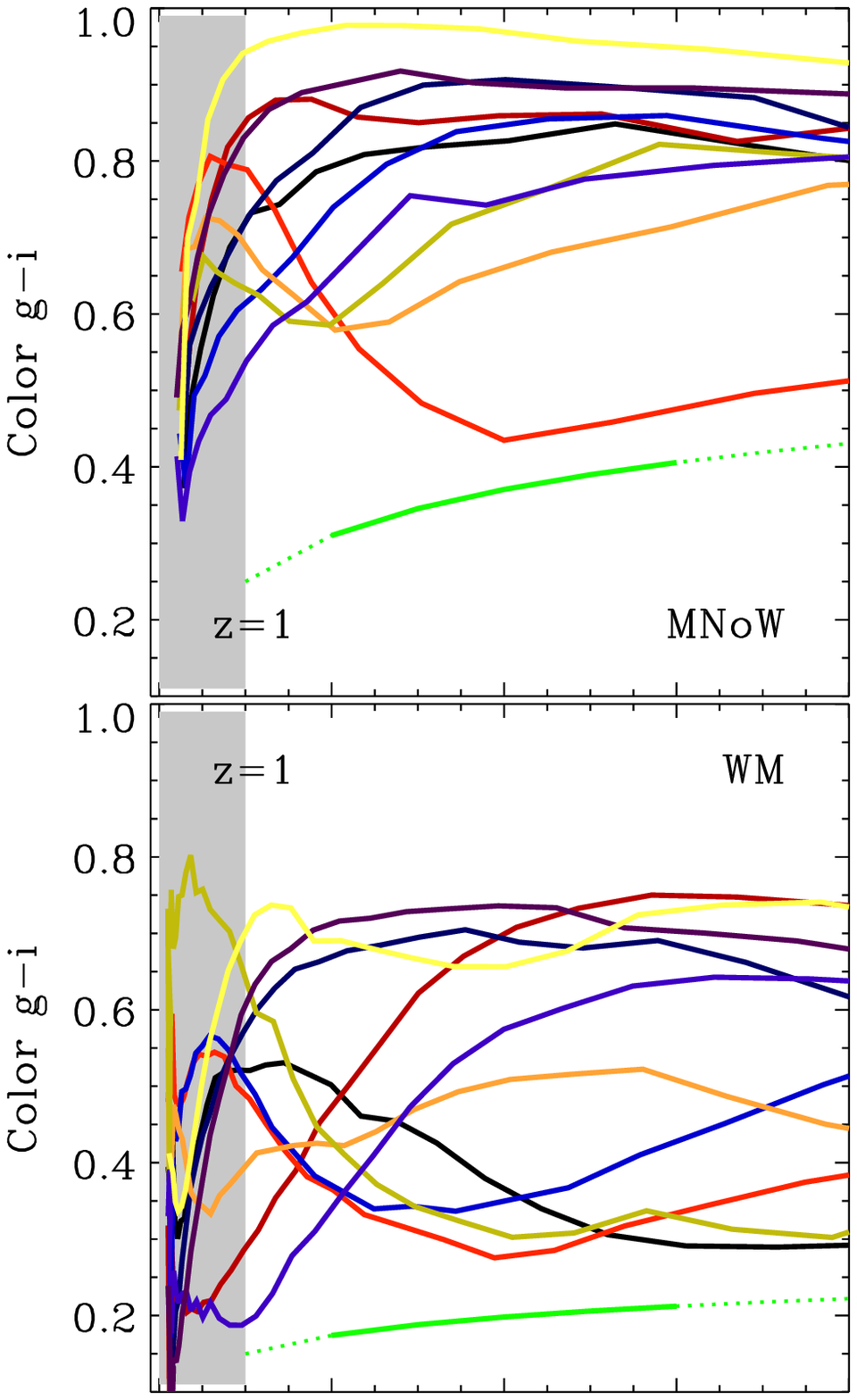, width=0.31\textwidth}\hspace{-0.5cm}
\epsfig{file=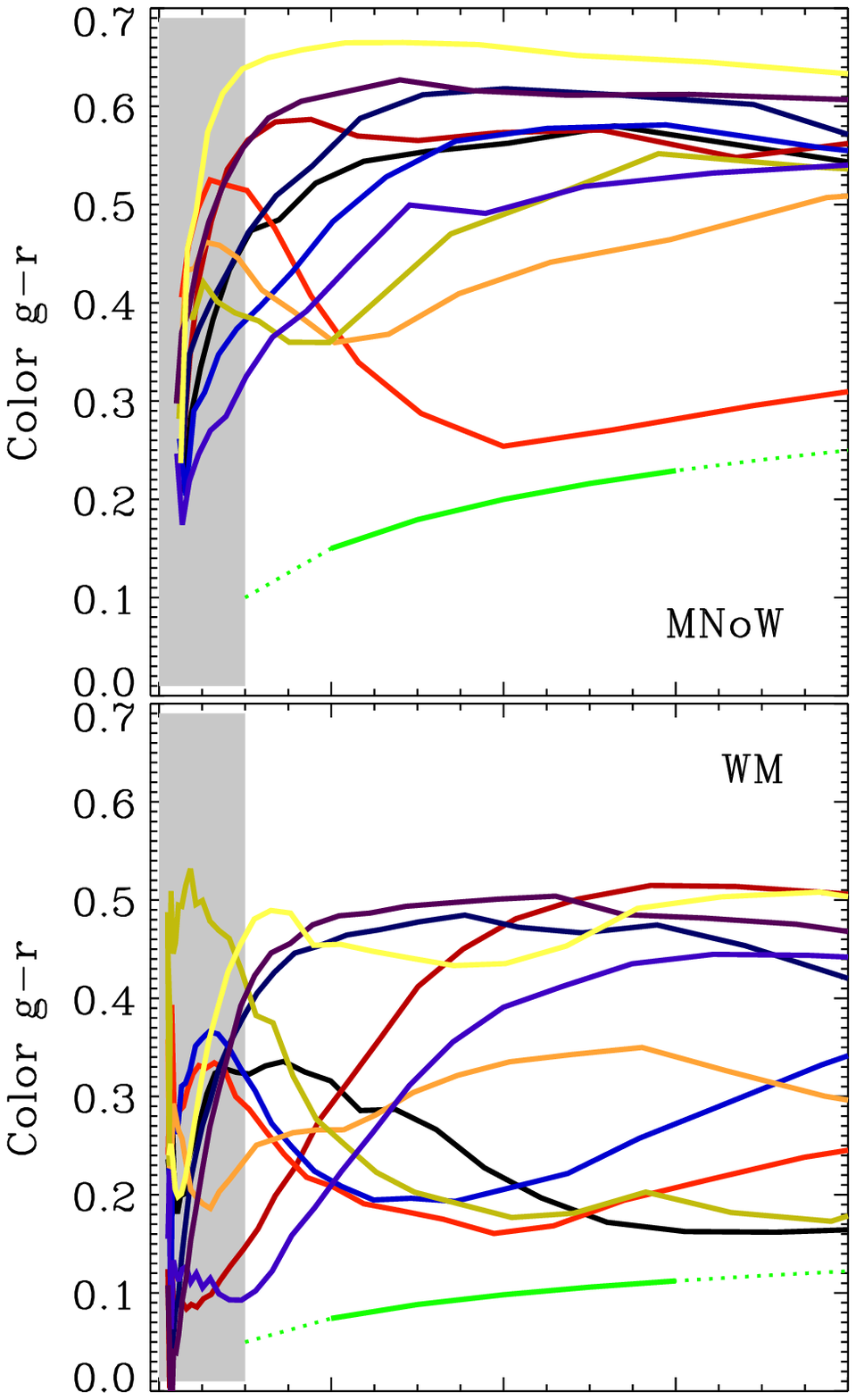, width=0.31\textwidth}\hspace{-0.6cm}
\epsfig{file=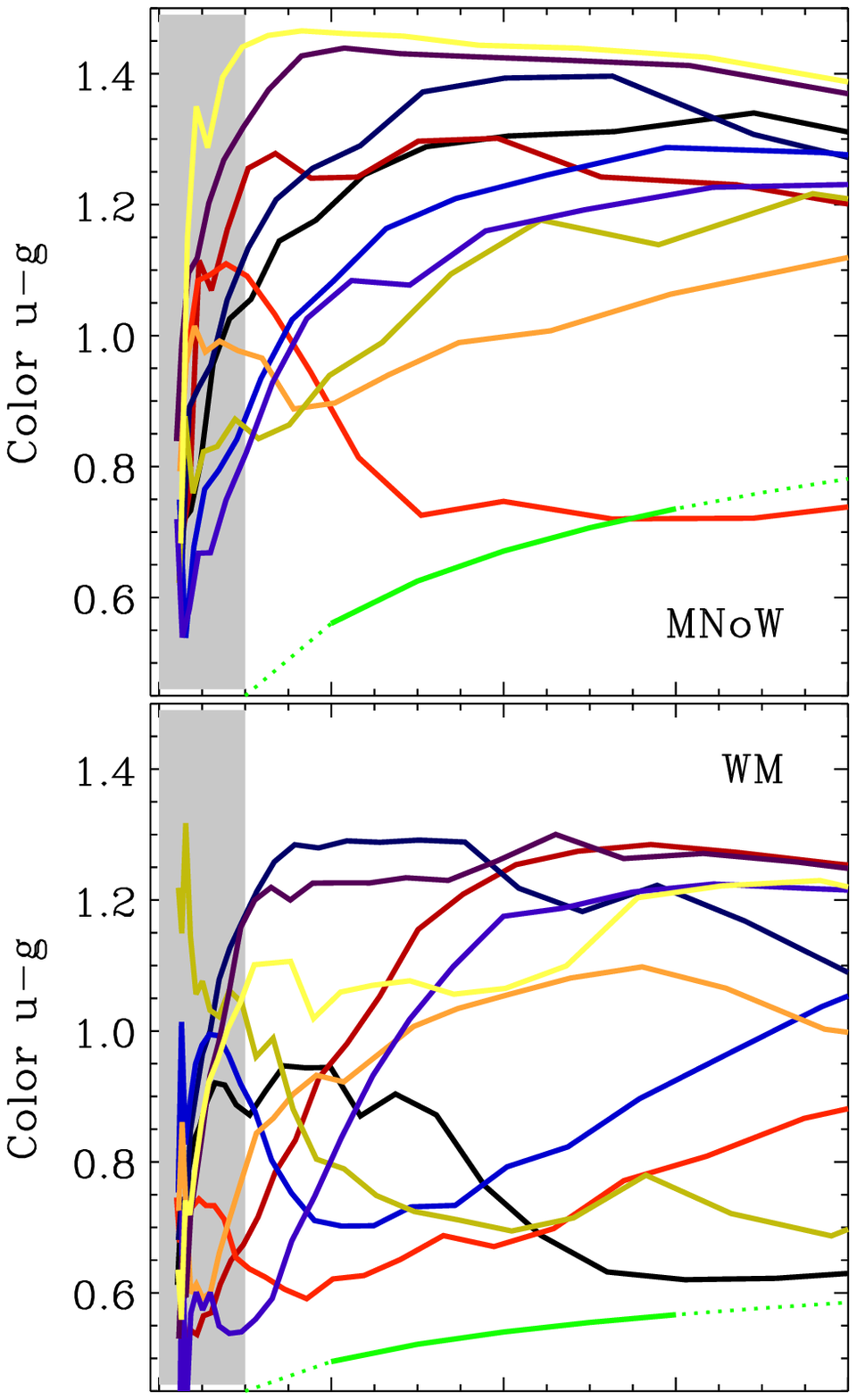, width=0.31\textwidth}
\end{minipage}
\begin{minipage}[b]{0.9\linewidth}\vspace{-0.8cm}
\centering
\epsfig{file=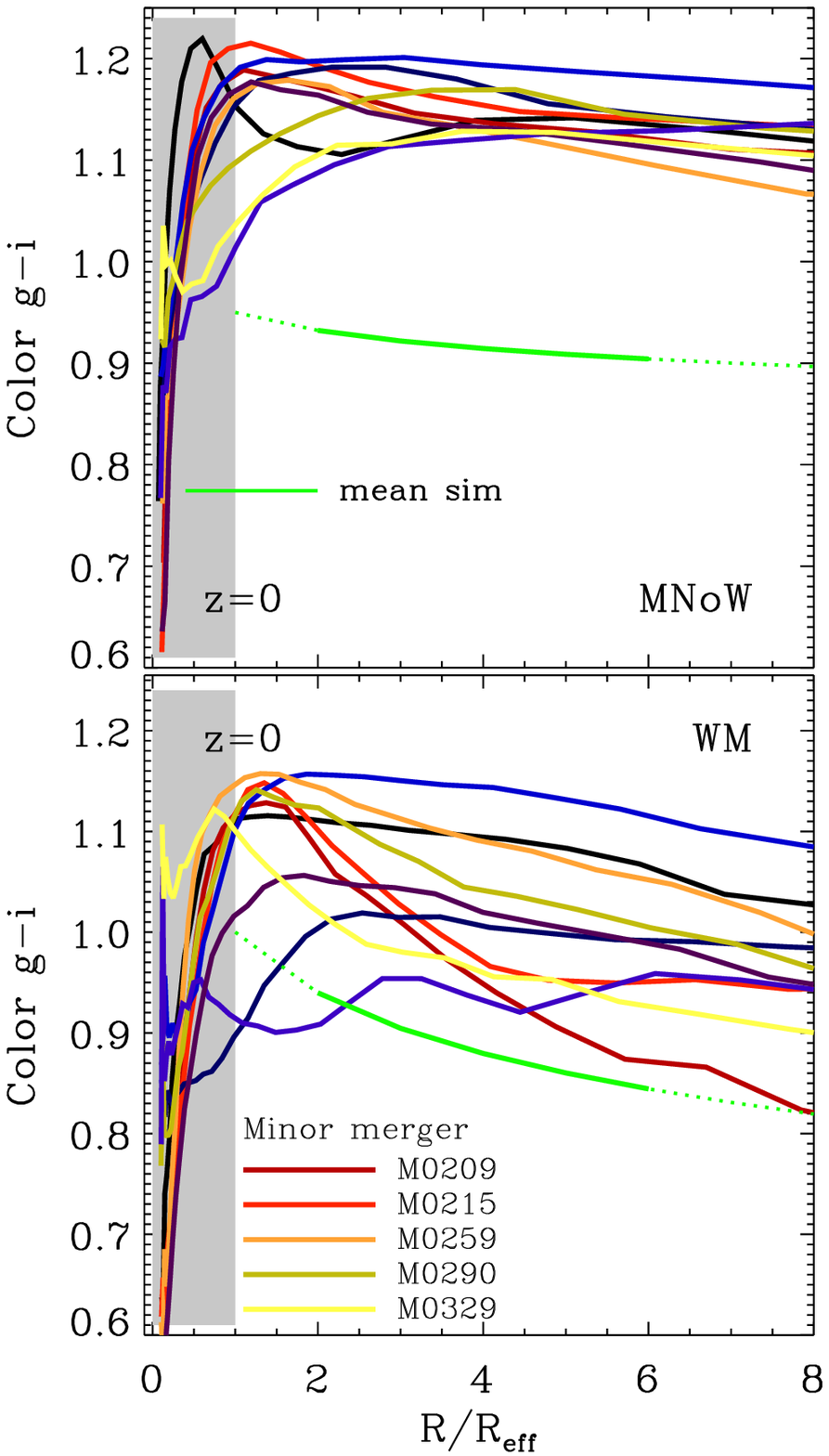, width=0.31\textwidth}\hspace{-0.5cm}
\epsfig{file=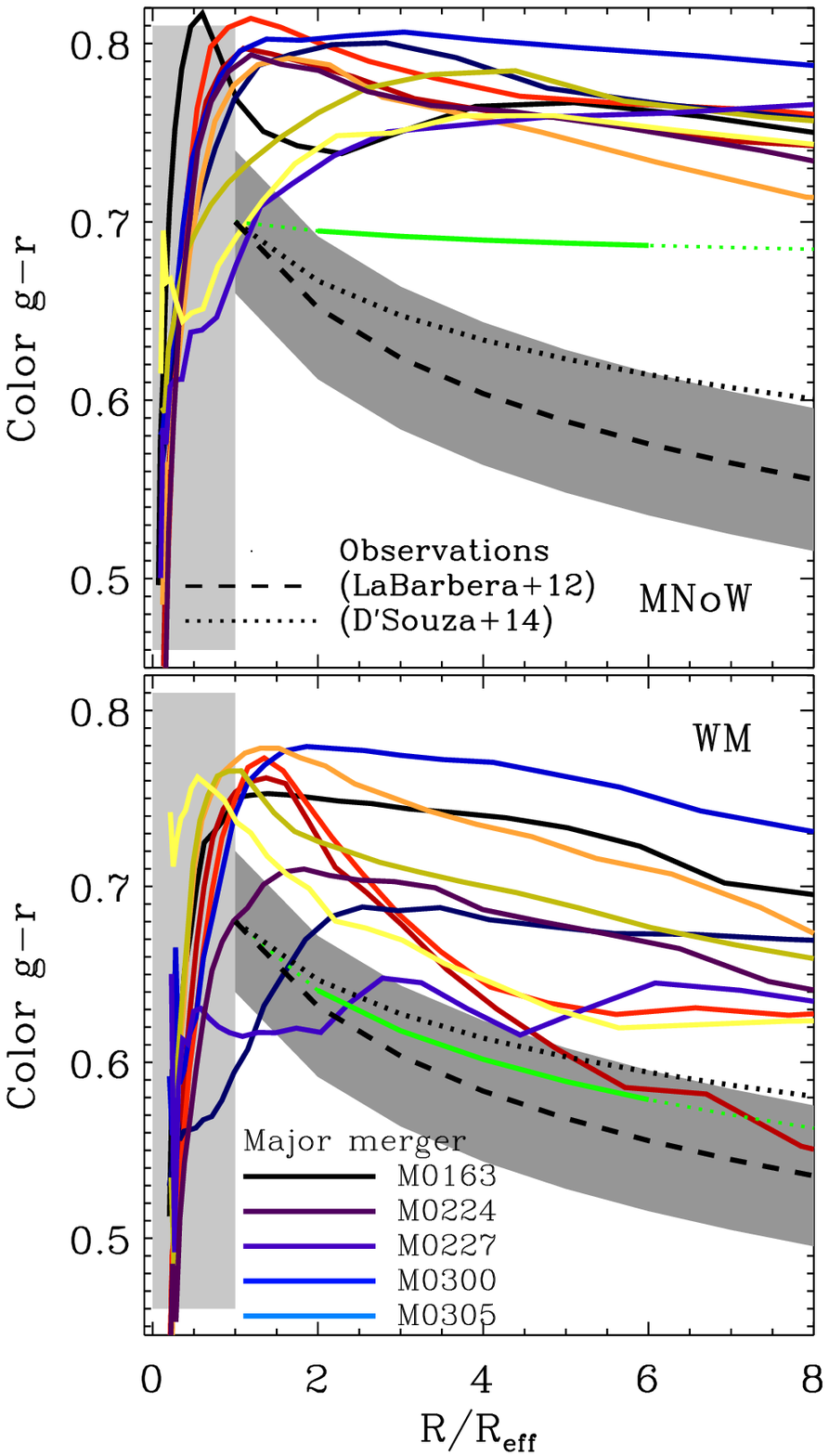, width=0.31\textwidth}\hspace{-0.6cm}
\epsfig{file=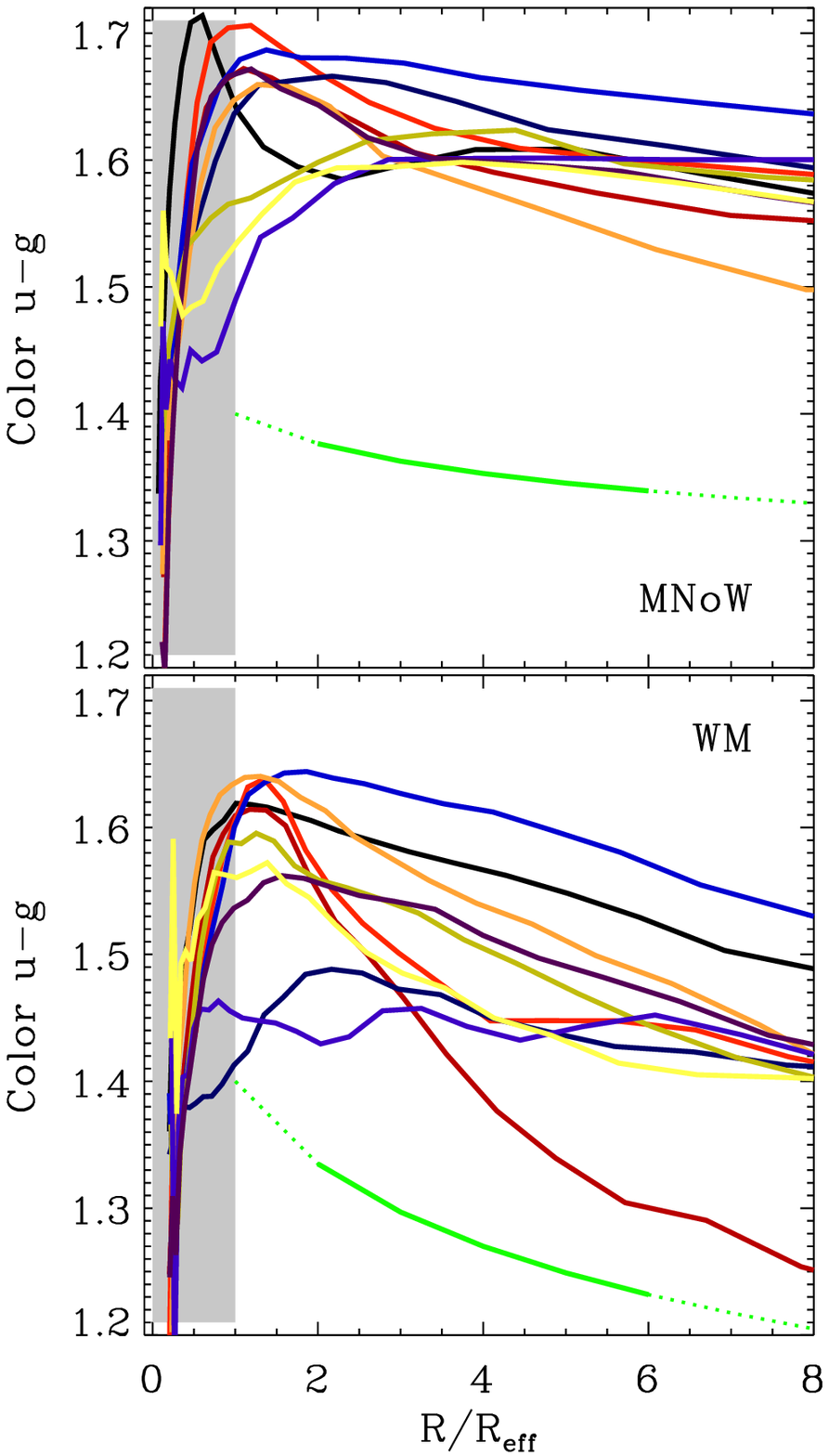, width=0.31\textwidth}
\end{minipage}
\vspace{-0.3cm}
 \caption{Colour gradients (g-i, g-r, and u-g, from left to right) at
   $z=2,1,0$ (from top to bottom) for the ten galaxies (different
   colours) in the MNoW and WM models. Galaxies with major mergers
   since $z=2$ are indicated by black-blue colours (those have
   typically the flattest colour gradients at $z=0$), while galaxies
   with only minor mergers since $z=2$ are illustrated by red-yellow
   colors (having steeper gradients).  
   At $r > 2 R_{\mathrm{eff}}$, present-day WM galaxies show negative
   colour gradients in the outer parts, which are steeper than for the
   MNoW galaxies.
}\label{Colorgrads}
\end{figure*}

\begin{table*}
\centering
\begin{tabular}{ | p{0.55cm} || p{0.5cm} p{0.5cm} p{0.5cm} p{0.5cm}
    p{0.5cm} p{0.5cm} p{0.5cm} p{0.5cm} p{0.5cm} p{0.5cm} p{0.5cm} p{0.5cm} p{0.5cm}
    p{0.5cm} p{0.5cm} p{0.5cm} p{0.5cm} p{0.5cm} |} 
\hline \multicolumn{18}{c}{ {\bf{MNoW}} }\\ \hline
{\bf{ID}}  & \multicolumn{2}{c}{$\nabla
(gi,z0)$} & \multicolumn{2}{c}{$\nabla
(gi,z1)$} & \multicolumn{2}{c}{$\nabla
(gi,z2)$}  & \multicolumn{2}{c}{$\nabla
(gr,z0)$} & \multicolumn{2}{c}{$\nabla
(gr,z1)$} & \multicolumn{2}{c}{$\nabla
(gr,z2)$} &  \multicolumn{2}{c}{$\nabla
(ug,z0)$} & \multicolumn{2}{c}{$\nabla
(ug,z1)$} & \multicolumn{2}{c}{$\nabla
(ug,z2)$}\\ \vspace{-0.5cm}
 &  {\scriptsize{1e-1}} & {\scriptsize{1e-3}} &
 {\scriptsize{1e-1}} & {\scriptsize{1e-2}} &
 {\scriptsize{1e-1}} & {\scriptsize{1e-2}} &
 {\scriptsize{1e-1}} & {\scriptsize{1e-3}} &
 {\scriptsize{1e-1}} & {\scriptsize{1e-3}} &
 {\scriptsize{1e-1}} & {\scriptsize{1e-2}} &
{\scriptsize{1e-1}} & {\scriptsize{1e-3}} &
{\scriptsize{1e-1}} & {\scriptsize{1e-2}} &
{\scriptsize{1e-1}} & {\scriptsize{1e-2}} \\
&  {\scriptsize{mag}} & {\scriptsize{mag}} &
 {\scriptsize{mag}} & {\scriptsize{mag}} &
 {\scriptsize{mag}} & {\scriptsize{mag}} &
 {\scriptsize{mag}} & {\scriptsize{mag}} &
 {\scriptsize{mag}} & {\scriptsize{mag}} &
 {\scriptsize{mag}} & {\scriptsize{mag}} &
{\scriptsize{mag}} & {\scriptsize{mag}} &
{\scriptsize{mag}} & {\scriptsize{mag}} &
{\scriptsize{mag}} & {\scriptsize{mag}} \\
 &  {\scriptsize{/dex}} & {\scriptsize{/kpc}} &
 {\scriptsize{/dex}} & {\scriptsize{/kpc}} &
 {\scriptsize{/dex}} & {\scriptsize{/kpc}} &
 {\scriptsize{/dex}} & {\scriptsize{/kpc}} &
 {\scriptsize{/dex}} & {\scriptsize{/kpc}} &
 {\scriptsize{/dex}} & {\scriptsize{/kpc}} &
{\scriptsize{/dex}} & {\scriptsize{/kpc}} &
{\scriptsize{/dex}} & {\scriptsize{/kpc}} &
{\scriptsize{/dex}} & {\scriptsize{/kpc}} \\
\hline \hline 
{ M0163} & $+1.8$ & $+0.0$ & $+1.9$ & $+1.2$ & $-4.6$ & $-5.4$&
$+1.2$& $+1.0$& $+1.6$ & $+9.7$ & $-3.2$ & $-3.8$ & $+0.9$ & $+0.8$ & $+8.7$ & $+5.5$ & $-2.6$ & $-3.0$ \\

{ M0209} & $-1.0$ & $-2.3$ & $+1.1$ & $+1.1$ & $-3.4$& $-5.1$& 
$-0.7$& $-1.7$& $+0.9$ & $+9.2$ & $-2.6$ & $-3.8$ & $-1.6$ & $-4.1$ &$+0.7$ & $+1.3$ & $-4.7$ & $-6.9$ \\

{ M0215} & $-1.2$ & $-2.5$ & $-2.3$ & $-1.1$ & $+2.8$& $+3.7$& 
$-0.8$& $-1.9$& $-1.5$ & $-7.6$ & $+2.0$ & $+2.7$ & $-1.8$ & $-4.2$ & $-5.0$ & $-2.9$ & $+2.6$ & $+3.4$ \\

{ M0224} & $-1.1$ & $-2.1$ &$-0.9$ & $-0.6$ & $-1.6$& $-1.4$& 
$-0.7$& $-1.4$& $-0.6$ & $-4.4$ & $-1.0$ & $-0.9$ & $-1.3$ & $-2.7$ & $-1.8$ & $-1.2$ & $-0.4$ & $-0.5$\\

{ M0227} & $-0.6$ & $-0.9$ &$+4.6$ & $+2.9$ & $+3.4$& $+1.9$& 
$-0.4$& $-0.7$& $+3.6$ & $+23$ & $+2.9$ & $+1.6$ & $-1.0$ & $-1.8$ & $+10$ & $+6.4$ & $+5.4$ & $+3.1$\\

{ M0259} & $-1.8$ & $-4.3$ &$+7.0$ & $+5.3$ & $+0.4$& $+0.5$& 
$-1.2$& $-3.3$& $+5.4$ & $+41$ & $+0.5$ & $+0.5$ & $-2.5$ & $-7.0$ & $+10$ & $+7.8$ & $+1.3$ & $+1.4$\\

{ M0290} & $-0.5$ & $-1.4$ &$+9.9$ & $+7.5$ & $-4.9$& $-4.3$& 
$+0.7$& $+1.3$& $+9.5$ & $+58$ & $-3.6$ & $-3.2$ & $+0.7$ & $+1.1$ & $+25$ & $+15$ & $-3.7$ & $-3.3$\\

{ M0300} & $+0.7$ & $+1.2$ &$-1.1$ & $-1.0$ & $+2.9$& $+2.7$& 
$+0.4$& $+0.8$& $-0.7$ & $-6.6$ & $+2.2$ & $+2.0$ & $+0.3$ & $+0.6$ & $-8.7$ & $-5.8$ & $+3.4$ & $+2.9$\\

{ M0305} & $-1.1$ & $-1.1$ & $-1.7$ & $-1.1$ & $-0.1$& $+0.0$& 
$-0.8$& $-0.9$&$-1.3$ & $-8.1$ & $+0.0$ & $+0.1$ & $-1.8$ & $-1.9$ & $-1.6$ & $-1.0$ & $-0.3$ & $-0.3$ \\

{ M0329} & $+0.4$ & $+0.7$ & $-0.5$ & $-0.4$ & $+3.6$& $+5.1$& 
$+0.6$& $+1.1$& $-0.3$ & $-2.3$ & $+2.8$ & $+3.9$ & $+0.1$ & $+0.3$ & $-0.6$ & $-0.4$ & $+1.6$ & $+2.2$\\ \hline \hline
{\bf Mean} & $-0.6$ & $-1.3$& $+2.0$ & $+1.4$ & $-0.1$ & $-0.2$
&$-0.2$ & $-0.5$ & $+1.7$ & $+11$ &
$+0.0$ & $-0.0$ & $-0.8$ & $-1.9$ & $+3.7$ & $+2.5$ & $+0.8$ & $+0.1$ \\ \hline
\end{tabular}
\caption{Halo ID and the slopes for the fitted
  g-i  $\nabla_{l/k}(gi)$, g-r $\nabla_{l/k}(gr)$ and u-g
  $\nabla_{l/k}(ug)$ color gradients at $z=0,1,2$ of the
  \textit{central} galaxies in the MNoW run. The gradients (fitted
  between $ 2-6 \times R_{\mathrm{eff}} $) are given in 1e-1~mag$/$dex
  and 1e-3 or 1e-2~mag$/$kpc.}    
\label{sim_coltab_MNoW}
\end{table*}

\begin{table*}
\centering
\begin{tabular}{ | p{0.55cm} || p{0.5cm} p{0.5cm} p{0.5cm} p{0.5cm}
    p{0.5cm} p{0.5cm} p{0.5cm} p{0.5cm} p{0.5cm} p{0.5cm} p{0.5cm} p{0.5cm} p{0.5cm}
    p{0.5cm} p{0.5cm} p{0.5cm} p{0.5cm} p{0.5cm} |} 
\hline \multicolumn{18}{c}{ {\bf{WM}} }\\ \hline
{\bf{ID}}  & \multicolumn{2}{c}{$\nabla
(gi,z0)$} & \multicolumn{2}{c}{$\nabla
(gi,z1)$} & \multicolumn{2}{c}{$\nabla
(gi,z2)$}  & \multicolumn{2}{c}{$\nabla
(gr,z0)$} & \multicolumn{2}{c}{$\nabla
(gr,z1)$} & \multicolumn{2}{c}{$\nabla
(gr,z2)$} &  \multicolumn{2}{c}{$\nabla
(ug,z0)$} & \multicolumn{2}{c}{$\nabla
(ug,z1)$} & \multicolumn{2}{c}{$\nabla
(ug,z2)$}\\ \vspace{-0.5cm}
 &  {\scriptsize{1e-1}} & {\scriptsize{1e-3}} &
 {\scriptsize{1e-1}} & {\scriptsize{1e-2}} &
 {\scriptsize{1e-1}} & {\scriptsize{1e-3}} &
 {\scriptsize{1e-1}} & {\scriptsize{1e-3}} &
 {\scriptsize{1e-1}} & {\scriptsize{1e-3}} &
 {\scriptsize{1e-1}} & {\scriptsize{1e-3}} &
{\scriptsize{1e-1}} & {\scriptsize{1e-3}} &
{\scriptsize{1e-1}} & {\scriptsize{1e-2}} &
{\scriptsize{1e-1}} & {\scriptsize{1e-2}} \\
&  {\scriptsize{mag}} & {\scriptsize{mag}} &
 {\scriptsize{mag}} & {\scriptsize{mag}} &
 {\scriptsize{mag}} & {\scriptsize{mag}} &
 {\scriptsize{mag}} & {\scriptsize{mag}} &
 {\scriptsize{mag}} & {\scriptsize{mag}} &
 {\scriptsize{mag}} & {\scriptsize{mag}} &
{\scriptsize{mag}} & {\scriptsize{mag}} &
{\scriptsize{mag}} & {\scriptsize{mag}} &
{\scriptsize{mag}} & {\scriptsize{mag}} \\
 &  {\scriptsize{/dex}} & {\scriptsize{/kpc}} &
 {\scriptsize{/dex}} & {\scriptsize{/kpc}} &
 {\scriptsize{/dex}} & {\scriptsize{/kpc}} &
 {\scriptsize{/dex}} & {\scriptsize{/kpc}} &
 {\scriptsize{/dex}} & {\scriptsize{/kpc}} &
 {\scriptsize{/dex}} & {\scriptsize{/kpc}} &
{\scriptsize{/dex}} & {\scriptsize{/kpc}} &
{\scriptsize{/dex}} & {\scriptsize{/kpc}} &
{\scriptsize{/dex}} & {\scriptsize{/kpc}} \\
\hline \hline 
{ M0163} & $-0.8$ & $-1.5$ & $-5.3$& $-5.3$& $-0.9$& $-7.1$& $-0.4$&
$-0.9$& $-4.3$& $-42.$& $-0.6$& $-5.2$& $-1.4$& $-2.5$& $-10$& $-10.$& $-1.3$& $-1.1$\\ 
{ M0209} & $-5.2$ & $-14$ & $+4.5$& $+2.0$& $+0.3$& $+5.4$& $-3.1$&
$-9.9$& $+3.6$& $+16$& $+0.3$& $+4.9$& $-5.8$& $-19$& $+4.7$& $+2.1$& $+0.6$& $+0.9$\\ 
{ M0215} & $-3.6$ & $-7.9$ & $-0.7$& $-0.3$& $-0.5$& $-5.5$& $-2.8$&
$-6.9$& $-0.3$& $-0.6$& $-0.4$& $-4.8$& $-3.5$& $-8.2$& $+1.5$& $+0.9$& $-2.4$& $-3.2$\\ 
{ M0224} & $-1.6$ & $-4.0$ & $-1.0$& $-1.1$& $+1.2$& $+10$& $-0.1$&
$-0.7$& $-0.8$& $-8.7$& $+0.9$& $+7.2$& $-1.3$& $-3.8$& $-6.2$& $-6.5$& $+1.9$& $+1.8$\\ 
{ M0227} & $-0.6$ & $-1.3$ & $+2.3$& $+1.5$& $+1.5$& $+5.2$& $-0.4$&
$-1.1$& $+2.1$& $+14$& $+1.1$& $+3.9$& $-1.3$& $-2.8$& $+6.7$& $+4.0$& $+1.8$& $+0.6$\\ 
{ M0259} & $-1.7$ & $-4.2$ & $+1.9$& $+0.3$& $+0.0$& $+1.1$& $-1.0$&
$-3.2$& $+1.7$& $+3.2$& $+0.0$& $+1.1$& $-2.4$& $-7.1$& $+3.8$& $+0.7$& $+1.1$& $+0.8$\\ 
{ M0290} & $-2.2$ & $-4.5$ & $-1.2$& $-0.5$& $-2.5$& $-26$& $-0.8$&
$-1.9$& $-0.7$& $-3.1$& $-1.9$& $-20$& $-2.3$& $-4.8$& $-0.1$& $-0.1$& $-2.9$& $-6.2$\\ 
{ M0300} & $-1.4$ & $-4.3$& $+5.5$& $+1.4$& $-0.5$& $-3.5$& $-0.9$&
$-3.2$& $+4.3$& $+11$& $-0.4$& $-2.8$& $-0.5$& $-1.9$& $+7.0$& $+1.7$& $-0.8$& $-0.5$\\ 
{ M0305} & $-2.3$ & $-4.8$ & $-0.1$& $-0.0$& $-1.3$& $-12$& $-0.6$&
$-1.9$& $+0.4$& $+2.2$& $-0.9$& $-8.1$& $-1.2$& $-3.1$& $+4.8$& $+3.1$& $-1.4$& $-1.1$\\ 
{ M0329} & $-2.1$ & $-4.7$& $+1.8$& $+1.4$& $+1.1$& $+8.2$& $-1.6$&
$-4.0$& $+1.7$& $+13$& $+1.0$& $+7.6$& $-1.9$& $-4.6$& $+3.8$& $+2.8$& $+2.1$& $+1.7$\\  \hline \hline
{\bf Mean} & $-2.0$ & $-5.1$& $+0.7$ & $+0.5$ & $-0.2$ & $-2.4$
&$-1.3$ & $-3.4$ & $+0.8$ & $+0.5$ & $-0.1$ & $-1.6$ & $-2.2$ & $-5.8$
& $+1.5$ & $+0.1$ & $-0.1$ & $-0.5$ \\ \hline
\end{tabular}
\caption{Halo ID and the slopes for the fitted
  g-i  $\nabla_{l/k}(gi)$, g-r $\nabla_{l/k}(gr)$ and u-g
  $\nabla_{l/k}(ug)$ color gradients at $z=0,1,2$ of the
  \textit{central} galaxies in the WM run. The gradients (fitted
  between $ 2-6 \times R_{\mathrm{eff}} $) are given in 1e-1~mag$/$dex
  and   1e-3 or 1e-2~mag$/$kpc.}    
\label{sim_coltab_WM}
\end{table*}

Fig. \ref{Colorgrads} shows the g-i (left panels), the g-r (middle
panels) and the u-g (right panels) colour gradients (solid lines) for
the MNoW and the WM galaxies (as indicated in the legend) at
$z=0,1,2$. For a given model and redshift, the overall behaviour for
the gradients with different colours is very similar and the
corresponding slopes (between $2-6\ R_{\mathrm{eff}}$) are
given in tables  \ref{sim_coltab_MNoW} and \ref{sim_coltab_WM}. 

Irrespectively of redshift and colour, WM galaxies are
generally slightly bluer than MNoW galaxies as a consequence of their
younger stellar populations with lower metallicity. 
In addition, all galaxies of both models are at $z=1$ and $z=0$
significantly bluer at the centre, revealing steep positive colour
gradients at $< 1\ R_{\mathrm{eff}}$ (marked by the grey shaded area
in Fig. \ref{Colorgrads}). This is an obvious consequence of the
positive stellar age gradients at these innermost radii (as seen in
Fig. \ref{Agegrads_only10_insacc_mm_zall}). Massive galaxies being so
blue in their centres is an unrealistic artefact of our models most
likely due to missing AGN feedback as already discussed before.

In observational literature (\citealp{2009MNRAS.395..554F, Suh10}),
there is found an interesting correlation in the sense that early-type
galaxies with blue cores (positive colour gradients) are mostly blue
overall, and associated to recent SF. In contrast, the vast majority
of massive early-type galaxies (which we have in our simulated sample)
out to z~1 feature red overall colours and red cores (negative colour
gradients). It will be interesting to see whether simulations with AGN
feedback will help to produce such a correlation, which we postpone to
forthcoming studies.

Turning to larger radii $> 2\ R_{\mathrm{eff}}$ (which are not
expected to be significantly influenced by AGN feedback), at $z=2$,
the shape of the g-i, g-r and u-g colour gradients ranges from strongly
negative (outer parts are bluer) to strongly positive (outer parts are
redder), irrespectively of the model (first two rows of
Fig. \ref{Colorgrads}) so that the average gradients (green lines) are
nearly flat. 

In general, whether colour gradients are positive or negative is driven
by the interplay of the radial distribution of stellar metallicity and
ages: negative metallicity gradients imply more negative colour
gradients, while positive age gradients would result in more positive
colour gradients. Depending on which effect is stronger, leads to
either overall positive or negative colour gradients. At $z=2$, M0227
and M0300 WM galaxies (dark blue lines in second left panel of
Fig. \ref{Agegrads_only10_insacc_mm_zall}), for example, have strongly
positive age gradients outweighing the effect of their negative
metallicity gradients and thus, leading to overall positive slopes for
the colour distribution. Instead, M0209 (red line in second left panel
of Fig. \ref{Agegrads_only10_insacc_mm_zall}), for example, has both a
negative age gradient and metallicity gradient clearly resulting in a
strongly negative colour gradient (second row in Fig. \ref{Colorgrads}).

The situation changes towards $z=1$, where the two models start to
behave differently. Almost all of the MNoW galaxies are now redder at the
outskirts (positive average colour gradient, third row of
Fig. \ref{Colorgrads}). This is a consequence of the relatively shallow
metallicity gradients which can apparently not overcome the slightly
positive age gradients. Instead, the WM galaxies have again both
steeply positive and steeply negative gradients (resulting in an
overall only weakly positive mean slope, green lines in the fourth row
of Fig. \ref{Colorgrads}) depending on whether the positive age or the
negative metallicity gradient predominates.

Finally at $z=0$, the MNoW galaxies have only very shallow colour
gradients, either slightly decreasing or increasing with an average
slopes of $-0.05$~mag/dex, $-0.02$~mag/dex and $-0.08$~mag/dex 
for the g-i, g-r and u-g colors (see table \ref{sim_coltab_MNoW}),
respectively. The shallow colour gradients in the MNoW model stem from
the relatively flat age and only slightly negative metallicity
gradients. As discussed, the former originates from a relatively old
in-situ stellar population (compared to the accreted stars), and the
latter from the accretion of relatively metal-rich stellar systems in
the MNoW model. 

Instead with galactic winds, nearly all of the WM galaxies become
continuously bluer with increasing radius and thus, reveal
significantly steeper negative colour gradients than the MNoW galaxies
(with mean slopes of $-0.20$~mag/dex, $-0.13$~mag/dex and
$-0.22$~mag/dex for the g-i, g-r and the u-g colours, see table
\ref{sim_coltab_WM}). This behaviour is entirely driven by the steeper
metallicity gradients of the WM galaxies washing out any effect of the
slightly increasing age gradients. The result is also consistent with
studies of massive spheroidal galaxies from a visually classified
sample extracted from the Advanced Camera for Surveys/Hubble Space
Telescope (ACS/HST) images (\citealp{2009MNRAS.395..554F}), where a
comparison with a model, assuming a gradient caused by age, predicts a
large change of the colour gradient with redshift, at odds with the
observations. Instead, for a model assuming that the color gradient is
caused by metallicity, they find a good agreement with what is observed.

We want to emphasize that, as at $z=1$ and at $z=0$, both the positive
age gradients and the negative metallicity gradients are mainly driven
by the accretion of old and metal-poor stellar systems (see sections
\ref{metgradients} and \ref{agegradients}), also the steepening of the
colour gradients (either positive or negative) originate from
accretion, particularly in the WM model. 

\begin{figure}
\begin{center}
  \epsfig{file=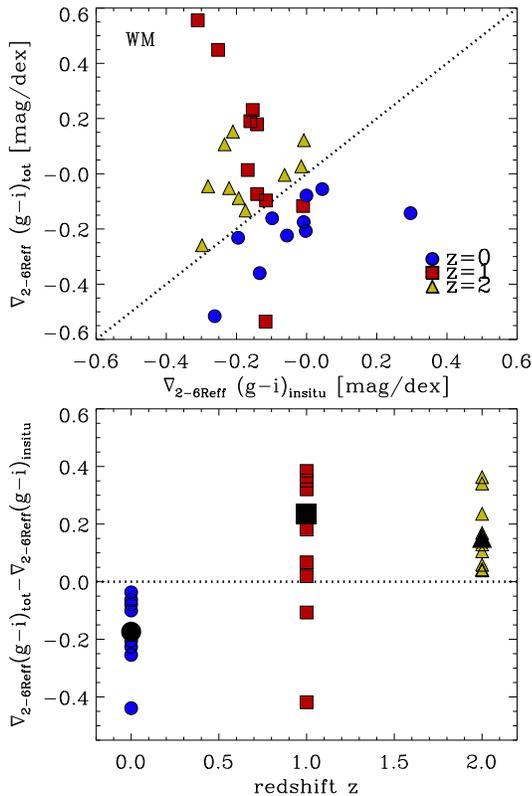, width=0.4\textwidth}
  \caption{Fitted g-i colour gradients (at $2<R/R_{\mathrm{eff}}<6$)
    of the total stellar component versus the ones of the in-situ
    stellar component at $z = 0, 1, $and $2$ (differently coloured
    symbols) for the ten massive galaxies in the WM model. The black
    dotted line indicates equal total and in-situ gradients.} 
\label{Insitu_tot_colorgrad}
\end{center}
\end{figure}

To demonstrate how much of the total colour gradients is really
caused by the accretion of stellar systems, the top panel of
Fig. \ref{Insitu_tot_colorgrad} shows - as an example for the g-i
colors only (but the g-r and u-g colours behave in the same way) - the
total colour gradients versus the in-situ ones at $z=0,1,2$ (indicated
by different colours) for the WM model. The bottom panel illustrates
the corresponding difference between the two gradients versus
redshift.  

At $z=2$, the in-situ gradients are all negative, while the total
gradients are both, negative and positive. This shows that 
accretion (of redder stellar populations due to their age) makes the 
g-i colour gradients on average more positive by +0.15~mag/dex. At
$z=1$, this effect is even amplified, the in-situ g-i colour gradients
(again only negative) become more positive by the accretion by
+0.24~mag/dex. This trend is, however, entirely reversed at $z=0$,
where the slopes of the in-situ distributions become more negative by
$-0.17$~mag/dex through accretion of bluer (as more metal-poor)
stellar systems. \textit{This demonstrates that the average negative
  g-i colour gradient of 0.2~mag/dex is almost entirely 
  driven by stellar accretion}. We will further discuss in section
\ref{Observations} that for being consistent with current, present-day
observations, the stellar accretion seems to play a crucial role.

%*****************************************************************************************************
%*****************************************************************************************************
\section{Effect of the merger history on the gradients}
\label{Mergerhistory}
%*****************************************************************************************************
%*****************************************************************************************************

We demonstrated in section \ref{metgradients}, \ref{agegradients} and
\ref{colorgradients} that the stellar feedback model has a significant
effect on stellar population gradients (e.g. a steepening at $z=0$),
particularly at lower redshifts. Apart from the strong influence of
the feedback model, however, recent observations also suggest a
significant effect of the individual past merger history on the
strength of the metallicity gradients: \citet{Kewley10}, for example,
show that metallicity gradients in close pairs are significantly
shallower than those in isolated galaxies suggesting a strong
relationship between the slope of the gradients and the galaxy (major)
mergers.   

To visually relate the recent merger histories of present-day WM 
galaxies\footnote{We discuss that only for the WM model, but the
  effects are the qualitatively same in the MNoW model.} with their
metallicity, age and colour gradients, the black-blue lines in
Figs. \ref{Metgrads_only10_insacc_mm_zall}, 
\ref{Agegrads_only10_insacc_mm_zall} and \ref{Colorgrads}
indicate galaxies which experienced at least one major galaxy merger
since $z=1$, while the red-yellow lines illustrate those having
undergone only minor galaxy mergers (using the results shown in
Fig. \ref{Minmajmerg_evol}). This already shows ``visually'' that
present-day galaxies having experienced a recent major merger have
typically flatter gradients than those with a more quiet merger
history. 

\begin{figure*}
\centering
\begin{minipage}{1.0\textwidth}
\epsfig{file=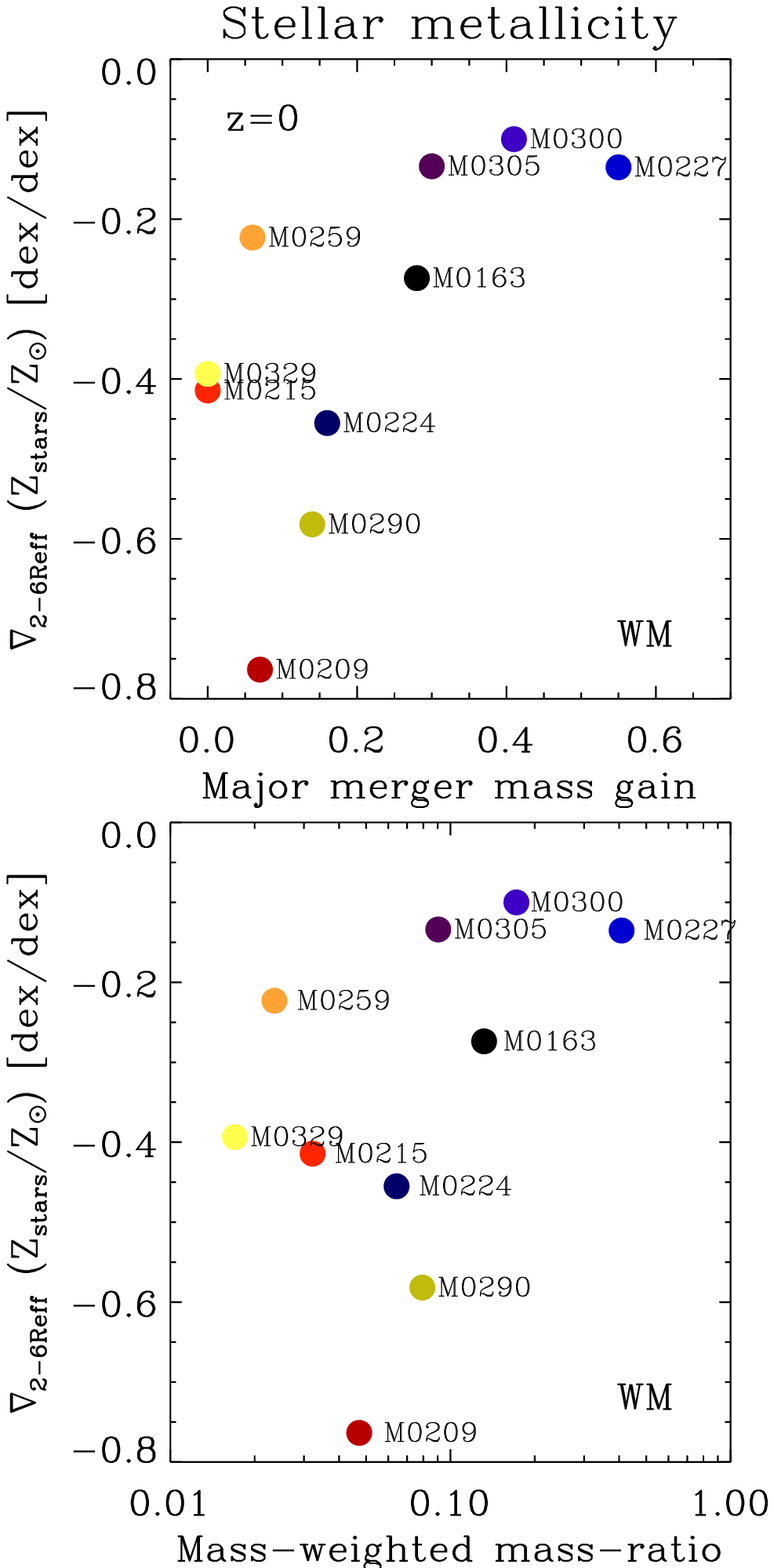,
  width=0.34\textwidth}\hspace{-0.35cm}
\epsfig{file=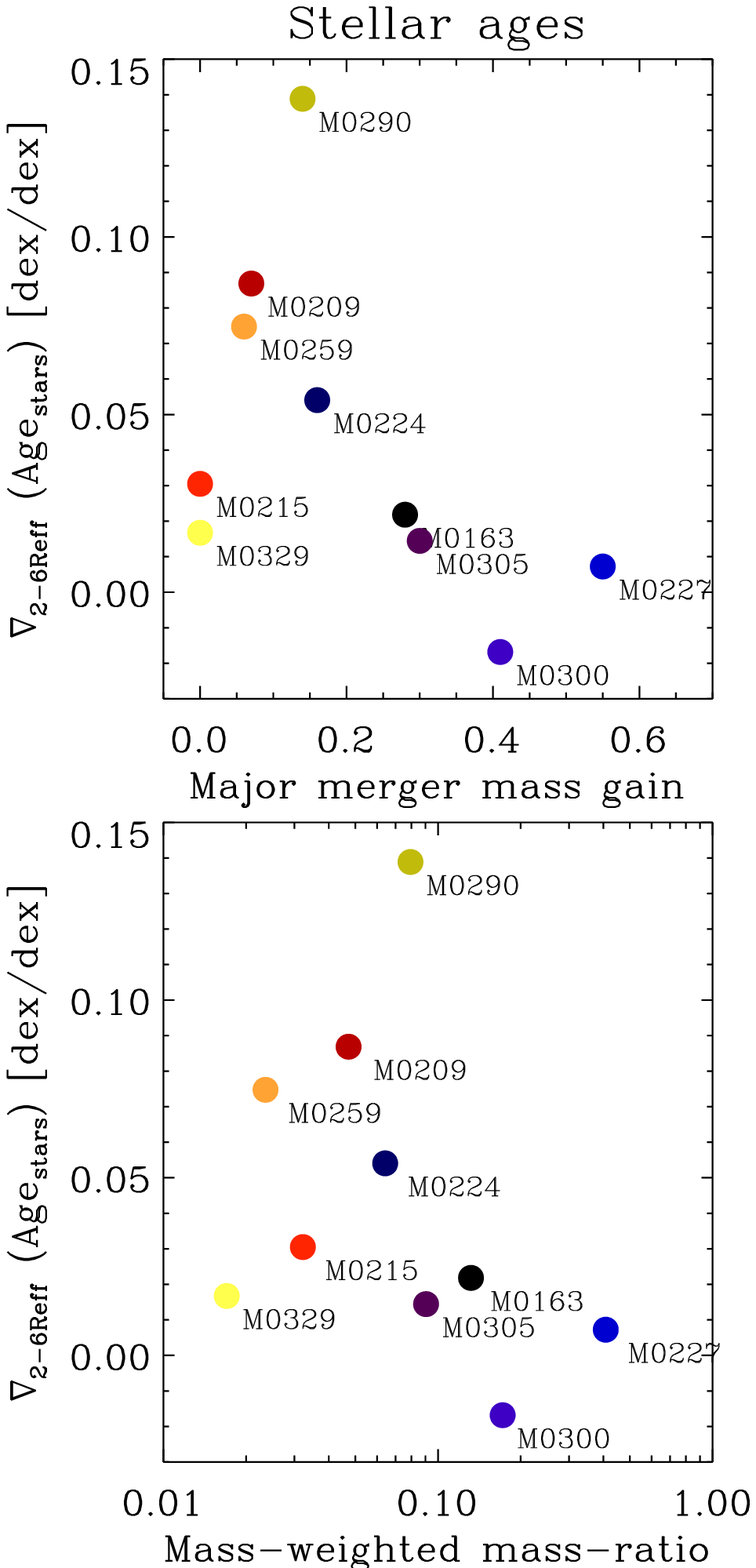, width=0.34\textwidth}\hspace{-0.3cm}
\epsfig{file=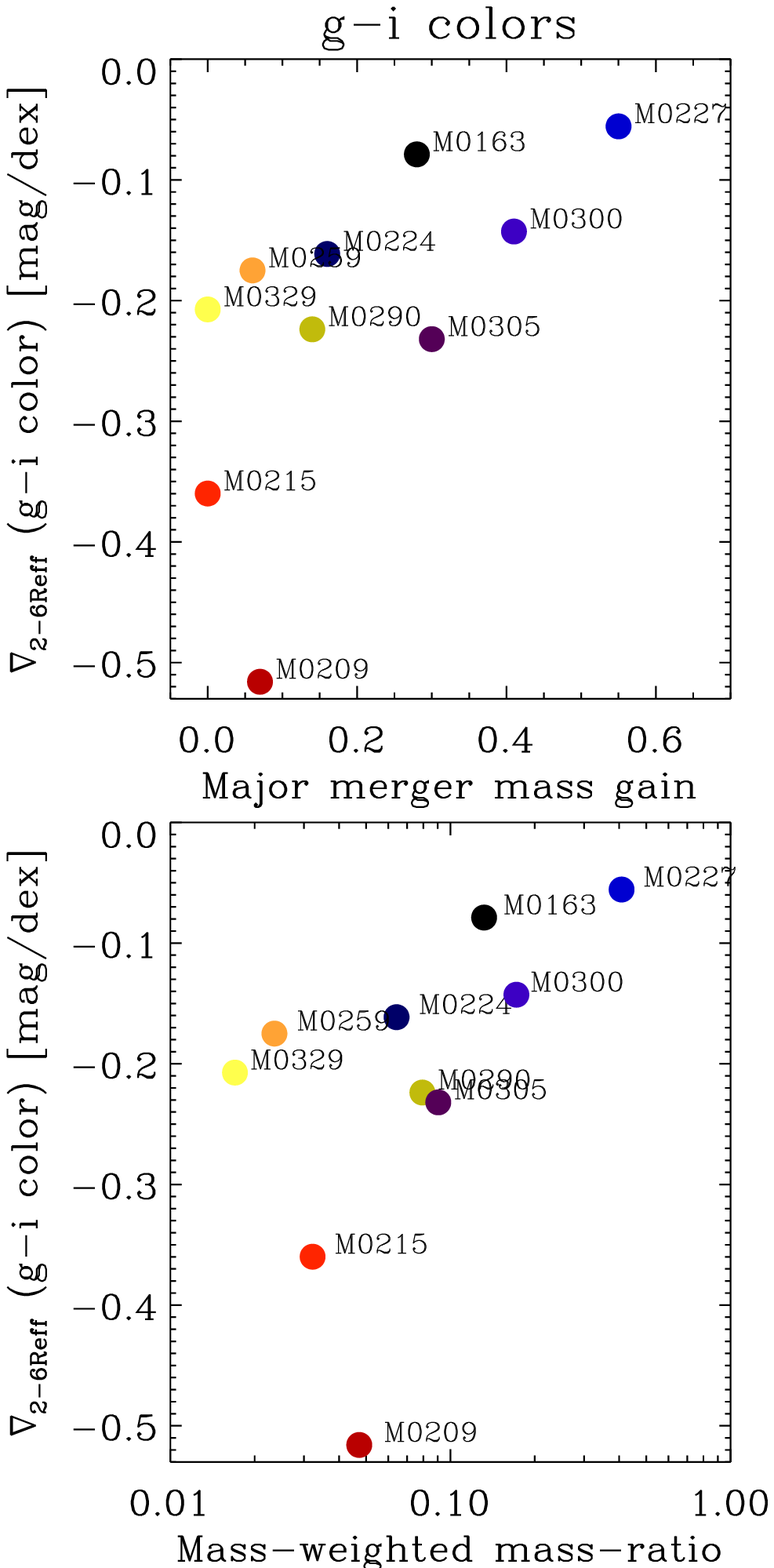, width=0.34\textwidth}
  \caption{\textit{Left panels:}Stellar metallicity gradients (between $2-6
    R_{\mathrm{eff}}$) versus the mass gain through major mergers (top
    panel) and versus the mass-weighted merger mass-ratio (bottom
    panels) for the ten re-simulated galaxies in the WM simulations at
    $z=0$. The points are colour-coded according to their past major
    mergers as in Fig. \ref{Metgrads_only10_insacc_mm_zall}. The
    higher the major merger mass gain or the mass-weighted mass-ratio,
    the smaller is the slope of the metallicity
    gradient. \textit{Middle panels:} The same as in the left panels,
    but for stellar age at $z=0$. The higher the major merger mass
    gain, the flatter are the age gradients. \textit{Right panels:}
    The same as in the left panels but for g-i colours at $z=0$. The
    higher the major merger mass gain or the mass-weighted mass-ratio,
    the smaller is the slope of the colour gradient. This trend is
    entirely driven by the negative metallicity gradients.}    
 {\label{Slope_major}}
\end{minipage}
\end{figure*}

Fig. \ref{Slope_major} quantifies the connection between the
galaxy merger history and the steepness of the metallicity (left
panels), age (middle panels) and g-i colour gradients (right panels):
we show the fitted metallicity/age/g-i colour gradients at $z=0$ for WM
galaxies (shown in table \ref{sim_tab_WM} and \ref{sim_coltab_WM})
versus the mass gain by major mergers (top panel) and versus the
mass-weighted merger mass-ratio (bottom panel). The mass gain by major
mergers considers the entire stellar mass which was brought into the main
galaxy by major mergers since $z=2$ normalised to the present-day
stellar mass. The mass-weighted mass-ratio (see \citealp{Oser12}), a
measure for the ``strength'' of a merger, is computed according to     
\begin{equation}
\sum_i \left(
  \frac{M_{\mathrm{sat,i}}}{M_{\mathrm{cent}}(z=0)-M_{\mathrm{cent}}(z=2)}
  \times \mu_i \right),
\end{equation}
where $\mu_i$ is the merger mass-ratio, $M_{\mathrm{sat,i}}$ the
stellar mass of the infalling satellite and $M_{\mathrm{cent}}(z=0) -
M_{\mathrm{cent}}(z=2)$ the total growth in stellar mass since $z=2$.   

As expected, the stellar age gradients (middle panels) tend to be
negatively related with both quantities which are shown, as galaxies
having undergone major mergers have a zero or even slightly negative
age gradient, while those galaxies having experienced only minor
mergers can have a broad range of slopes (from very shallow to more
positive). 

Instead, the fitted metallicity and g-i color gradients strongly
correlate with the past merger history: for a major merger
mass gain above 20~\% or a mass-weighted mass-ratio above 0.1, the
metallicity and colour gradients are flatter than $-0.3$~dex/dex and
$-0.2$~mag/dex, respectively. Instead, for lower x-values, the
metallicity and colour gradients are mostly more negative than
$-0.4$~dex/dex and $-0.2$~mag/dex, respectively\footnote{Note,
  however, that for higher redshifts $z=1$ and $z=2$, we do not find
  any clear correlation between the merger history and the slope of
  the metallicity, age and colour gradients (not explicitly shown),
  most likely because the mergers are typically more gas-rich inducing
  a lot of central star formation.}.

For the metallicity gradients, this is due to the fact that the
accreted stars show a huge variety of metallicities from
$Z_{\mathrm{stars}}/Z_\odot \sim -0.6$ to $Z_{\mathrm{stars}}/Z_\odot
\sim +0.1$ (see bottom middle panel of
Fig. \ref{Metgrads_only10_insacc_mm_zall}) depending on the exact 
merger history: in case of a recent major merger, the accreted
metallicity is significantly larger than without a major merger (as
more massive galaxies have higher metallicity, see left column of
Fig. \ref{Mgal_sat_only10}). The higher metallicity of the accreted
stars, together with the different mixing behaviour (violent
relaxation) in case of major mergers, flattens the total metallicity
gradients. Interestingly, accreted metallicity gradients can also have 
negative slopes, indicating the metal poor stars living in the outer
regions of accreted satellites are stripped at larger radii (see
e.g. \citealp{2012MNRAS.425.3119H, 2013MNRAS.429.2924H}). 

The above explanation is also valid for the relation between colour
gradients and the individual merger history, as at $z=0$, the
negative colour gradients closely follow the metallicity gradients
(outweighing the shallow positive age gradients). 

The result that major mergers flatten metallicity gradients is
consistent with earlier studies of e.g. \citet{2004MNRAS.347..740K}
who performed chemo-dynamical simulations of 124 elliptical galaxies in
a cosmological context. For the evolution of the metallicity gradients
they find a destruction of steep gradients by mergers to an extent
dependent on the progenitor mass ratio. Also \citet{Rupke10} show,
using equal mass merger simulations, that the gradients flatten
shortly after the first per-centre passage by radial inflow of
low-metallicity gas from the outskirts of the two merging (disk)
galaxies.  

Fig. \ref{Slope_major} also shows that - when considering colour and
metallicity gradients, the ``M0259'' galaxy seems to be an
``outlier'' having a shallow gradient despite of the small
contribution by major mergers. However, its merger history is
extremely quiet with only one minor merger after $z=1$ (see
Fig. \ref{Minmajmerg_evol}) so that the metallicity is dominated by 
that of  in-situ formed stars out to large radii (see bottom right
panel of Fig. \ref{Metgrads_only10_insacc_mm_zall}) leading to a
shallow total metallicity and colour gradient. This also demonstrates
that both the occurrence of major mergers and the absence of
  minor mergers can  significantly flatten the metallicity gradients,
while a couple of minor mergers typically help to steepen the gradients.

Overall, this result bears an important implication for observations,
as it can help to re-construct the past assembly history for observed
present-day metallicity and colour gradients:
The relation between the steepness of the gradients and the
  individual merger history implies that observed massive galaxies
  having steep outer gradients most likely have not experienced any
  major merger event after $z=1$, but instead have had numerous minor
  mergers.  

%*****************************************************************************************************
%*****************************************************************************************************
\section{Discussion}
\label{Discussion}
%*****************************************************************************************************
%*****************************************************************************************************

%*****************************************************************************************************
%*****************************************************************************************************
\subsection{Origin of insitu formed stars at large radii}
\label{insitularger}
%*****************************************************************************************************
%*****************************************************************************************************

We have seen in section \ref{metgradients} that -- at least for
  the wind model -- insitu star formation can dominate the stellar 
populations in central galaxies out to $\sim$4-5 $R_{\mathrm{eff}}$ at 
$z=0$ (see the right panels in
Figs. \ref{Insitu_tot_metgrad} and \ref{Insitu_tot_agegrad}). At first
sight, it is not clear whether those star particles formed already at
such large radii or whether they instead formed in the central
region, but were then migrated out to larger radii. Investigating
the origin of insitu formed stars presently residing at large radii is
particularly important for exploring the significance of missing AGN
feedback on our analysis (see next subsection). 

\begin{figure}
\begin{center}
  \epsfig{file=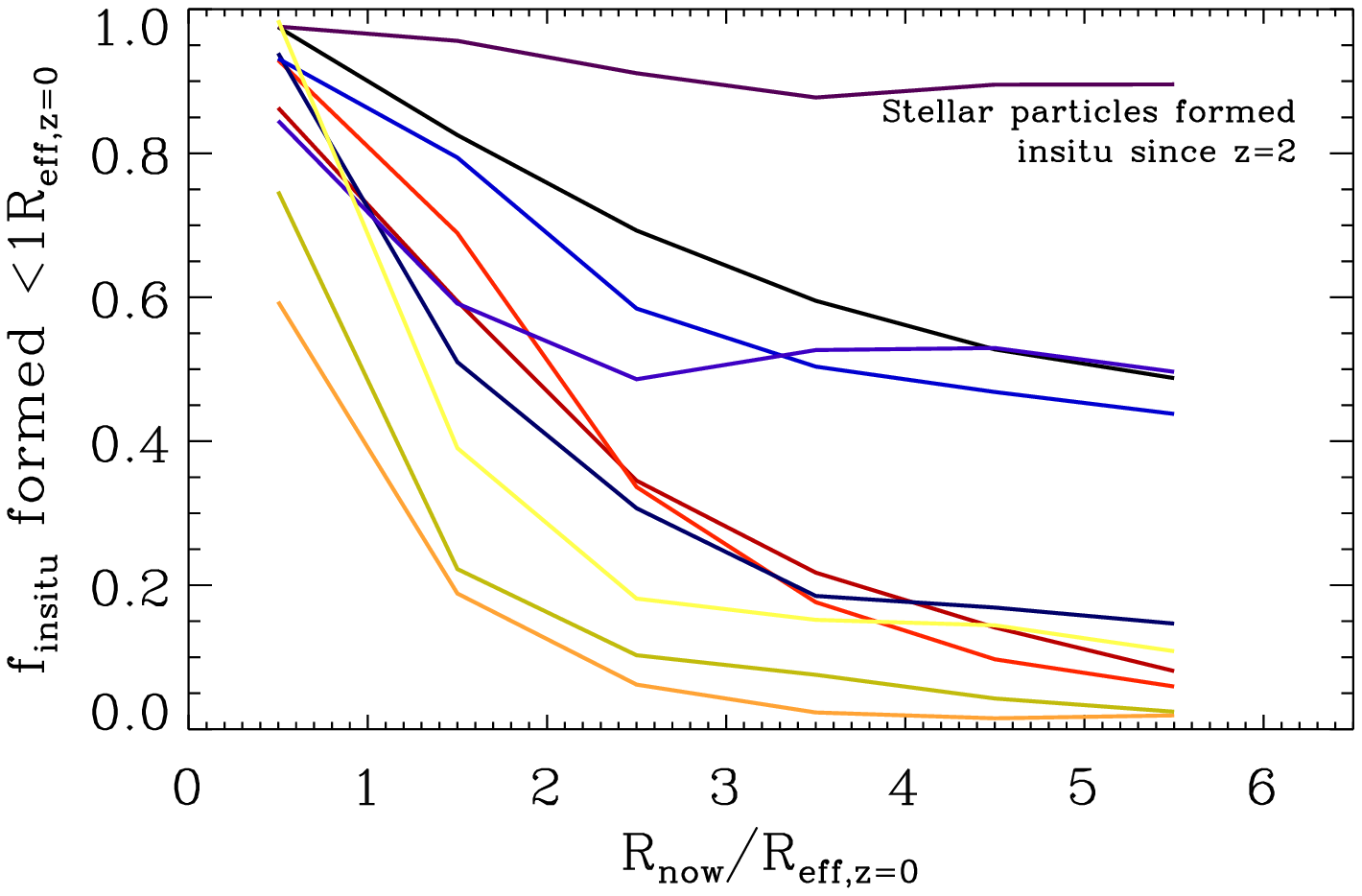, width=0.45\textwidth}
  \epsfig{file=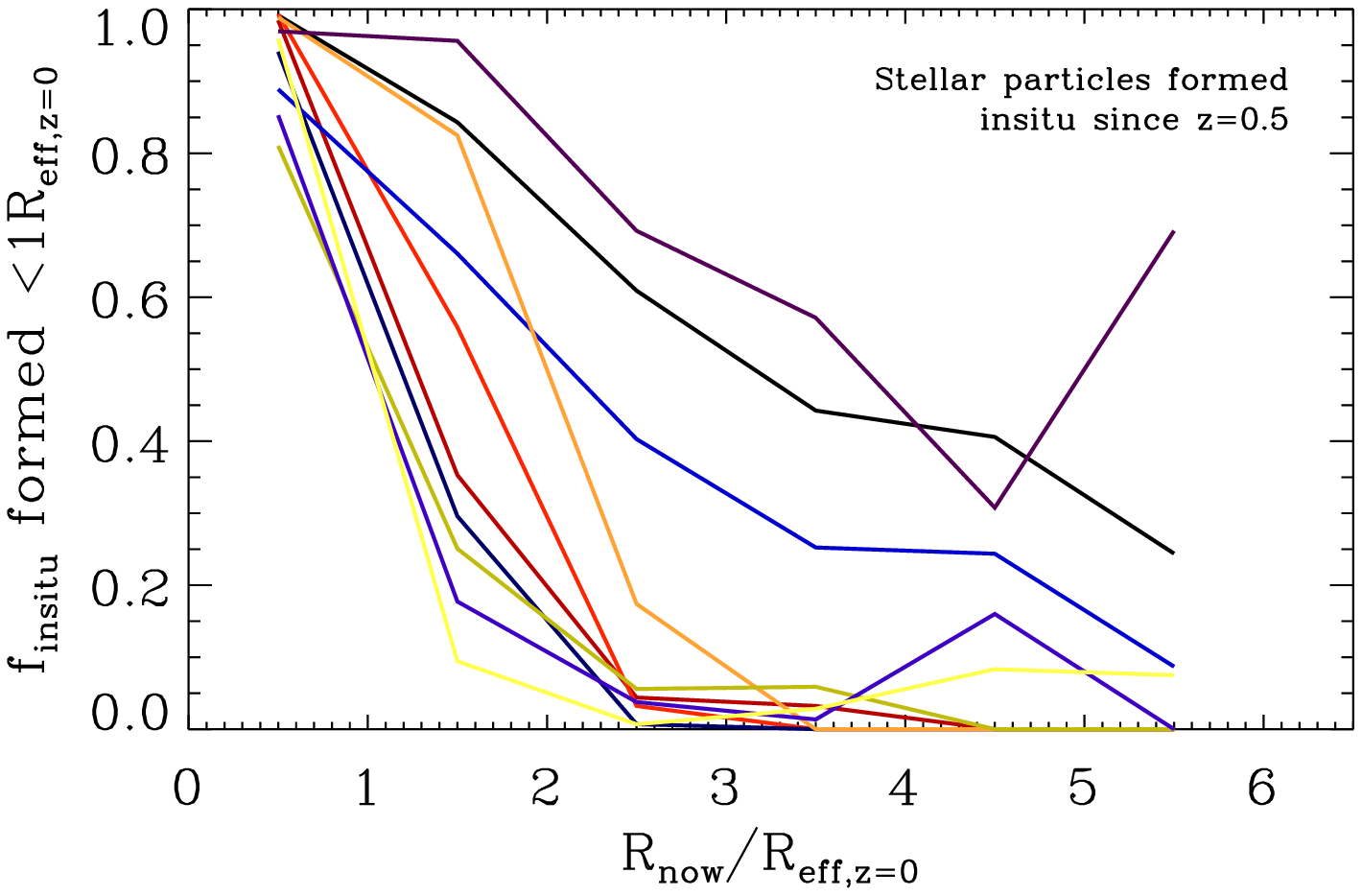, width=0.45\textwidth}
  \caption{ Fraction of present-day in-situ stars that have formed
    inside the present-day (physical) effective radius
    ($<R_{\mathrm{eff,z=0}}$) and have migrated outwards versus the
    present-day radial distance to the galaxy center for the ten
    massive galaxies in the wind model (differently colored lines).
    In the top panel, we consider insitu star formation since $z=2$,
    while in the bottom panel only since $z=0.5$. The color coding is
    the same as throughout the paper, e.g. in
    Fig. \ref{Metgrads_only10_insacc_mm_zall} and therefore,
    visualising the connection with the amount of major mergers
    experienced since $z=2$ (blue-lila colors). Galaxies, which have
    recently experienced a major merger, can have a large fraction of
    centrally formed insitu stars having migrated even out to radii
    of $6 \times R_{\mathrm{eff}}$.}   
\label{Scatteredinsitu}
\end{center}
\end{figure}

 Fig. \ref{Scatteredinsitu} shows the fraction of present-day
  in-situ stars that have formed inside the present-day (physical)
  effective radius  ($<R_{\mathrm{eff,z=0}}$) and have migrated
  outwards versus the present-day radial distance to the galaxy
  center. In the top panel, we consider insitu star formation since
  $z=2$, while in the bottom panel only since $z=0.5$.  

  The majority of the centrally formed stars at
  higher redshifts are still residing in the central part
  ($<1 \times R_{\mathrm{eff,z=0}}$) at $z=0$. In particular for
  galaxies not having experienced any recent major mergers 
  (see yellow-red lines), the fraction of migrated stars is below 30\%
  (since z=2) and 15\% (since z=0.5). Therefore, the majority of
  in-situ stars at the investigated radii were also born at large
  radii.  

  The in-situ component of galaxies with one or more recent
  major mergers (since $z=2$ or since $z=0.5$, see blue-lila lines)
  can , however, be dominated by migrated stars  (up to 90\% in the
  most extreme  case).  But for those galaxies the global stellar
  populations gradients are widely dominated by accretion of stars
  beyond $2 \times R_{\mathrm{eff,z=0}}$ weakening the overall effect
  of stellar migration.

%*****************************************************************************************************
%*****************************************************************************************************
\subsection{Possible effects of AGN feedback on stellar population gradients}
\label{AGNfb}
%*****************************************************************************************************
%*****************************************************************************************************

 The study of \citet{2013MNRAS.436.2929H} has already
   indicated that -- compared to abundance matching models -- in 
   the simulations  (in particular in the wind model) SFRs
   are over-estimated in central galaxies residing in massive halos
   ($>3 \times 10^{12} M_\odot$) \textit{only after} $z=0.5$, which
   can be, therefore, expected to be mainly suppressed by AGN
   feedback. Also the study of \citet{Choi14}, investigating the
   impact of mechanical and radiative AGN feedback in cosmological
   zoom simulations, illustrates that the SFRs particularly decline
   after z=0.6 with respect to the simulations without (or
   inefficient) AGN feedback (see their Fig. 8). We may therefore
   speculate that additional feedback from black holes will
   preferentially affect the \textit{present-day} stellar population
   gradients, but less those at higher redshifts $z=1,2$. 

In addition, as we have shown in the last subsection, stellar
  particles formed insitu within $1 \times R_{\mathrm{eff}}$ since
  $z=2, 0.5$, which are most likely affected by AGN feedback, are
  hardly migrated beyond $2-3 \times R_{\mathrm{eff}}$ at present-day,
  at least for those galaxies which have \textit{not} experienced a
  recent major merger. The fraction of the corresponding migrated
  stars is lower than 30\% or 15\% when considering insitu star
  formation since $z=2$ or $z=0.5$, respectively. We, therefore, 
  conclude that AGN feedback will hardly affect the (steep)
  metallicity and color gradients of galaxies without any major
  mergers. This might be especially true for AGN feedback affecting
  the late insitu star formation after $z=0.5$.

Instead, energy release from AGN may play a more important role
  for the stellar population gradients of galaxies 
  having undergone major mergers. As their global stellar populations
  are mostly dominated by accreted stellar systems (beyond $2 \times
  R_{\mathrm{eff}}$), insitu formed stars (and thus, the migration of
  stars) play a less significant role for the outer population
  gradients. Moreover, as that gradients are widely flat, we would
  only expect a decrease of the \textit{global} stellar metallicity
  content or color (making them less metal-rich and thus, bluer) but
  no significant change of slopes themselves. Overall, we may,
  therefore, conclude that additional AGN feedback should not
  significantly alter the outer stellar population gradients beyond 
  $2 \times R_{\mathrm{eff}}$. However, a further, more detailed
  quantification of how AGN feedback may affect insitu star formation
  and the corresponding stellar population distributions clearly goes
  beyond the scope of this study.

Furthermore, one may wonder whether additional AGN feedback 
  would also affect the contribution from \textit{accreted} systems to
  the global stellar population gradients.  In Fig. 2 of
  \citet{2013MNRAS.436.2929H}  we have shown -- as already stated
  above -- that only \textit{central} galaxies residing in halos with
  masses above $3 \times 10^{12} M_\odot$ seem to be mainly affected
  by the over-cooling problem after z=0.5.  Satellite galaxies instead,
  residing in less massive halos are in reasonably good agreement with
  SFRs derived from abundance matching predictions, even if for the
  lowest mass satellites, SFRs might be too large due to still too
  inefficient galactic wind feedback in our simulations (see
  \citealp{Dave13} for further discussion). The latter point implies
  that a model having stronger galactic winds might lead to even
  stronger metallicity gradients (due to even more efficiently
  suppressed metal enrichment in lower mass satellite
  galaxies). Overall, we do not expect that AGN feedback will have a
  significant impact on suppressing star formation in satellite
  galaxies. 

%*****************************************************************************************************
%*****************************************************************************************************
\subsection{Comparison with observations}
\label{Observations}
%*****************************************************************************************************
%*****************************************************************************************************

In general, comparing our simulated, present-day metallicity, age and
colour gradients at large radii (beyond $2\ R_{\mathrm{eff}}$) is
difficult as most observational studies focus on inner gradients out
to only $\sim 1\ R_{\mathrm{eff}}$ which are easier and more
precise to measure. In the following, we will mainly compare the
observational results of \citet{Coccato10,LaBarbera12, Greene12,
  Greene13, Pastorello14, DSouza14} and \citet{Raskutti14} for
ellipticals to our simulations, even if most of these studies do not
measure gradients significantly beyond $3\ R_{\mathrm{eff}}$ (expect
of \citealp{LaBarbera12}). Therefore, we want to emphasize that our
simulations provide predictions of outer gradients for future, more
elaborate observational surveys which will (hopefully) strengthen our
(preliminary) conclusions below. 

However, as mentioned before, due to our possibly by a factor of
  1.5-2 too small galaxy sizes, the \textit{absolute range} within
  which the slopes are fitted changes when adopting more realistic
  effective radii (the slopes remain unchanged, though). For example,
  artificially increasing our present-day effective radii by a factor
  of 2 would lead to a decrease of the fitted radial range of 1-4
  $R_{\mathrm{eff}}$ (instead of 2-6 $R_{\mathrm{eff}}$). For a fair
  comparison with observations we will keep this in mind in the
  following discussion.   

\subsubsection{Colour gradients}

Starting with the colour gradients, in the photometric study by
\citet{LaBarbera12}, they investigated colour gradients of early-type
galaxies ($M_{\mathrm{stellar}} > 3 \times 10^{10} M_\odot $) even out
to $8 \times R_{\mathrm{eff}}$ using the SDSS-based Spider survey. They
measure average slopes for g-i colours of $-0.06 \pm 0.04$~mag/dex and 
for g-r colours of $-0.16 \pm 0.04$~mag/dex (see black dashed line and
grey shaded areas in Fig. \ref{Colorgrads}). Note that the g-r
  color gradients out to $4 \times R_{\mathrm{eff}}$ which might be
  more consistent to compare to our simulations hardly change
  ($-0.13$~mag/dex).  The observed g-i colours, however, are still
affected by a ``red-halo'', i.e. wings in the PSF (point spread
function) cause a spurious red colour excess at large radii, typically
flattening the gradients. As discussed in their paper, the outer g-i
gradients are, therefore, not as trustable as the g-r
gradients. Compared to observed, reliable g-r colour gradients, MNoW
galaxies have on average too flat gradients ($-0.02$~mag/dex), while
those of the WM galaxies are reasonably steep ($-0.13$~mag/dex) -- in
good agreement with the observations.

The recent work of \citet{DSouza14} show g-r colour profiles out to
100~kpc for stacked ellipticals (to be more precise, high
concentration galaxies) with masses between $10^{10}-10^{11.4}
M_\odot$ using roughly 45.500 galaxies from the SDSS survey. They find
values between $-0.11 - 0.14$~mag/dex with little mass trend, in
perfect agreement with the g-r colour gradients predicted by our
simulations (see black dotted lines in Fig. \ref{Colorgrads}, where we
have used the slope of $-0.11$~mag/dex measured for stellar masses of
$10^{11.2} M_\odot$).  

In addition, the on-going MegaCam survey of Atlas3D galaxies (Duc et
al., 2014, MNRAS in press) will soon provide observational constraints
on the outer colour gradients for a large sample of more than 200
early-type galaxies (Karabal et al., in prep). Preliminary results on
a sub-sample of it (24 objects) indicate a mean  g-r colour slope of
$−0.03$~mag/dex, much flatter than predicted by our WM model. These
preliminary results are \textit{not} corrected for artificial red
halos contaminated by their galactic nuclei (wings in the PSF),
typically flattening the gradients (even if  the galaxies in the
subsample have at first sight no visible contamination). Nevertheless,
correctly accounting for this is expected to significantly steepen the
gradients.    

Overall, this indicates that strong stellar winds are crucial for
producing massive galaxies with realistic (steep enough)
\textit{colour} gradients at large radii mainly driven by the accretion
of older, bluer stellar systems (in-situ gradients are not sufficient). 

\subsubsection{Metallicity gradients}

Turning to the metallicity gradients, in a recent spectroscopic study,  
\citet{Pastorello14}, using the SLUGGS survey, investigated
metallicity gradients up to $3.0 R_{\mathrm{eff}}$. Their two dozen
early-type galaxies have masses in the range $3 \times 10^{10} <
M_{\mathrm{stellar}} < 3 \times 10^{11} M_{\odot}$. For comparable stellar
masses, their galaxies reveal slopes between $-1.15 - +0.18$~dex/dex
using stellar population  models of \citet{Vazdekis03}.  

In the study by \citet{LaBarbera12}, where metallicities are derived
from colors using different stellar population synthesis models, they
find for massive galaxies with $10^{11} < M_{\mathrm{stellar}} < 7
\times 10^{11} M_\odot $ outer metallicity gradients ($1 - 8 \times
R_{\mathrm{eff}}$) in the range of $-0.74 - -0.29$ dex/dex depending
on the stellar population model. For the same stellar population 
  model we are using (\citealp{Bruzual03}), they find a mean
  metallicity gradient of $-0.29 \pm 0.12$~dex/dex measured out to $8 
  \times R_{\mathrm{eff}}$ (illustrated by the black dashed lines and the
  grey shaded area in
  Fig. \ref{Metgrads_only10_insacc_mm_zall}). Metallicity gradients
  measured only out to $4 \times R_{\mathrm{eff}}$ would result in a
  mean slope of $-0.4$~dex/dex in even better agreement with our
  simulated result.

In addition, in a recent study of \citet{Montes14} analysing the
metallicity distribution of M87, they find at radii larger than $>
R_{\mathrm{eff}}$ a metallicity gradient of $-0.26$~dex/dex more
consistent with gradients of the WM galaxies. \citet{Coccato10},
however, investigating metallicity gradients of two Ellipticals in the
Coma cluster, find a shallower gradient of $-0.1$~dex/dex.

Overall, our WM galaxies are able to cover such a broad range of
slopes ($-0.8 - -0.1$~dex/dex) much better than the MNoW galaxies
whose slopes are clearly too flat (within the range  of $-0.25 -
+0.03$~dex/dex). The average metallicity gradient of the WM galaxies
($-0.35$~dex/dex) is in excellent agreement with the one of
\citet{LaBarbera12} and the comparable mass galaxies from
\citet{Pastorello14}.  

As expected a priori, this implies that a strong stellar feedback is a
key mechanism to be consistent with observed steep metallicity
gradients in massive galaxies in the local Universe. In addition, we
demonstrated that in the WM model, the in-situ gradients are much
flatter (by roughly 0.2~dex) than the total gradients and they would,
therefore, be insufficient to reproduce the observational data. This
convincingly highlights the crucial role of the stellar accretion of
metal-poor systems in establishing realistically steep metallicity
gradients. 

\subsubsection{Age gradients}

We saw in section \ref{agegradients}, that both models predict
\textit{strong positive} inner age gradients at $< 2\
R_{\mathrm{eff}}$ in tension with recent observations of
\citet{Greene13, Raskutti14} (who study the radial dependences of
stellar populations of 33 massive, near-by elliptical galaxies) and
also of \citet{Gonzalez14}, analysing CALIFA galaxies (finding roughly
flat age gradientes), but in better qualitative agreement with
\citet{LaBarbera12}. As, however, the former observations are based on
spectral templates, which typically result in a higher accuracy in the
inner region than a derivation from colours (which instead allows to
measure stellar populations out to larger radii), the inner drop in
the simulations might be unrealisticly strong entirely driven by
recently (in-situ) formed new, young stars (bottom, middle panels of
Fig. \ref{Agegrads_only10_insacc_mm_zall}). \footnote{We also want to 
  mention that the strong age gradients by \citet{LaBarbera12} are
  mostly found for group galaxies (i.e. mainly satellites, as  group
  centrals are a minor fraction of the group galaxies), while field
  galaxies (i.e. centrals) have also only mildly positive or even
  inverted age gradients. Spectroscopic studies are based on smaller
  samples, and the definition of the environment is far less clear.}

The tension with observations indicates some weaknesses of our models:
in both runs, we do not account for any feedback from AGN which is,
however, expected to particularly reduce the late star formation in
the central parts of a galaxy and, therefore, to reduce or even
alleviate the positive inner gradients. Moreover, the WM galaxies
(where the in-situ formed stars tend to be even younger than in the
MNoW galaxies), suffer from too much late re-accretion of previously
expelled gas leading to (too) high star formation rates at low
redshifts for massive galaxies (see e.g. \citealp{2013MNRAS.436.2929H,
  Oppenheimer10}).   

Instead, the almost flat, but slightly positive outer age gradients
are consistent with recent observations of \citet{Greene13},
\citet{Raskutti14} and of \citet{Gonzalez14} out to 3
$R_{\mathrm{eff}}$, while \citet{LaBarbera12} predict steeper positive
age gradients out to 8 $R_{\mathrm{eff}}$ up to 0.5~dex/dex in tension
with our results (see black dashed lines in Fig.
\ref{Agegrads_only10_insacc_mm_zall}). However, we have to keep 
in mind that the latter observational results have also a large
scatter in stellar age at such large radii due to their derivation from
photometry (colours). Here, future, more detailed observational
studies will be necessary to draw more reliable conclusions.

%*****************************************************************************************************
%*****************************************************************************************************
\section{Summary}
\label{summary}
%*****************************************************************************************************
%*****************************************************************************************************

In this study, we investigate the origin of metallicity, age and
color gradients in cosmological, zoom simulations of ten massive
galaxies ($6 \times 10^{12} < M_{\mathrm{halo}} < 2 \times 10^{13}
M_\odot$) at large radii ($2-6\ R_{\mathrm{eff}}$) with a particular
focus on the role of accretion of stars in major and minor mergers. We
present the differential effect of strong stellar feedback driving
galactic winds and the influence of the individual merger history
(distinguishing  between minor and major mergers) on the steepness of
the population gradients. 

The zoom simulations include radiative cooling, star formation, metal
enrichment due to SNII, SNIa and AGB stars and depending on the
simulation set, an empirically motivated model for momentum-driven
galactic winds (\citealp{Oppenheimer06, Oppenheimer08}). While
metallicity and stellar ages are a direct output of the simulation, we
employed metal- and age-dependent models for the spectral evolution of
stellar populations (\citealp{Bruzual03}) to also derive photometric galaxy
properties as the g-i, g-r and u-g colours. Our main results can be
summarised as follows:   

\begin{itemize}
\item{Galactic winds significantly delay star formation and thus,
    metal enrichment of a galaxy, leading to a significantly
    different evolution of the mass-metallicity and mass-age relation
    compared to simulations without winds. At a given stellar mass
    galaxies simulated with winds have lower metallicities and the
    effect is stronger for low mass galaxies and at lower redshift.   
    While in the model without winds the relation flattens towards
    $z=0$ (all galaxies and their satellites have super-solar
    metallicity), in the wind model, the lower mass (satellite)
    galaxies still have sub-solar metallicity resulting in a good match
    to the present-day observed mass-metallicity relation. 
    These results are in agreement with previous cosmological
    simulations of large cosmological volumes and individual zooms
    using a similar simple purely empirical wind model
    (e.g. \citealp{Oppenheimer06, Oppenheimer08, 2013MNRAS.436.2929H,  
      Marinacci13, Angles13}).

    As many of the (low metallicity in the wind model) satellites are
    accreted onto the central galaxies and the accreted stars
    typically become more dominant at large galacto-centric radii, this
    has important consequences for the outer population gradients. }
\item{{The outer negative metallicity gradients (at radii
    $r>2\ R_{\mathrm{eff}}$) of present-day massive galaxies are
    mainly determined by the accretion of stars (added from minor
    mergers) with lower metallicity at redshifts $z \leq 1$. Accreted
    low metallicity stars become more and more dominant at large radii
    and the metallicity gradients of in-situ formed stars in the wind
    model are enhanced by $\sim 0.2$~dex/dex by accretion of
    metal-poor systems. } 

    {Overall, the model with galactic winds predicts much steeper
      total metallicity gradients (on average $-0.35$~dex/dex at $z=0$
      and $-0.45$~dex/dex at $z=1$) as the accreted stellar systems
      are significantly more metal-poor and despite of the fact that
      much less stellar mass in total is accreted compared to the
      no-wind model. } 

    {The steep gradients in the wind model tend to be more
    consistent with recent observations of massive early-type galaxies
    in the local Universe by \citet{LaBarbera12} and
    \citet{Pastorello14}. This clearly supports our \textit{a priori}
    expectation that both the stellar accretion and strong
    galactic-scale winds - reducing the metallicity in the accreted
    systems - are the physical mechanisms setting the outer
    metallicity gradients. } }  
\item{{The age gradients at large radii are flat or slightly
    positive as the accreted systems are in general older than the
    central galaxies independent of the model (see also
    \citealp{Oser10}).  The wind model, however, predicts on average
    more positive gradients than the model without winds, particularly
    at $z=1$ (mean slope of $1.91$~dex/dex as opposed to
    $0.03$~dex/dex). }
    
    {At $z=0$, also the age gradients in the wind model
    are rather flat (with a mean positive slope of $0.04$~dex/dex)
    which are consistent with recent observations of massive
    elliptical galaxies by \citealp{Greene13}, but too shallow
    compared to the study of \citealp{LaBarbera12} (where ages are
    derived from photometric properties). }   }
\item{{Gradients in galaxy colour (g-i, g-r, u-g) are the result of the
    superposition of age and metallicity distributions. At higher
    redshifts $z=1,2$, we find galaxies with both strongly positive
    and strongly negative colour gradients (g-i, g-r, u-g) depending on
    whether positive age or negative metallicity gradients have a
    stronger impact on the populations. At $z=0$, negative metallicity
    gradients outweigh the relatively shallow positive age gradients
    leading to overall negative colour gradients in both models. }
    
    {In the wind model, the colour gradients are significantly
      steeper than in the model without winds (reaching average slopes
      of $-0.2$, $-0.13$ and $-0.22$~mag/dex for g-i, g-r and u-g
      colours, respectively) more consistent with the broad range of
      observed gradients of e.g. \citet{LaBarbera12}, but too flat
      compared to preliminary results from Karabal et
      al. for discussed reasons (in prep., Atlas3D survey).} 
    {Overall, the colour gradients for the in-situ component only
      appear too shallow to be consistent with observations.  } }  
\item{{The strength of the gradients strongly depends on the
      individual merger history, where a recent major merger (since
      $z=1$) typically flattens the gradients
      (\citealp{White78, 2004MNRAS.347..740K, DiMatteo09,
        Rupke10,  Navarro13}), while minor mergers significantly increase the
      slopes to more negative values for colours and metallicities or
      to more positive values for ages (\citealp{1983MNRAS.204..219V,
        2012MNRAS.425.3119H, 2013MNRAS.429.2924H})}}    
\end{itemize}

Overall, we can conclude that stellar accretion (via minor mergers) of
low mass satellites results in steep outer negative metallicity and
colour gradients and slightly positive age gradients  successfully
matching the broad range of observed colour and metallicity profiles of
local galaxies at large radii. Nevertheless, a drawback of our
cosmological zoom simulations is the missing AGN feedback which is
expected to suppress late central star formation in massive galaxies
(currently overestimated leading to unrealistically young and blue
cores of our galaxies). However, we want to emphasise that the
  stellar population distributions of galaxies investigated and
  discussed only at large radii are not expected to be significantly
  affected by such additional physical processes.  Instead, they are
  supposed to primarily influence the central regions (at $r<2 \times
  R_{\mathrm{eff}}$), which were not the focus of this study. 

In addition, our sample of massive galaxies consists of only ten
galaxies so that the results cannot be interpreted in a statistical
sense, we only highlight the effect of the accretion of the stellar
populations of metal poor galaxies, which is a direct consequence of
strong stellar feedback at all cosmic epochs in the simulations. We
note that this effect will only be of importance for massive
early-type galaxies in the range of $10^{11} M_\odot$, i.e. massive
centrals in galaxy groups. The stellar populations of lower mass
early-type galaxies are expected to be dominated by in-situ star
formation with significantly lower fractions of accreted stars. Direct
evidence for this is the high fraction of very flattened axi-symmetric
systems with disk-like morphologies and kinematics (\citealp{Rix99, 
Cappellari11a, 2011MNRAS.413..813C, Emsellem11, Weijmans14}). Here the
abundance is most likely set by processes similar to inside-out
forming spiral galaxies.  The expected decreasing importance of
stellar accretion with decreasing stellar mass is  supported by
abundance matching estimates (\citealp{2013MNRAS.428.3121M,
  2013ApJ...770...57B, Yang13}), semi-analytical models
(\citealp{Guo08}) and direct simulations (\citealp{Gabor12,
  2010ApJ...712...88L, Lackner12}).    

In forthcoming studies we plan to significantly increase
our sample of re-simulated galaxies to obtain a statistically more
relevant analysis for stellar population gradients and to include
models for black hole growth and AGN feedback to explore their effects
on metallicity, age and colour distributions in the central region of
a galaxy. This will provide additional constraints on different models
for AGN feedback which are still only poorly understood.

\section*{Acknowledgements}
We thank Ignacio Ferreras, Jenny Greene, Francesco La Barbera, 
Pierluigi Monaco and Nicola Pastorello for fruitful discussions. In
particular, we are very grateful to Francesco La Barbera for providing
us with fits to observed color gradients. 

MH acknowledges financial support from the European Research Council
under the European Community's Seventh Framework Programme
(FP7/2007-2013)/ERC grant agreement n. 202781 and from the European
Research Council via an Advanced Grant under grant agreement
no. 321323 NEOGAL. TN acknowledges support from the DFG Cluster of
Excellence 'Origin and Structure of the Universe'. DAF thanks the ARC
for support via DP130100388.

\bibliographystyle{mn2e}
\bibliography{Literaturdatenbank}

\begin{thebibliography}{}

\bibitem[\protect\citeauthoryear{{Angl{\'e}s-Alc{\'a}zar}, {Dav{\'e}},
  {{\"O}zel} \& {Oppenheimer}}{{Angl{\'e}s-Alc{\'a}zar}
  et~al.}{2014}]{Angles13}
{Angl{\'e}s-Alc{\'a}zar} D.,  {Dav{\'e}} R.,  {{\"O}zel} F.,    {Oppenheimer}
  B.~D.,  2014, \apj, 782, 84

\bibitem[\protect\citeauthoryear{{Annibali}, {Bressan}, {Rampazzo}, {Zeilinger}
  \& {Danese}}{{Annibali} et~al.}{2007}]{Annibali07}
{Annibali} F.,  {Bressan} A.,  {Rampazzo} R.,  {Zeilinger} W.~W.,    {Danese}
  L.,  2007, \aap, 463, 455

\bibitem[\protect\citeauthoryear{{Behroozi}, {Wechsler} \& {Conroy}}{{Behroozi}
  et~al.}{2013}]{2013ApJ...770...57B}
{Behroozi} P.~S.,  {Wechsler} R.~H.,    {Conroy} C.,  2013, \apj, 770, 57

\bibitem[\protect\citeauthoryear{{Bell}, {Naab}, {McIntosh}, {Somerville},
  {Caldwell}, {Barden}, {Wolf}, {Rix}, {Beckwith}, {Borch}, {H{\"a}ussler},
  {Heymans}, {Jahnke}, {Jogee}, {Koposov}, {Meisenheimer}, {Peng}, {Sanchez} \&
  {Wisotzki}}{{Bell} et~al.}{2006}]{2006ApJ...640..241B}
{Bell} E.~F.,  {Naab} T.,  {McIntosh} D.~H.,  {Somerville} R.~S.,  {Caldwell}
  J.~A.~R.,  {Barden} M.,  {Wolf} C.,  {Rix} H.-W.,  {Beckwith} S.~V.,  {Borch}
  A.,  {H{\"a}ussler} B.,  {Heymans} C.,  {Jahnke} K.,  {Jogee} S.,  {Koposov}
  S.,  {Meisenheimer} K.,  {Peng} C.~Y.,  {Sanchez} S.~F.,    {Wisotzki} L.,
  2006, \apj, 640, 241

\bibitem[\protect\citeauthoryear{{Bezanson}, {van Dokkum}, {Tal}, {Marchesini},
  {Kriek}, {Franx} \& {Coppi}}{{Bezanson} et~al.}{2009}]{2009ApJ...697.1290B}
{Bezanson} R.,  {van Dokkum} P.~G.,  {Tal} T.,  {Marchesini} D.,  {Kriek} M.,
  {Franx} M.,    {Coppi} P.,  2009, \apj, 697, 1290

\bibitem[\protect\citeauthoryear{{Boylan-Kolchin}, {Ma} \&
  {Quataert}}{{Boylan-Kolchin} et~al.}{2005}]{2005MNRAS.362..184B}
{Boylan-Kolchin} M.,  {Ma} C.-P.,    {Quataert} E.,  2005, \mnras, 362, 184

\bibitem[\protect\citeauthoryear{{Boylan-Kolchin}, {Ma} \&
  {Quataert}}{{Boylan-Kolchin} et~al.}{2006}]{2006MNRAS.369.1081B}
{Boylan-Kolchin} M.,  {Ma} C.-P.,    {Quataert} E.,  2006, \mnras, 369, 1081

\bibitem[\protect\citeauthoryear{{Boylan-Kolchin}, {Ma} \&
  {Quataert}}{{Boylan-Kolchin} et~al.}{2008}]{2008MNRAS.383...93B}
{Boylan-Kolchin} M.,  {Ma} C.-P.,    {Quataert} E.,  2008, \mnras, 383, 93

\bibitem[\protect\citeauthoryear{{Bruzual} \& {Charlot}}{{Bruzual} \&
  {Charlot}}{2003}]{Bruzual03}
{Bruzual} G.,  {Charlot} S.,  2003, \mnras, 344, 1000

\bibitem[\protect\citeauthoryear{{Buitrago}, {Trujillo}, {Conselice},
  {Bouwens}, {Dickinson} \& {Yan}}{{Buitrago}
  et~al.}{2008}]{2008ApJ...687L..61B}
{Buitrago} F.,  {Trujillo} I.,  {Conselice} C.~J.,  {Bouwens} R.~J.,
  {Dickinson} M.,    {Yan} H.,  2008, \apjl, 687, L61

\bibitem[\protect\citeauthoryear{{Cappellari}}{{Cappellari}}{2011a}]{2011MNRAS%
.413..813C}
{Cappellari} M. e.~a.,  2011a, \mnras, 413, 813

\bibitem[\protect\citeauthoryear{{Cappellari}}{{Cappellari}}{2011b}]{Cappellar%
i11a}
{Cappellari} M. e.~a.,  2011b, \mnras, 416, 1680

\bibitem[\protect\citeauthoryear{{Carollo}, {Danziger} \& {Buson}}{{Carollo}
  et~al.}{1993}]{Carollo93}
{Carollo} C.~M.,  {Danziger} I.~J.,    {Buson} L.,  1993, \mnras, 265, 553

\bibitem[\protect\citeauthoryear{{Cenarro} \& {Trujillo}}{{Cenarro} \&
  {Trujillo}}{2009}]{2009ApJ...696L..43C}
{Cenarro} A.~J.,  {Trujillo} I.,  2009, \apjl, 696, L43

\bibitem[\protect\citeauthoryear{{Chabrier}}{{Chabrier}}{2003}]{Chabrier03}
{Chabrier} G.,  2003, \pasp, 115, 763

\bibitem[\protect\citeauthoryear{{Chiappini}, {Matteucci} \&
  {Romano}}{{Chiappini} et~al.}{2001}]{Chiappini01}
{Chiappini} C.,  {Matteucci} F.,    {Romano} D.,  2001, \apj, 554, 1044

\bibitem[\protect\citeauthoryear{{Choi}, {Ostriker}, {Naab}, {Oser} \&
  {Moster}}{{Choi} et~al.}{2014}]{Choi14}
{Choi} E.,  {Ostriker} J.~P.,  {Naab} T.,  {Oser} L.,    {Moster} B.~P.,  2014,
  ArXiv e-prints

\bibitem[\protect\citeauthoryear{{Cimatti}, {Cassata}, {Pozzetti}, {Kurk},
  {Mignoli}, {Renzini}, {Daddi}, {Bolzonella}, {Brusa}, {Rodighiero},
  {Dickinson}, {Franceschini}, {Zamorani}, {Berta}, {Rosati} \&
  {Halliday}}{{Cimatti} et~al.}{2008}]{2008A&A...482...21C}
{Cimatti} A.,  {Cassata} P.,  {Pozzetti} L.,  {Kurk} J.,  {Mignoli} M.,
  {Renzini} A.,  {Daddi} E.,  {Bolzonella} M.,  {Brusa} M.,  {Rodighiero} G.,
  {Dickinson} M.,  {Franceschini} A.,  {Zamorani} G.,  {Berta} S.,  {Rosati}
  P.,    {Halliday} C.,  2008, \aap, 482, 21

\bibitem[\protect\citeauthoryear{{Coccato}, {Gerhard} \& {Arnaboldi}}{{Coccato}
  et~al.}{2010}]{Coccato10}
{Coccato} L.,  {Gerhard} O.,    {Arnaboldi} M.,  2010, \mnras, 407, L26

\bibitem[\protect\citeauthoryear{{Cole}, {Lacey}, {Baugh} \& {Frenk}}{{Cole}
  et~al.}{2000}]{Cole00}
{Cole} S.,  {Lacey} C.~G.,  {Baugh} C.~M.,    {Frenk} C.~S.,  2000, \mnras,
  319, 168

\bibitem[\protect\citeauthoryear{{Conroy}, {van Dokkum} \& {Kravtsov}}{{Conroy}
  et~al.}{2014}]{Conroy14}
{Conroy} C.,  {van Dokkum} P.,    {Kravtsov} A.,  2014, ArXiv e-prints

\bibitem[\protect\citeauthoryear{{Cooper}, {D'Souza}, {Kauffmann}, {Wang},
  {Boylan-Kolchin}, {Guo}, {Frenk} \& {White}}{{Cooper}
  et~al.}{2013}]{Cooper13}
{Cooper} A.~P.,  {D'Souza} R.,  {Kauffmann} G.,  {Wang} J.,  {Boylan-Kolchin}
  M.,  {Guo} Q.,  {Frenk} C.~S.,    {White} S.~D.~M.,  2013, \mnras, 434, 3348

\bibitem[\protect\citeauthoryear{{Cresci}, {Mannucci}, {Maiolino}, {Marconi},
  {Gnerucci} \& {Magrini}}{{Cresci} et~al.}{2010}]{Cresci10}
{Cresci} G.,  {Mannucci} F.,  {Maiolino} R.,  {Marconi} A.,  {Gnerucci} A.,
  {Magrini} L.,  2010, \nat, 467, 811

\bibitem[\protect\citeauthoryear{{Daddi}, {Renzini}, {Pirzkal}, {Cimatti},
  {Malhotra}, {Stiavelli}, {Xu}, {Pasquali}, {Rhoads}, {Brusa}, {di Serego
  Alighieri}, {Ferguson}, {Koekemoer}, {Moustakas}, {Panagia} \&
  {Windhorst}}{{Daddi} et~al.}{2005}]{2005ApJ...626..680D}
{Daddi} E.,  {Renzini} A.,  {Pirzkal} N.,  {Cimatti} A.,  {Malhotra} S.,
  {Stiavelli} M.,  {Xu} C.,  {Pasquali} A.,  {Rhoads} J.~E.,  {Brusa} M.,  {di
  Serego Alighieri} S.,  {Ferguson} H.~C.,  {Koekemoer} A.~M.,  {Moustakas}
  L.~A.,  {Panagia} N.,    {Windhorst} R.~A.,  2005, \apj, 626, 680

\bibitem[\protect\citeauthoryear{{Dalla Vecchia} \& {Schaye}}{{Dalla Vecchia}
  \& {Schaye}}{2008}]{DallaVecchia08}
{Dalla Vecchia} C.,  {Schaye} J.,  2008, \mnras, 387, 1431

\bibitem[\protect\citeauthoryear{{Damjanov}, {McCarthy}, {Abraham},
  {Glazebrook}, {Yan}, {Mentuch}, {LeBorgne}, {Savaglio}, {Crampton},
  {Murowinski}, {Juneau}, {Carlberg}, {J{\o}rgensen}, {Roth}, {Chen} \&
  {Marzke}}{{Damjanov} et~al.}{2009}]{2009ApJ...695..101D}
{Damjanov} I.,  {McCarthy} P.~J.,  {Abraham} R.~G.,  {Glazebrook} K.,  {Yan}
  H.,  {Mentuch} E.,  {LeBorgne} D.,  {Savaglio} S.,  {Crampton} D.,
  {Murowinski} R.,  {Juneau} S.,  {Carlberg} R.~G.,  {J{\o}rgensen} I.,  {Roth}
  K.,  {Chen} H.-W.,    {Marzke} R.~O.,  2009, \apj, 695, 101

\bibitem[\protect\citeauthoryear{{Dav{\'e}}, {Katz}, {Oppenheimer}, {Kollmeier}
  \& {Weinberg}}{{Dav{\'e}} et~al.}{2013}]{Dave13}
{Dav{\'e}} R.,  {Katz} N.,  {Oppenheimer} B.~D.,  {Kollmeier} J.~A.,
  {Weinberg} D.~H.,  2013, \mnras, 434, 2645

\bibitem[\protect\citeauthoryear{{Davies}, {Sadler} \& {Peletier}}{{Davies}
  et~al.}{1993}]{Davies93}
{Davies} R.~L.,  {Sadler} E.~M.,    {Peletier} R.~F.,  1993, \mnras, 262, 650

\bibitem[\protect\citeauthoryear{{Davis}, {Efstathiou}, {Frenk} \&
  {White}}{{Davis} et~al.}{1985}]{Davis85}
{Davis} M.,  {Efstathiou} G.,  {Frenk} C.~S.,    {White} S.~D.~M.,  1985, \apj,
  292, 371

\bibitem[\protect\citeauthoryear{{De Lucia} \& {Blaizot}}{{De Lucia} \&
  {Blaizot}}{2007}]{2007MNRAS.375....2D}
{De Lucia} G.,  {Blaizot} J.,  2007, \mnras, 375, 2

\bibitem[\protect\citeauthoryear{{De Lucia}, {Springel}, {White}, {Croton} \&
  {Kauffmann}}{{De Lucia} et~al.}{2006}]{2006MNRAS.366..499D}
{De Lucia} G.,  {Springel} V.,  {White} S.~D.~M.,  {Croton} D.,    {Kauffmann}
  G.,  2006, \mnras, 366, 499

\bibitem[\protect\citeauthoryear{{de Vaucouleurs}}{{de
  Vaucouleurs}}{1961}]{Vaucouleurs61}
{de Vaucouleurs} G.,  1961, \apjs, 5, 233

\bibitem[\protect\citeauthoryear{{Di Matteo}, {Jog}, {Lehnert}, {Combes} \&
  {Semelin}}{{Di Matteo} et~al.}{2009}]{2009A&A...501L...9D}
{Di Matteo} P.,  {Jog} C.~J.,  {Lehnert} M.~D.,  {Combes} F.,    {Semelin} B.,
  2009, \aap, 501, L9

\bibitem[\protect\citeauthoryear{{Di Matteo}, {Pipino}, {Lehnert}, {Combes} \&
  {Semelin}}{{Di Matteo} et~al.}{2009}]{DiMatteo09}
{Di Matteo} P.,  {Pipino} A.,  {Lehnert} M.~D.,  {Combes} F.,    {Semelin} B.,
  2009, \aap, 499, 427

\bibitem[\protect\citeauthoryear{{D'Souza}, {Kauffman}, {Wang} \&
  {Vegetti}}{{D'Souza} et~al.}{2014}]{DSouza14}
{D'Souza} R.,  {Kauffman} G.,  {Wang} J.,    {Vegetti} S.,  2014, \mnras, 443,
  1433

\bibitem[\protect\citeauthoryear{{Dubois}, {Gavazzi}, {Peirani} \&
  {Silk}}{{Dubois} et~al.}{2013}]{Dubois13}
{Dubois} Y.,  {Gavazzi} R.,  {Peirani} S.,    {Silk} J.,  2013, \mnras, 433,
  3297

\bibitem[\protect\citeauthoryear{{Eigenthaler} \& {Zeilinger}}{{Eigenthaler} \&
  {Zeilinger}}{2013}]{Eigenthaler13}
{Eigenthaler} P.,  {Zeilinger} W.~W.,  2013, \aap, 553, A99

\bibitem[\protect\citeauthoryear{{Emsellem}}{{Emsellem}}{2011}]{Emsellem11}
{Emsellem} E. e.~a.,  2011, \mnras, 414, 888

\bibitem[\protect\citeauthoryear{{Feldmann}, {Carollo} \& {Mayer}}{{Feldmann}
  et~al.}{2011}]{2011ApJ...736...88F}
{Feldmann} R.,  {Carollo} C.~M.,    {Mayer} L.,  2011, \apj, 736, 88

\bibitem[\protect\citeauthoryear{{Feldmann}, {Carollo}, {Mayer}, {Renzini},
  {Lake}, {Quinn}, {Stinson} \& {Yepes}}{{Feldmann}
  et~al.}{2010}]{2010ApJ...709..218F}
{Feldmann} R.,  {Carollo} C.~M.,  {Mayer} L.,  {Renzini} A.,  {Lake} G.,
  {Quinn} T.,  {Stinson} G.~S.,    {Yepes} G.,  2010, \apj, 709, 218

\bibitem[\protect\citeauthoryear{{Ferreras}, {Lisker}, {Pasquali} \&
  {Kaviraj}}{{Ferreras} et~al.}{2009}]{2009MNRAS.395..554F}
{Ferreras} I.,  {Lisker} T.,  {Pasquali} A.,    {Kaviraj} S.,  2009, \mnras,
  395, 554

\bibitem[\protect\citeauthoryear{{Ferreras}, {Trujillo},
  {M{\'a}rmol-Queralt{\'o}}, {P{\'e}rez-Gonz{\'a}lez}, {Cava}, {Barro},
  {Cenarro}, {Hern{\'a}n-Caballero}, {Cardiel},
  {Rodr{\'{\i}}guez-Zaur{\'{\i}}n} \& {Cebri{\'a}n}}{{Ferreras}
  et~al.}{2014}]{Ferreras14}
{Ferreras} I.,  {Trujillo} I.,  {M{\'a}rmol-Queralt{\'o}} E.,
  {P{\'e}rez-Gonz{\'a}lez} P.~G.,  {Cava} A.,  {Barro} G.,  {Cenarro} J.,
  {Hern{\'a}n-Caballero} A.,  {Cardiel} N.,  {Rodr{\'{\i}}guez-Zaur{\'{\i}}n}
  J.,    {Cebri{\'a}n} M.,  2014, \mnras, 444, 906

\bibitem[\protect\citeauthoryear{{Foster}, {Proctor}, {Forbes}, {Spolaor},
  {Hopkins} \& {Brodie}}{{Foster} et~al.}{2009}]{Foster09}
{Foster} C.,  {Proctor} R.~N.,  {Forbes} D.~A.,  {Spolaor} M.,  {Hopkins}
  P.~F.,    {Brodie} J.~P.,  2009, \mnras, 400, 2135

\bibitem[\protect\citeauthoryear{{Franx} \& {Illingworth}}{{Franx} \&
  {Illingworth}}{1990}]{Franx90}
{Franx} M.,  {Illingworth} G.,  1990, \apjl, 359, L41

\bibitem[\protect\citeauthoryear{{Franx}, {van Dokkum}, {Schreiber}, {Wuyts},
  {Labb{\'e}} \& {Toft}}{{Franx} et~al.}{2008}]{2008ApJ...688..770F}
{Franx} M.,  {van Dokkum} P.~G.,  {Schreiber} N.~M.~F.,  {Wuyts} S.,
  {Labb{\'e}} I.,    {Toft} S.,  2008, \apj, 688, 770

\bibitem[\protect\citeauthoryear{{Fu}, {Kauffmann}, {Huang}, {Yates}, {Moran},
  {Heckman}, {Dav{\'e}}, {Guo} \& {Henriques}}{{Fu} et~al.}{2013}]{Fu13}
{Fu} J.,  {Kauffmann} G.,  {Huang} M.-l.,  {Yates} R.~M.,  {Moran} S.,
  {Heckman} T.~M.,  {Dav{\'e}} R.,  {Guo} Q.,    {Henriques} B.~M.~B.,  2013,
  \mnras, 434, 1531

\bibitem[\protect\citeauthoryear{{Gabor} \& {Dav{\'e}}}{{Gabor} \&
  {Dav{\'e}}}{2012}]{Gabor12}
{Gabor} J.~M.,  {Dav{\'e}} R.,  2012, \mnras, 427, 1816

\bibitem[\protect\citeauthoryear{{Gallazzi}, {Charlot}, {Brinchmann}, {White}
  \& {Tremonti}}{{Gallazzi} et~al.}{2005}]{Gallazzi05}
{Gallazzi} A.,  {Charlot} S.,  {Brinchmann} J.,  {White} S.~D.~M.,
  {Tremonti} C.~A.,  2005, \mnras, 362, 41

\bibitem[\protect\citeauthoryear{{Genel}, {Genzel}, {Bouch{\'e}}, {Sternberg},
  {Naab}, {Schreiber}, {Shapiro}, {Tacconi}, {Lutz}, {Cresci}, {Buschkamp},
  {Davies} \& {Hicks}}{{Genel} et~al.}{2008}]{2008ApJ...688..789G}
{Genel} S.,  {Genzel} R.,  {Bouch{\'e}} N.,  {Sternberg} A.,  {Naab} T.,
  {Schreiber} N.~M.~F.,  {Shapiro} K.~L.,  {Tacconi} L.~J.,  {Lutz} D.,
  {Cresci} G.,  {Buschkamp} P.,  {Davies} R.~I.,    {Hicks} E.~K.~S.,  2008,
  \apj, 688, 789

\bibitem[\protect\citeauthoryear{{Gonzalez}, {Faber} \& {Worthey}}{{Gonzalez}
  et~al.}{1993}]{Gonzalez93}
{Gonzalez} J.~J.,  {Faber} S.~M.,    {Worthey} G.,  1993, in American
  Astronomical Society Meeting Abstracts Vol.~25 of Bulletin of the American
  Astronomical Society, {Age and Metallicity of Elliptical Galaxies}.
p.~1355

\bibitem[\protect\citeauthoryear{{Gonz{\'a}lez Delgado}}{{Gonz{\'a}lez
  Delgado}}{2014}]{Gonzalez14}
{Gonz{\'a}lez Delgado} R.~M. e.~a.,  2014, \aap, 562, A47

\bibitem[\protect\citeauthoryear{{Greene}, {Murphy}, {Comerford}, {Gebhardt} \&
  {Adams}}{{Greene} et~al.}{2012}]{Greene12}
{Greene} J.~E.,  {Murphy} J.~D.,  {Comerford} J.~M.,  {Gebhardt} K.,    {Adams}
  J.~J.,  2012, \apj, 750, 32

\bibitem[\protect\citeauthoryear{{Greene}, {Murphy}, {Graves}, {Gunn},
  {Raskutti}, {Comerford} \& {Gebhardt}}{{Greene} et~al.}{2013}]{Greene13}
{Greene} J.~E.,  {Murphy} J.~D.,  {Graves} G.~J.,  {Gunn} J.~E.,  {Raskutti}
  S.,  {Comerford} J.~M.,    {Gebhardt} K.,  2013, \apj, 776, 64

\bibitem[\protect\citeauthoryear{{Guo} \& {White}}{{Guo} \&
  {White}}{2008a}]{2008MNRAS.384....2G}
{Guo} Q.,  {White} S.~D.~M.,  2008a, \mnras, 384, 2

\bibitem[\protect\citeauthoryear{{Guo} \& {White}}{{Guo} \&
  {White}}{2008b}]{Guo08}
{Guo} Q.,  {White} S.~D.~M.,  2008b, \mnras, 384, 2

\bibitem[\protect\citeauthoryear{{Haardt} \& {Madau}}{{Haardt} \&
  {Madau}}{2001}]{Haardt01}
{Haardt} F.,  {Madau} P.,  2001, in {Neumann} D.~M.,  {Tran} J.~T.~V.,  eds,
  Clusters of Galaxies and the High Redshift Universe Observed in X-rays
  {Modelling the UV/X-ray cosmic background with CUBA}

\bibitem[\protect\citeauthoryear{{Hilz}, {Naab} \& {Ostriker}}{{Hilz}
  et~al.}{2013}]{2013MNRAS.429.2924H}
{Hilz} M.,  {Naab} T.,    {Ostriker} J.~P.,  2013, \mnras, 429, 2924

\bibitem[\protect\citeauthoryear{{Hilz}, {Naab}, {Ostriker}, {Thomas},
  {Burkert} \& {Jesseit}}{{Hilz} et~al.}{2012}]{2012MNRAS.425.3119H}
{Hilz} M.,  {Naab} T.,  {Ostriker} J.~P.,  {Thomas} J.,  {Burkert} A.,
  {Jesseit} R.,  2012, \mnras, 425, 3119

\bibitem[\protect\citeauthoryear{{Hirschmann}, {Khochfar}, {Burkert}, {Naab},
  {Genel} \& {Somerville}}{{Hirschmann} et~al.}{2010}]{Hirschmann10}
{Hirschmann} M.,  {Khochfar} S.,  {Burkert} A.,  {Naab} T.,  {Genel} S.,
  {Somerville} R.~S.,  2010, \mnras, 407, 1016

\bibitem[\protect\citeauthoryear{{Hirschmann}, {Naab}, {Dav{\'e}},
  {Oppenheimer}, {Ostriker}, {Somerville}, {Oser}, {Genzel}, {Tacconi},
  {F{\"o}rster-Schreiber}, {Burkert} \& {Genel}}{{Hirschmann}
  et~al.}{2013}]{2013MNRAS.436.2929H}
{Hirschmann} M.,  {Naab} T.,  {Dav{\'e}} R.,  {Oppenheimer} B.~D.,  {Ostriker}
  J.~P.,  {Somerville} R.~S.,  {Oser} L.,  {Genzel} R.,  {Tacconi} L.~J.,
  {F{\"o}rster-Schreiber} N.~M.,  {Burkert} A.,    {Genel} S.,  2013, \mnras,
  436, 2929

\bibitem[\protect\citeauthoryear{{Hirschmann}, {Naab}, {Somerville}, {Burkert}
  \& {Oser}}{{Hirschmann} et~al.}{2012}]{Hirschmann12}
{Hirschmann} M.,  {Naab} T.,  {Somerville} R.~S.,  {Burkert} A.,    {Oser} L.,
  2012, \mnras, 419, 3200

\bibitem[\protect\citeauthoryear{{Hopkins}, {Bundy}, {Croton}, {Hernquist},
  {Keres}, {Khochfar}, {Stewart}, {Wetzel} \& {Younger}}{{Hopkins}
  et~al.}{2010}]{2010ApJ...715..202H}
{Hopkins} P.~F.,  {Bundy} K.,  {Croton} D.,  {Hernquist} L.,  {Keres} D.,
  {Khochfar} S.,  {Stewart} K.,  {Wetzel} A.,    {Younger} J.~D.,  2010, \apj,
  715, 202

\bibitem[\protect\citeauthoryear{{Hopkins}, {Bundy}, {Murray}, {Quataert},
  {Lauer} \& {Ma}}{{Hopkins} et~al.}{2009}]{2009MNRAS.398..898H}
{Hopkins} P.~F.,  {Bundy} K.,  {Murray} N.,  {Quataert} E.,  {Lauer} T.~R.,
  {Ma} C.-P.,  2009, \mnras, 398, 898

\bibitem[\protect\citeauthoryear{{Jogee}}{{Jogee}}{2009}]{2009ApJ...697.1971J}
{Jogee} S. e.~a.,  2009, \apj, 697, 1971

\bibitem[\protect\citeauthoryear{{Johansson}, {Naab} \& {Ostriker}}{{Johansson}
  et~al.}{2009}]{2009ApJ...697L..38J}
{Johansson} P.~H.,  {Naab} T.,    {Ostriker} J.~P.,  2009, \apjl, 697, L38

\bibitem[\protect\citeauthoryear{{Johansson}, {Naab} \& {Ostriker}}{{Johansson}
  et~al.}{2012}]{Johansson12}
{Johansson} P.~H.,  {Naab} T.,    {Ostriker} J.~P.,  2012, \apj, 754, 115

\bibitem[\protect\citeauthoryear{{Jones}, {Ellis}, {Richard} \&
  {Jullo}}{{Jones} et~al.}{2013}]{Jones13}
{Jones} T.,  {Ellis} R.~S.,  {Richard} J.,    {Jullo} E.,  2013, \apj, 765, 48

\bibitem[\protect\citeauthoryear{{Kauffmann}}{{Kauffmann}}{1996}]{1996MNRAS.28%
1..487K}
{Kauffmann} G.,  1996, \mnras, 281, 487

\bibitem[\protect\citeauthoryear{{Kauffmann}, {Charlot} \& {White}}{{Kauffmann}
  et~al.}{1996}]{1996MNRAS.283L.117K}
{Kauffmann} G.,  {Charlot} S.,    {White} S.~D.~M.,  1996, \mnras, 283, L117

\bibitem[\protect\citeauthoryear{{Kawata} \& {Gibson}}{{Kawata} \&
  {Gibson}}{2003}]{Kawata03}
{Kawata} D.,  {Gibson} B.~K.,  2003, \mnras, 346, 135

\bibitem[\protect\citeauthoryear{{Kewley}, {Rupke}, {Zahid}, {Geller} \&
  {Barton}}{{Kewley} et~al.}{2010}]{Kewley10}
{Kewley} L.~J.,  {Rupke} D.,  {Zahid} H.~J.,  {Geller} M.~J.,    {Barton}
  E.~J.,  2010, \apjl, 721, L48

\bibitem[\protect\citeauthoryear{{Khochfar} \& {Silk}}{{Khochfar} \&
  {Silk}}{2006}]{2006ApJ...648L..21K}
{Khochfar} S.,  {Silk} J.,  2006, \apjl, 648, L21

\bibitem[\protect\citeauthoryear{{Kobayashi}}{{Kobayashi}}{2004}]{2004MNRAS.34%
7..740K}
{Kobayashi} C.,  2004, \mnras, 347, 740

\bibitem[\protect\citeauthoryear{{Kormendy}, {Fisher}, {Cornell} \&
  {Bender}}{{Kormendy} et~al.}{2009}]{2009ApJS..182..216K}
{Kormendy} J.,  {Fisher} D.~B.,  {Cornell} M.~E.,    {Bender} R.,  2009, \apjs,
  182, 216

\bibitem[\protect\citeauthoryear{{La Barbera} \& {de Carvalho}}{{La Barbera} \&
  {de Carvalho}}{2009}]{LaBarbera09}
{La Barbera} F.,  {de Carvalho} R.~R.,  2009, \apjl, 699, L76

\bibitem[\protect\citeauthoryear{{La Barbera}, {de Carvalho}, {Gal},
  {Busarello}, {Merluzzi}, {Capaccioli} \& {Djorgovski}}{{La Barbera}
  et~al.}{2005}]{LaBarbera05}
{La Barbera} F.,  {de Carvalho} R.~R.,  {Gal} R.~R.,  {Busarello} G.,
  {Merluzzi} P.,  {Capaccioli} M.,    {Djorgovski} S.~G.,  2005, \apjl, 626,
  L19

\bibitem[\protect\citeauthoryear{{La Barbera}, {Ferreras}, {de Carvalho},
  {Bruzual}, {Charlot}, {Pasquali} \& {Merlin}}{{La Barbera}
  et~al.}{2012}]{LaBarbera12}
{La Barbera} F.,  {Ferreras} I.,  {de Carvalho} R.~R.,  {Bruzual} G.,
  {Charlot} S.,  {Pasquali} A.,    {Merlin} E.,  2012, \mnras, 426, 2300

\bibitem[\protect\citeauthoryear{{Lackner}, {Cen}, {Ostriker} \&
  {Joung}}{{Lackner} et~al.}{2012a}]{2012MNRAS.425..641L}
{Lackner} C.~N.,  {Cen} R.,  {Ostriker} J.~P.,    {Joung} M.~R.,  2012a,
  \mnras, 425, 641

\bibitem[\protect\citeauthoryear{{Lackner}, {Cen}, {Ostriker} \&
  {Joung}}{{Lackner} et~al.}{2012b}]{Lackner12}
{Lackner} C.~N.,  {Cen} R.,  {Ostriker} J.~P.,    {Joung} M.~R.,  2012b,
  \mnras, 425, 641

\bibitem[\protect\citeauthoryear{{Lackner} \& {Ostriker}}{{Lackner} \&
  {Ostriker}}{2010}]{2010ApJ...712...88L}
{Lackner} C.~N.,  {Ostriker} J.~P.,  2010, \apj, 712, 88

\bibitem[\protect\citeauthoryear{{Laporte}, {White}, {Naab} \& {Gao}}{{Laporte}
  et~al.}{2013}]{Laporte13}
{Laporte} C.~F.~P.,  {White} S.~D.~M.,  {Naab} T.,    {Gao} L.,  2013, \mnras,
  435, 901

\bibitem[\protect\citeauthoryear{{Laporte}, {White}, {Naab}, {Ruszkowski} \&
  {Springel}}{{Laporte} et~al.}{2012}]{2012MNRAS.424..747L}
{Laporte} C.~F.~P.,  {White} S.~D.~M.,  {Naab} T.,  {Ruszkowski} M.,
  {Springel} V.,  2012, \mnras, 424, 747

\bibitem[\protect\citeauthoryear{{Limongi} \& {Chieffi}}{{Limongi} \&
  {Chieffi}}{2005}]{Limongi05}
{Limongi} M.,  {Chieffi} A.,  2005, in {Turatto} M.,  {Benetti} S.,  {Zampieri}
  L.,   {Shea} W.,  eds, 1604-2004: Supernovae as Cosmological Lighthouses
  Vol.~342 of Astronomical Society of the Pacific Conference Series,
  {Presupernova Evolution and Explosive Nucleosynthesis of Massive Stars at
  Various Metallicities from Z=0 to $Z=Z_{\sun}$}.
p.~122

\bibitem[\protect\citeauthoryear{{Longhetti}, {Saracco}, {Severgnini}, {Della
  Ceca}, {Mannucci}, {Bender}, {Drory}, {Feulner} \& {Hopp}}{{Longhetti}
  et~al.}{2007}]{2007MNRAS.374..614L}
{Longhetti} M.,  {Saracco} P.,  {Severgnini} P.,  {Della Ceca} R.,  {Mannucci}
  F.,  {Bender} R.,  {Drory} N.,  {Feulner} G.,    {Hopp} U.,  2007, \mnras,
  374, 614

\bibitem[\protect\citeauthoryear{{Lotz}}{{Lotz}}{2008}]{2008ApJ...672..177L}
{Lotz} J.~M. e.~a.,  2008, \apj, 672, 177

\bibitem[\protect\citeauthoryear{{MacArthur}, {Courteau}, {Bell} \&
  {Holtzman}}{{MacArthur} et~al.}{2004}]{MacArthur04}
{MacArthur} L.~A.,  {Courteau} S.,  {Bell} E.,    {Holtzman} J.~A.,  2004,
  \apjs, 152, 175

\bibitem[\protect\citeauthoryear{{Makino} \& {Hut}}{{Makino} \&
  {Hut}}{1997}]{1997ApJ...481...83M}
{Makino} J.,  {Hut} P.,  1997, \apj, 481, 83

\bibitem[\protect\citeauthoryear{{Man}, {Toft}, {Zirm}, {Wuyts} \& {van der
  Wel}}{{Man} et~al.}{2012}]{2012ApJ...744...85M}
{Man} A.~W.~S.,  {Toft} S.,  {Zirm} A.~W.,  {Wuyts} S.,    {van der Wel} A.,
  2012, \apj, 744, 85

\bibitem[\protect\citeauthoryear{{Marinacci}, {Pakmor} \&
  {Springel}}{{Marinacci} et~al.}{2013}]{Marinacci13}
{Marinacci} F.,  {Pakmor} R.,    {Springel} V.,  2013, ArXiv e-prints

\bibitem[\protect\citeauthoryear{{Martin}}{{Martin}}{2005}]{Martin05}
{Martin} C.~L.,  2005, \apj, 621, 227

\bibitem[\protect\citeauthoryear{{Matteucci} \& {Francois}}{{Matteucci} \&
  {Francois}}{1989}]{Matteucci89}
{Matteucci} F.,  {Francois} P.,  1989, \mnras, 239, 885

\bibitem[\protect\citeauthoryear{{Maulbetsch}, {Avila-Reese}, {Col{\'{\i}}n},
  {Gottl{\"o}ber}, {Khalatyan} \& {Steinmetz}}{{Maulbetsch}
  et~al.}{2007}]{Maulbetsch07}
{Maulbetsch} C.,  {Avila-Reese} V.,  {Col{\'{\i}}n} P.,  {Gottl{\"o}ber} S.,
  {Khalatyan} A.,    {Steinmetz} M.,  2007, \apj, 654, 53

\bibitem[\protect\citeauthoryear{{McClure} \& {Racine}}{{McClure} \&
  {Racine}}{1969}]{McClure69}
{McClure} R.~D.,  {Racine} R.,  1969, \aj, 74, 1000

\bibitem[\protect\citeauthoryear{{McKee} \& {Ostriker}}{{McKee} \&
  {Ostriker}}{1977}]{McKee77}
{McKee} C.~F.,  {Ostriker} J.~P.,  1977, \apj, 218, 148

\bibitem[\protect\citeauthoryear{{Mehlert}, {Saglia}, {Bender} \&
  {Wegner}}{{Mehlert} et~al.}{2000}]{Mehlert00}
{Mehlert} D.,  {Saglia} R.~P.,  {Bender} R.,    {Wegner} G.,  2000, \aaps, 141,
  449

\bibitem[\protect\citeauthoryear{{Mehlert}, {Thomas}, {Saglia}, {Bender} \&
  {Wegner}}{{Mehlert} et~al.}{2003}]{Mehlert03}
{Mehlert} D.,  {Thomas} D.,  {Saglia} R.~P.,  {Bender} R.,    {Wegner} G.,
  2003, \aap, 407, 423

\bibitem[\protect\citeauthoryear{{Mihos}, {Harding}, {Rudick} \&
  {Feldmeier}}{{Mihos} et~al.}{2013}]{Mihos13}
{Mihos} J.~C.,  {Harding} P.,  {Rudick} C.~S.,    {Feldmeier} J.~J.,  2013,
  \apjl, 764, L20

\bibitem[\protect\citeauthoryear{{Moll{\'a}} \& {D{\'{\i}}az}}{{Moll{\'a}} \&
  {D{\'{\i}}az}}{2005}]{Molla05}
{Moll{\'a}} M.,  {D{\'{\i}}az} A.~I.,  2005, \mnras, 358, 521

\bibitem[\protect\citeauthoryear{{Molla}, {Ferrini} \& {Diaz}}{{Molla}
  et~al.}{1997}]{Molla97}
{Molla} M.,  {Ferrini} F.,    {Diaz} A.~I.,  1997, \apj, 475, 519

\bibitem[\protect\citeauthoryear{{Montes}, {Trujillo}, {Prieto} \&
  {Acosta-Pulido}}{{Montes} et~al.}{2014}]{Montes14}
{Montes} M.,  {Trujillo} I.,  {Prieto} M.~A.,    {Acosta-Pulido} J.~A.,  2014,
  \mnras, 439, 990

\bibitem[\protect\citeauthoryear{{Moster}, {Naab} \& {White}}{{Moster}
  et~al.}{2013}]{2013MNRAS.428.3121M}
{Moster} B.~P.,  {Naab} T.,    {White} S.~D.~M.,  2013, \mnras, 428, 3121

\bibitem[\protect\citeauthoryear{{Moster}, {Somerville}, {Maulbetsch}, {van den
  Bosch}, {Macci{\`o}}, {Naab} \& {Oser}}{{Moster} et~al.}{2010}]{Moster10}
{Moster} B.~P.,  {Somerville} R.~S.,  {Maulbetsch} C.,  {van den Bosch} F.~C.,
  {Macci{\`o}} A.~V.,  {Naab} T.,    {Oser} L.,  2010, \apj, 710, 903

\bibitem[\protect\citeauthoryear{{Murray}, {Quataert} \& {Thompson}}{{Murray}
  et~al.}{2005}]{Murray05}
{Murray} N.,  {Quataert} E.,    {Thompson} T.~A.,  2005, \apj, 618, 569

\bibitem[\protect\citeauthoryear{{Naab}, {Johansson} \& {Ostriker}}{{Naab}
  et~al.}{2009}]{2009ApJ...699L.178N}
{Naab} T.,  {Johansson} P.~H.,    {Ostriker} J.~P.,  2009, \apjl, 699, L178

\bibitem[\protect\citeauthoryear{{Naab}, {Johansson}, {Ostriker} \&
  {Efstathiou}}{{Naab} et~al.}{2007}]{Naab07}
{Naab} T.,  {Johansson} P.~H.,  {Ostriker} J.~P.,    {Efstathiou} G.,  2007,
  \apj, 658, 710

\bibitem[\protect\citeauthoryear{{Naab}, {Khochfar} \& {Burkert}}{{Naab}
  et~al.}{2006}]{2006ApJ...636L..81N}
{Naab} T.,  {Khochfar} S.,    {Burkert} A.,  2006, \apjl, 636, L81

\bibitem[\protect\citeauthoryear{{Naab}, {Oser}, {Emsellem}, {Cappellari},
  {Krajnovic}, {McDermid}, {Alatalo}, {Bayet}, {Blitz} \& {Bois}}{{Naab}
  et~al.}{2013}]{2013arXiv1311.0284N}
{Naab} T.,  {Oser} L.,  {Emsellem} E.,  {Cappellari} M.,  {Krajnovic} D.,
  {McDermid} R.~M.,  {Alatalo} K.,  {Bayet} E.,  {Blitz} L.,    {Bois} M.,
  2013, ArXiv e-prints

\bibitem[\protect\citeauthoryear{{Naab} \& {Ostriker}}{{Naab} \&
  {Ostriker}}{2006}]{Naab06}
{Naab} T.,  {Ostriker} J.~P.,  2006, \mnras, 366, 899

\bibitem[\protect\citeauthoryear{{Naab}}{{Naab}}{2014}]{Naab14}
{Naab} T. e.~a.,  2014, \mnras, 444, 3357

\bibitem[\protect\citeauthoryear{{Navarro-Gonz{\'a}lez}, {Ricciardelli},
  {Quilis} \& {Vazdekis}}{{Navarro-Gonz{\'a}lez} et~al.}{2013}]{Navarro13}
{Navarro-Gonz{\'a}lez} J.,  {Ricciardelli} E.,  {Quilis} V.,    {Vazdekis} A.,
  2013, \mnras, 436, 3507

\bibitem[\protect\citeauthoryear{{Newman}, {Ellis}, {Bundy} \& {Treu}}{{Newman}
  et~al.}{2012}]{2012ApJ...746..162N}
{Newman} A.~B.,  {Ellis} R.~S.,  {Bundy} K.,    {Treu} T.,  2012, \apj, 746,
  162

\bibitem[\protect\citeauthoryear{{Nipoti}, {Treu} \& {Bolton}}{{Nipoti}
  et~al.}{2009}]{2009ApJ...703.1531N}
{Nipoti} C.,  {Treu} T.,    {Bolton} A.~S.,  2009, \apj, 703, 1531

\bibitem[\protect\citeauthoryear{{Nipoti}, {Treu}, {Leauthaud}, {Bundy},
  {Newman} \& {Auger}}{{Nipoti} et~al.}{2012}]{2012MNRAS.422.1714N}
{Nipoti} C.,  {Treu} T.,  {Leauthaud} A.,  {Bundy} K.,  {Newman} A.~B.,
  {Auger} M.~W.,  2012, \mnras, 422, 1714

\bibitem[\protect\citeauthoryear{{Oogi} \& {Habe}}{{Oogi} \&
  {Habe}}{2013}]{2013MNRAS.428..641O}
{Oogi} T.,  {Habe} A.,  2013, \mnras, 428, 641

\bibitem[\protect\citeauthoryear{{Oppenheimer} \& {Dav{\'e}}}{{Oppenheimer} \&
  {Dav{\'e}}}{2006}]{Oppenheimer06}
{Oppenheimer} B.~D.,  {Dav{\'e}} R.,  2006, \mnras, 373, 1265

\bibitem[\protect\citeauthoryear{{Oppenheimer} \& {Dav{\'e}}}{{Oppenheimer} \&
  {Dav{\'e}}}{2008}]{Oppenheimer08}
{Oppenheimer} B.~D.,  {Dav{\'e}} R.,  2008, \mnras, 387, 577

\bibitem[\protect\citeauthoryear{{Oppenheimer}, {Dav{\'e}}, {Kere{\v s}},
  {Fardal}, {Katz}, {Kollmeier} \& {Weinberg}}{{Oppenheimer}
  et~al.}{2010}]{Oppenheimer10}
{Oppenheimer} B.~D.,  {Dav{\'e}} R.,  {Kere{\v s}} D.,  {Fardal} M.,  {Katz}
  N.,  {Kollmeier} J.~A.,    {Weinberg} D.~H.,  2010, \mnras, 406, 2325

\bibitem[\protect\citeauthoryear{{Oser}, {Naab}, {Ostriker} \&
  {Johansson}}{{Oser} et~al.}{2012}]{Oser12}
{Oser} L.,  {Naab} T.,  {Ostriker} J.~P.,    {Johansson} P.~H.,  2012, \apj,
  744, 63

\bibitem[\protect\citeauthoryear{{Oser}, {Ostriker}, {Naab}, {Johansson} \&
  {Burkert}}{{Oser} et~al.}{2010}]{Oser10}
{Oser} L.,  {Ostriker} J.~P.,  {Naab} T.,  {Johansson} P.~H.,    {Burkert} A.,
  2010, \apj, 725, 2312

\bibitem[\protect\citeauthoryear{{Pasquali}, {Gallazzi}, {Fontanot}, {van den
  Bosch}, {De Lucia}, {Mo} \& {Yang}}{{Pasquali} et~al.}{2010}]{Pasquali10}
{Pasquali} A.,  {Gallazzi} A.,  {Fontanot} F.,  {van den Bosch} F.~C.,  {De
  Lucia} G.,  {Mo} H.~J.,    {Yang} X.,  2010, \mnras, 407, 937

\bibitem[\protect\citeauthoryear{{Pastorello}, {Forbes}, {Foster}, {Brodie},
  {Usher}, {Romanowsky}, {Strader} \& {Arnold}}{{Pastorello}
  et~al.}{2014}]{Pastorello14}
{Pastorello} N.,  {Forbes} D.~A.,  {Foster} C.,  {Brodie} J.~P.,  {Usher} C.,
  {Romanowsky} A.~J.,  {Strader} J.,    {Arnold} J.~A.,  2014, ArXiv e-prints

\bibitem[\protect\citeauthoryear{{Patel}, {van Dokkum}, {Franx}, {Quadri},
  {Muzzin}, {Marchesini}, {Williams}, {Holden} \& {Stefanon}}{{Patel}
  et~al.}{2013}]{Patel13}
{Patel} S.~G.,  {van Dokkum} P.~G.,  {Franx} M.,  {Quadri} R.~F.,  {Muzzin} A.,
   {Marchesini} D.,  {Williams} R.~J.,  {Holden} B.~P.,    {Stefanon} M.,
  2013, \apj, 766, 15

\bibitem[\protect\citeauthoryear{{Peletier}, {Davies} \&
  {Illingworth}}{{Peletier} et~al.}{1990}]{Peletier90}
{Peletier} R.,  {Davies} R.~L.,    {Illingworth} G.,  1990, {Structure and
  colour gradients in elliptical galaxies.}.
pp 267--269

\bibitem[\protect\citeauthoryear{{Pilkington}, {Few}, {Gibson}, {Calura},
  {Michel-Dansac}, {Thacker}, {Moll{\'a}}, {Matteucci}, {Rahimi}, {Kawata},
  {Kobayashi}, {Brook}, {Stinson}, {Couchman}, {Bailin} \&
  {Wadsley}}{{Pilkington} et~al.}{2012}]{Pilkington12}
{Pilkington} K.,  {Few} C.~G.,  {Gibson} B.~K.,  {Calura} F.,  {Michel-Dansac}
  L.,  {Thacker} R.~J.,  {Moll{\'a}} M.,  {Matteucci} F.,  {Rahimi} A.,
  {Kawata} D.,  {Kobayashi} C.,  {Brook} C.~B.,  {Stinson} G.~S.,  {Couchman}
  H.~M.~P.,  {Bailin} J.,    {Wadsley} J.,  2012, \aap, 540, A56

\bibitem[\protect\citeauthoryear{{Prantzos} \& {Silk}}{{Prantzos} \&
  {Silk}}{1998}]{Prantzos98}
{Prantzos} N.,  {Silk} J.,  1998, \apj, 507, 229

\bibitem[\protect\citeauthoryear{{Queyrel}, {Contini}, {Kissler-Patig},
  {Epinat}, {Amram}, {Garilli}, {Le F{\`e}vre}, {Moultaka}, {Paioro}, {Tasca},
  {Tresse}, {Vergani}, {L{\'o}pez-Sanjuan} \& {Perez-Montero}}{{Queyrel}
  et~al.}{2012}]{Queyrel12}
{Queyrel} J.,  {Contini} T.,  {Kissler-Patig} M.,  {Epinat} B.,  {Amram} P.,
  {Garilli} B.,  {Le F{\`e}vre} O.,  {Moultaka} J.,  {Paioro} L.,  {Tasca} L.,
  {Tresse} L.,  {Vergani} D.,  {L{\'o}pez-Sanjuan} C.,    {Perez-Montero} E.,
  2012, \aap, 539, A93

\bibitem[\protect\citeauthoryear{{Raskutti}, {Greene} \& {Murphy}}{{Raskutti}
  et~al.}{2014}]{Raskutti14}
{Raskutti} S.,  {Greene} J.,    {Murphy} J.,  2014, ArXiv e-prints

\bibitem[\protect\citeauthoryear{{Rawle}, {Smith} \& {Lucey}}{{Rawle}
  et~al.}{2010}]{Rawle10}
{Rawle} T.~D.,  {Smith} R.~J.,    {Lucey} J.~R.,  2010, \mnras, 401, 852

\bibitem[\protect\citeauthoryear{{Rix}, {Carollo} \& {Freeman}}{{Rix}
  et~al.}{1999}]{Rix99}
{Rix} H.-W.,  {Carollo} C.~M.,    {Freeman} K.,  1999, \apjl, 513, L25

\bibitem[\protect\citeauthoryear{{Rupke}, {Veilleux} \& {Sanders}}{{Rupke}
  et~al.}{2005}]{Rupke05}
{Rupke} D.~S.,  {Veilleux} S.,    {Sanders} D.~B.,  2005, \apjs, 160, 115

\bibitem[\protect\citeauthoryear{{Rupke}, {Kewley} \& {Barnes}}{{Rupke}
  et~al.}{2010}]{Rupke10}
{Rupke} D.~S.~N.,  {Kewley} L.~J.,    {Barnes} J.~E.,  2010, \apjl, 710, L156

\bibitem[\protect\citeauthoryear{{S{\'a}nchez-Bl{\'a}zquez}, {Forbes},
  {Strader}, {Brodie} \& {Proctor}}{{S{\'a}nchez-Bl{\'a}zquez}
  et~al.}{2007}]{Sanchez07}
{S{\'a}nchez-Bl{\'a}zquez} P.,  {Forbes} D.~A.,  {Strader} J.,  {Brodie} J.,
  {Proctor} R.,  2007, \mnras, 377, 759

\bibitem[\protect\citeauthoryear{{Saracco}, {Gargiulo} \&
  {Longhetti}}{{Saracco} et~al.}{2012}]{2012MNRAS.422.3107S}
{Saracco} P.,  {Gargiulo} A.,    {Longhetti} M.,  2012, \mnras, 422, 3107

\bibitem[\protect\citeauthoryear{{Scannapieco} \& {Bildsten}}{{Scannapieco} \&
  {Bildsten}}{2005}]{Scannapieco05}
{Scannapieco} E.,  {Bildsten} L.,  2005, \apjl, 629, L85

\bibitem[\protect\citeauthoryear{{Serra}}{{Serra}}{2014}]{Serra14}
{Serra} P. e.~a.,  2014, ArXiv e-prints

\bibitem[\protect\citeauthoryear{{Spergel}, {Verde}, {Peiris}, {Komatsu},
  {Nolta}, {Bennett}, {Halpern}, {Hinshaw}, {Jarosik}, {Kogut}, {Limon},
  {Meyer}, {Page}, {Tucker}, {Weiland}, {Wollack} \& {Wright}}{{Spergel}
  et~al.}{2003}]{Spergel03}
{Spergel} D.~N.,  {Verde} L.,  {Peiris} H.~V.,  {Komatsu} E.,  {Nolta} M.~R.,
  {Bennett} C.~L.,  {Halpern} M.,  {Hinshaw} G.,  {Jarosik} N.,  {Kogut} A.,
  {Limon} M.,  {Meyer} S.~S.,  {Page} L.,  {Tucker} G.~S.,  {Weiland} J.~L.,
  {Wollack} E.,    {Wright} E.~L.,  2003, \apjs, 148, 175

\bibitem[\protect\citeauthoryear{{Spitoni} \& {Matteucci}}{{Spitoni} \&
  {Matteucci}}{2011}]{Spitoni11}
{Spitoni} E.,  {Matteucci} F.,  2011, \aap, 531, A72

\bibitem[\protect\citeauthoryear{{Spolaor}, {Kobayashi}, {Forbes}, {Couch} \&
  {Hau}}{{Spolaor} et~al.}{2010}]{Spolaor10}
{Spolaor} M.,  {Kobayashi} C.,  {Forbes} D.~A.,  {Couch} W.~J.,    {Hau}
  G.~K.~T.,  2010, \mnras, 408, 272

\bibitem[\protect\citeauthoryear{{Springel} \& {Hernquist}}{{Springel} \&
  {Hernquist}}{2003}]{Springel03}
{Springel} V.,  {Hernquist} L.,  2003, \mnras, 339, 289

\bibitem[\protect\citeauthoryear{{Springel}, {White}, {Jenkins}, {Frenk},
  {Yoshida}, {Gao}, {Navarro}, {Thacker}, {Croton}, {Helly}, {Peacock}, {Cole},
  {Thomas}, {Couchman}, {Evrard}, {Colberg} \& {Pearce}}{{Springel}
  et~al.}{2005}]{Springel05}
{Springel} V.,  {White} S.~D.~M.,  {Jenkins} A.,  {Frenk} C.~S.,  {Yoshida} N.,
   {Gao} L.,  {Navarro} J.,  {Thacker} R.,  {Croton} D.,  {Helly} J.,
  {Peacock} J.~A.,  {Cole} S.,  {Thomas} P.,  {Couchman} H.,  {Evrard} A.,
  {Colberg} J.,    {Pearce} F.,  2005, \nat, 435, 629

\bibitem[\protect\citeauthoryear{{Springel}, {Yoshida} \& {White}}{{Springel}
  et~al.}{2001}]{Springel01GAD}
{Springel} V.,  {Yoshida} N.,    {White} S.~D.~M.,  2001, New Astronomy, 6, 79

\bibitem[\protect\citeauthoryear{{Steinmetz} \& {Mueller}}{{Steinmetz} \&
  {Mueller}}{1994}]{Steinmetz94}
{Steinmetz} M.,  {Mueller} E.,  1994, \aap, 281, L97

\bibitem[\protect\citeauthoryear{{Suh}, {Jeong}, {Oh}, {Yi}, {Ferreras} \&
  {Schawinski}}{{Suh} et~al.}{2010}]{Suh10}
{Suh} H.,  {Jeong} H.,  {Oh} K.,  {Yi} S.~K.,  {Ferreras} I.,    {Schawinski}
  K.,  2010, \apjs, 187, 374

\bibitem[\protect\citeauthoryear{{Sutherland} \& {Dopita}}{{Sutherland} \&
  {Dopita}}{1993}]{Sutherland93}
{Sutherland} R.~S.,  {Dopita} M.~A.,  1993, \apjs, 88, 253

\bibitem[\protect\citeauthoryear{{Szomoru}, {Franx} \& {van Dokkum}}{{Szomoru}
  et~al.}{2012}]{2012ApJ...749..121S}
{Szomoru} D.,  {Franx} M.,    {van Dokkum} P.~G.,  2012, \apj, 749, 121

\bibitem[\protect\citeauthoryear{{Thomas}, {Maraston}, {Bender} \& {Mendes de
  Oliveira}}{{Thomas} et~al.}{2005}]{Thomas05}
{Thomas} D.,  {Maraston} C.,  {Bender} R.,    {Mendes de Oliveira} C.,  2005,
  \apj, 621, 673

\bibitem[\protect\citeauthoryear{{Thomas}, {Maraston}, {Schawinski}, {Sarzi} \&
  {Silk}}{{Thomas} et~al.}{2010}]{Thomas10}
{Thomas} D.,  {Maraston} C.,  {Schawinski} K.,  {Sarzi} M.,    {Silk} J.,
  2010, \mnras, 404, 1775

\bibitem[\protect\citeauthoryear{{Toft}, {van Dokkum}, {Franx}, {Labbe},
  {F{\"o}rster Schreiber}, {Wuyts}, {Webb}, {Rudnick}, {Zirm}, {Kriek}, {van
  der Werf}, {Blakeslee}, {Illingworth}, {Rix}, {Papovich} \&
  {Moorwood}}{{Toft} et~al.}{2007}]{2007ApJ...671..285T}
{Toft} S.,  {van Dokkum} P.,  {Franx} M.,  {Labbe} I.,  {F{\"o}rster Schreiber}
  N.~M.,  {Wuyts} S.,  {Webb} T.,  {Rudnick} G.,  {Zirm} A.,  {Kriek} M.,  {van
  der Werf} P.,  {Blakeslee} J.~P.,  {Illingworth} G.,  {Rix} H.-W.,
  {Papovich} C.,    {Moorwood} A.,  2007, \apj, 671, 285

\bibitem[\protect\citeauthoryear{{Tortora}, {Napolitano}, {Cardone},
  {Capaccioli}, {Jetzer} \& {Molinaro}}{{Tortora} et~al.}{2010}]{Tortora10}
{Tortora} C.,  {Napolitano} N.~R.,  {Cardone} V.~F.,  {Capaccioli} M.,
  {Jetzer} P.,    {Molinaro} R.,  2010, \mnras, 407, 144

\bibitem[\protect\citeauthoryear{{Tran}, {van Dokkum}, {Franx}, {Illingworth},
  {Kelson} \& {Schreiber}}{{Tran} et~al.}{2005}]{2005ApJ...627L..25T}
{Tran} K.-V.~H.,  {van Dokkum} P.,  {Franx} M.,  {Illingworth} G.~D.,  {Kelson}
  D.~D.,    {Schreiber} N.~M.~F.,  2005, \apjl, 627, L25

\bibitem[\protect\citeauthoryear{{Trujillo}, {F{\"o}rster Schreiber},
  {Rudnick}, {Barden}, {Franx}, {Rix}, {Caldwell}, {McIntosh}, {Toft},
  {H{\"a}ussler}, {Zirm}, {van Dokkum} \& {Labb{\'e}}}{{Trujillo}
  et~al.}{2006}]{2006ApJ...650...18T}
{Trujillo} I.,  {F{\"o}rster Schreiber} N.~M.,  {Rudnick} G.,  {Barden} M.,
  {Franx} M.,  {Rix} H.-W.,  {Caldwell} J.~A.~R.,  {McIntosh} D.~H.,  {Toft}
  S.,  {H{\"a}ussler} B.,  {Zirm} A.,  {van Dokkum} P.~G.,    {Labb{\'e}} I.,
  2006, \apj, 650, 18

\bibitem[\protect\citeauthoryear{{van de Sande}, {Kriek}, {Franx}, {van
  Dokkum}, {Bezanson}, {Whitaker}, {Brammer}, {Labb{\'e}}, {Groot} \&
  {Kaper}}{{van de Sande} et~al.}{2011}]{2011ApJ...736L...9V}
{van de Sande} J.,  {Kriek} M.,  {Franx} M.,  {van Dokkum} P.~G.,  {Bezanson}
  R.,  {Whitaker} K.~E.,  {Brammer} G.,  {Labb{\'e}} I.,  {Groot} P.~J.,
  {Kaper} L.,  2011, \apjl, 736, L9

\bibitem[\protect\citeauthoryear{{van der Wel}, {Franx}, {van Dokkum}, {Rix},
  {Illingworth} \& {Rosati}}{{van der Wel} et~al.}{2005}]{2005ApJ...631..145V}
{van der Wel} A.,  {Franx} M.,  {van Dokkum} P.~G.,  {Rix} H.-W.,
  {Illingworth} G.~D.,    {Rosati} P.,  2005, \apj, 631, 145

\bibitem[\protect\citeauthoryear{{van der Wel}, {Holden}, {Zirm}, {Franx},
  {Rettura}, {Illingworth} \& {Ford}}{{van der Wel}
  et~al.}{2008}]{2008ApJ...688...48V}
{van der Wel} A.,  {Holden} B.~P.,  {Zirm} A.~W.,  {Franx} M.,  {Rettura} A.,
  {Illingworth} G.~D.,    {Ford} H.~C.,  2008, \apj, 688, 48

\bibitem[\protect\citeauthoryear{{van Dokkum}}{{van
  Dokkum}}{2005}]{2005AJ....130.2647V}
{van Dokkum} P.~G.,  2005, \aj, 130, 2647

\bibitem[\protect\citeauthoryear{{van Dokkum}, {Franx}, {Kriek}, {Holden},
  {Illingworth}, {Magee}, {Bouwens}, {Marchesini}, {Quadri}, {Rudnick},
  {Taylor} \& {Toft}}{{van Dokkum} et~al.}{2008}]{2008ApJ...677L...5V}
{van Dokkum} P.~G.,  {Franx} M.,  {Kriek} M.,  {Holden} B.,  {Illingworth}
  G.~D.,  {Magee} D.,  {Bouwens} R.,  {Marchesini} D.,  {Quadri} R.,  {Rudnick}
  G.,  {Taylor} E.~N.,    {Toft} S.,  2008, \apjl, 677, L5

\bibitem[\protect\citeauthoryear{{van Dokkum}, {Whitaker}, {Brammer}, {Franx},
  {Kriek}, {Labb{\'e}}, {Marchesini}, {Quadri}, {Bezanson}, {Illingworth},
  {Muzzin}, {Rudnick}, {Tal} \& {Wake}}{{van Dokkum}
  et~al.}{2010}]{2010ApJ...709.1018V}
{van Dokkum} P.~G.,  {Whitaker} K.~E.,  {Brammer} G.,  {Franx} M.,  {Kriek} M.,
   {Labb{\'e}} I.,  {Marchesini} D.,  {Quadri} R.,  {Bezanson} R.,
  {Illingworth} G.~D.,  {Muzzin} A.,  {Rudnick} G.,  {Tal} T.,    {Wake} D.,
  2010, \apj, 709, 1018

\bibitem[\protect\citeauthoryear{{van Zee}, {Salzer}, {Haynes}, {O'Donoghue} \&
  {Balonek}}{{van Zee} et~al.}{1998}]{vanZee98}
{van Zee} L.,  {Salzer} J.~J.,  {Haynes} M.~P.,  {O'Donoghue} A.~A.,
  {Balonek} T.~J.,  1998, \aj, 116, 2805

\bibitem[\protect\citeauthoryear{{Vazdekis}, {Cenarro}, {Gorgas}, {Cardiel} \&
  {Peletier}}{{Vazdekis} et~al.}{2003}]{Vazdekis03}
{Vazdekis} A.,  {Cenarro} A.~J.,  {Gorgas} J.,  {Cardiel} N.,    {Peletier}
  R.~F.,  2003, \mnras, 340, 1317

\bibitem[\protect\citeauthoryear{{Vila-Costas} \& {Edmunds}}{{Vila-Costas} \&
  {Edmunds}}{1992}]{Vila92}
{Vila-Costas} M.~B.,  {Edmunds} M.~G.,  1992, \mnras, 259, 121

\bibitem[\protect\citeauthoryear{{Villumsen}}{{Villumsen}}{1983}]{1983MNRAS.20%
4..219V}
{Villumsen} J.~V.,  1983, \mnras, 204, 219

\bibitem[\protect\citeauthoryear{{Weijmans}, {Cappellari}, {Bacon}, {de Zeeuw},
  {Emsellem}, {Falc{\'o}n-Barroso}, {Kuntschner}, {McDermid}, {van den Bosch}
  \& {van de Ven}}{{Weijmans} et~al.}{2009}]{Weijmans09}
{Weijmans} A.-M.,  {Cappellari} M.,  {Bacon} R.,  {de Zeeuw} P.~T.,  {Emsellem}
  E.,  {Falc{\'o}n-Barroso} J.,  {Kuntschner} H.,  {McDermid} R.~M.,  {van den
  Bosch} R.~C.~E.,    {van de Ven} G.,  2009, \mnras, 398, 561

\bibitem[\protect\citeauthoryear{{Weijmans}}{{Weijmans}}{2014}]{Weijmans14}
{Weijmans} A.-M. e.~a.,  2014, \mnras, 444, 3340

\bibitem[\protect\citeauthoryear{{White}}{{White}}{1978}]{1978MNRAS.184..185W}
{White} S.~D.~M.,  1978, \mnras, 184, 185

\bibitem[\protect\citeauthoryear{{White}}{{White}}{1979}]{1979MNRAS.189..831W}
{White} S.~D.~M.,  1979, \mnras, 189, 831

\bibitem[\protect\citeauthoryear{{White}}{{White}}{1980}]{1980MNRAS.191P...1W}
{White} S.~D.~M.,  1980, \mnras, 191, 1P

\bibitem[\protect\citeauthoryear{{White} \& {Rees}}{{White} \&
  {Rees}}{1978}]{White78}
{White} S.~D.~M.,  {Rees} M.~J.,  1978, \mnras, 183, 341

\bibitem[\protect\citeauthoryear{{Wu}, {Shao}, {Mo}, {Xia} \& {Deng}}{{Wu}
  et~al.}{2005}]{Wu05}
{Wu} H.,  {Shao} Z.,  {Mo} H.~J.,  {Xia} X.,    {Deng} Z.,  2005, \apj, 622,
  244

\bibitem[\protect\citeauthoryear{{Wyse} \& {Silk}}{{Wyse} \&
  {Silk}}{1989}]{Wyse89}
{Wyse} R.~F.~G.,  {Silk} J.,  1989, \apj, 339, 700

\bibitem[\protect\citeauthoryear{{Yang}, {Mo}, {van den Bosch}, {Bonaca}, {Li},
  {Lu}, {Lu} \& {Lu}}{{Yang} et~al.}{2013}]{Yang13}
{Yang} X.,  {Mo} H.~J.,  {van den Bosch} F.~C.,  {Bonaca} A.,  {Li} S.,  {Lu}
  Y.,  {Lu} Y.,    {Lu} Z.,  2013, ArXiv e-prints

\bibitem[\protect\citeauthoryear{{Zaritsky}, {Kennicutt} Jr. \&
  {Huchra}}{{Zaritsky} et~al.}{1994}]{Zaritsky94}
{Zaritsky} D.,  {Kennicutt} Jr. R.~C.,    {Huchra} J.~P.,  1994, \apj, 420, 87

\bibitem[\protect\citeauthoryear{{Zhang} \& {Thompson}}{{Zhang} \&
  {Thompson}}{2012}]{Zhang12}
{Zhang} D.,  {Thompson} T.~A.,  2012, \mnras, 424, 1170

\end{thebibliography}

\label{lastpage}

\end{document}